\documentclass[3p]{elsarticle}
\usepackage{amssymb}
\usepackage{amsfonts}
\usepackage{float}
\usepackage{tikz}
\usetikzlibrary{external}
\usepackage{graphicx}
\biboptions{numbers,sort&compress}
\usepackage{subfigure}
\usepackage{amsmath}
\usepackage{amsthm}
\usepackage{soul,color}
\usepackage{multirow}
\usepackage{booktabs}
\usepackage{tabularx}
\usepackage{adjustbox}
\usepackage{siunitx}
\usepackage{comment}
\sethlcolor{yellow}
\usepackage{xcolor}
\usepackage{url}
\usepackage{pgfplots}
\usepgfplotslibrary{patchplots}
\pgfplotsset{compat=newest, samples=015} 

\usepackage{etoolbox}
\makeatletter
\patchcmd{\ps@pprintTitle}
  {Preprint submitted to}
  {Accepted manuscript for}
  {}{}
\makeatother

\usetikzlibrary{shapes.geometric, arrows, calc, positioning}
\tikzstyle{startstop} = [rectangle, thick, rounded corners=2.5mm, minimum width=2cm, minimum height=5mm,text centered, draw=black]
\tikzstyle{io} = [trapezium, thick, trapezium left angle=70, trapezium right angle=110, text width=3.75cm, minimum height=0.5cm, text centered, draw=black]
\tikzstyle{process} = [rectangle, thick, minimum width=2.5cm, text width=4cm, minimum height=0.5cm, text centered, draw=black]
\tikzstyle{block} = [rectangle, thick, minimum width=0.5cm, minimum height=1cm, text centered, draw=black]
\tikzstyle{support} = [coordinate, join=by fuzzy]
\tikzstyle{decision} = [diamond, thick, minimum width=3cm, minimum height=0.25cm, text centered, draw=black]
\tikzstyle{dottedbox} = [rectangle, dotted, thick, minimum width=2.5cm, text width=2.8cm, minimum height=0.5cm, text centered, draw=black]
\tikzstyle{arrow} = [thick,->,>=stealth]
\tikzstyle{dottedarrow} = [thick, dotted,->,>=stealth]
\tikzstyle{blockx} = [draw, rectangle, dashed,
    minimum height=8.3cm, minimum width=13.2cm]
\tikzstyle{blocky} = [draw, rectangle, dashed,
    minimum height=21.3cm, minimum width=13.2cm]
\tikzstyle{blockz} = [draw, rectangle, dashed,
    minimum height=12.8cm, minimum width=10cm]
\tikzstyle{build} = [draw, fill=red!10, rectangle, 
   minimum height=2.5em, minimum width=10em]
\tikzstyle{blk} = [draw, fill=green!10, rectangle, 
    minimum height=2em, minimum width=3em]
\tikzstyle{blka} = [draw, fill=blue!10, rectangle, 
    minimum height=2em, minimum width=3em]
\tikzstyle{nn} = [coordinate]

\journal{Mechanical Systems and Signal Processing}

\begin{document}

\newtheorem{theorem}{Theorem}
\newtheorem{assumption}{Assumption}
\newtheorem{remark}{Remark}

\begin{frontmatter}

\title{A tuning algorithm for a sliding mode controller of buildings with ATMD \tnoteref{t1}}
\tnotetext[t1]{\copyright $<2021>$. This manuscript version is made available under the CC-BY-NC-ND 4.0 license \url{http://creativecommons.org/licenses/by-nc-nd/4.0/}. Published version: \url{https://doi.org/10.1016/j.ymssp.2020.107539}}

\author[First]{Antonio Concha}
\ead{aconcha@ucol.mx}
\author[Second]{Suresh Thenozhi}
\ead{suresh@uaq.mx}
\author[Third]{Ramón J. Betancourt}
\ead{rjimenez@ucol.mx}
\author[Fourth]{S. K. Gadi\corref{cor1}}
\ead{Research@SKGadi.com}

\address[First]{Facultad de Ingeniería Mecánica y Eléctrica, Universidad de Colima, Coquimatlán, Colima 28400, México}
\address[Second]{Facultad de Ingeniería, Universidad Autónoma de Querétaro, Santiago de Querétaro, Querétaro 76010, México}
\address[Third]{Facultad de Ingeniería Electromecánica, Universidad de Colima, Manzanillo, Colima 28860, México}
\address[Fourth]{Facultad de Ingeniería Mecánica y Eléctrica, Universidad Autónoma de Coahuila, Torreón, Coahuila 27276, México}
\cortext[cor1]{Corresponding author.}

\begin{abstract}
This paper proposes an automatic tuning algorithm for a sliding mode controller (SMC) based on the Ackermann's formula, that attenuates the structural vibrations of a seismically excited building equipped with an Active Tuned Mass Damper (ATMD) mounted on its top floor. The switching gain and sliding surface of the SMC are designed through the proposed tuning algorithm to suppress the structural vibrations by minimizing either the top floor displacement or the control force applied to the ATMD. Moreover, the tuning algorithm selects the SMC parameters to guarantee the following closed-loop characteristics: 1) the transient responses of the structure and the ATMD are sufficiently fast and damped; and 2) the control force, as well as the displacements and velocities of the building and ATMD are within acceptable limits under the frequency band of the earthquake excitation. The proposed SMC shows robustness against the unmodeled dynamics such as the friction of the damper. Experimental results on a reduced scale structure permits demonstrating the efficiency of the tuning algorithm for the SMC, which is compared with the traditional Linear Quadratic Regulator (LQR) and with the Optimal Sliding Mode Controller (OSMC).   
\end{abstract}


\end{frontmatter}


\section*{Highlights}
\begin{itemize}
    \item A tuning algorithm for a sliding mode vibration control of buildings is proposed.
    \item The tuned controller can minimize the top floor displacement or the control force. 
    \item A desired transient and frequency response of the closed-loop system is guaranteed. 
    \item Experimental results verify the effectiveness of the proposed tuning algorithm.
\end{itemize}

\section{Introduction}
Buildings can be subject to external forces such as earthquakes or winds, which can damage them \cite{msspreview, chen2019stochastic}. To protect the building against these natural hazards, a passive \cite{kim2019peak, pandey2019compliant}, or semi-active \cite{amjadian2019seismic, nguyen2018modeling}, or active \cite{paul2018} device can be added to the structure. A well-established approach is to employ Mass Dampers (MDs). Its passive version, known as Tuned Mass Damper (TMD), is composed of a moving mass attached to a spring and a viscous damper. The active version of the MD is named as Active Mass Damper (AMD), which is constructed by coupling an actuator to the moving mass. By adding an actuator to the TMD results in a hybrid device, that is called Active Tuned Mass Damper (ATMD) \cite{chesne2019}. Active vibration control techniques using AMD or ATMD have been of great interest in recent years, due to their ability to provide higher vibration attenuation than the TMD. 

Linear controllers are by far the most widely applied techniques for active vibration control using AMD or ATMD. They include the classical proportional-derivative (PD) or the proportional-integral-derivative (PID) controllers \cite{paul2018,kayabekir2020}; acceleration feedback regulators \cite{yang2017,talib2019}; state-feedback with variable gain \cite{younespour2015}; full-state-feedback control using displacement, velocity, and acceleration of the structure \cite{Chang:1995,Ankireddi:1996,xu2008modeling,li2010optimum}; Linear Quadratic Regulator (LQR) using the knowledge of the seismic excitation \cite{Ricciardelli:2003,li2019multi} or without it \cite{lei2020}; Linear Quadratic Gaussian (LQG) \cite{allaoua2019lqg}; feedforward and feedback optimal tracking controller (FFOTC) \cite{zhang2016}; and robust controllers like $H_{2}$, $H_{\infty}$ \cite{Spencer:1994,santos2007active} or $H_{2}/H_{\infty}$ \cite{xu2018}. On the other hand, intelligent techniques have also been applied for active control of structures. Yang et al. \cite{yang2006} designed a neural-network for system identification and vibration control of a structure with AMD. Thenozhi and Yu \cite{thenozhi2015} proposed Fuzzy PD/PID controllers for structures with friction uncertainty. Genetic algorithms were applied in \cite{li2000multi,banaei2020} for optimization of structural active control laws.  

An alternative to the aforementioned linear and intelligent control techniques is the sliding mode controller (SMC). It is widely accepted for structural control and is designed to drive the trajectories of the closed-loop system to a sliding surface, that is a linear combination of the system state, and it can include nonlinear or fractional terms \cite{fei2020fuzzy,fei2019exp}. Once that the trajectories have reached the sliding surface, the closed-loop system is robust against disturbances and unmodeled dynamics. Yang et al. \cite{Yang:1995} presented a continuous SMC based on a saturation function for seismically excited structures, where the authors showed with numerical simulations that this controller avoids the undesirable chattering effect. Adhikari et al. \cite{Adhikari:1997} designed a SMC based on the theory of compensators to prevent a large response in the building due to its interaction with the ATMD. On the other hand, Wang et al. \cite{Wang:2007} developed a fuzzy SMC that uses a Mamdani inference method to determine the behavior of the closed-loop system in the sliding mode. Moreover, Li et al. \cite{li2019adaptive} proposed a model reference SMC for a building with an ATMD at its top floor, where the reference model is the structure coupled to the TMD. Soleymani et al. \cite{soleymani2018} designed a SMC for a building modeled through a second-order reduced model, and they consider time delays in the control force applied to an AMD coupled to the structure. In addition, Mamat et al. \cite{mamat2020} presented an adaptive nonsingular terminal SMC employed in a three-story building equipped with an ATMD, that was simulated in Matlab/Simulink. Finally, Khatibinia et al. \cite{khatibinia2020optimal} developed an Optimal SMC, denoted as OSMC, that was obtained by transforming the model of a building with ATMD into the regular form. 

This article proposes an algorithm to automatically tune the sliding variable and switching gain of a SMC based on the Ackermann's formula, that is used for vibration control of a building containing an ATMD on its top floor. Considering that the first mode of the building response is dominant during the earthquake, the structure equipped with the ATMD is modeled as a fourth-order system. It is shown that the closed-loop system in the sliding mode is reduced to a third-order system, whose displacement of the dominant mode of the structure and that of the ATMD damper, as well as its control force behave as the outputs of dominant second-order filters, whose input is the seismic excitation. These filters have the advantage that are easier to design than the fourth-order filters presented in \cite{yang2017}. The aim of the proposed tuning algorithm is to design these dominant second-order filters to: 
\begin{itemize}
\renewcommand{\labelitemi}{$\bullet$}
\item minimize the displacement of the top floor of the structure as much as possible, or to minimize the control force applied to the ATMD while offering a great attenuation of this displacement.
\item produce sufficiently fast and damped transient responses of the ATMD and building.
\item guarantee that the Root Mean Square (RMS) values of the ATMD control force, displacements and velocities of both building and damper are within acceptable limits in the frequency band of the earthquake excitation. 
\end{itemize} 

Unlike the SMC techniques presented in \cite{Yang:1995,Adhikari:1997,Wang:2007,li2019adaptive,soleymani2018,mamat2020,khatibinia2020optimal}, this article proves that the seismic excitation signal is not a coupled disturbance, whose effect on the controller and on the movements of the ATMD and building is analyzed. Moreover, in contrast to \cite{Adhikari:1997}, that uses a compensator to filter out undesirable ATMD responses, the present work uses the dominant second-order filters to remove these undesirable responses, which are automatically designed using the proposed tuning algorithm. Thus, large responses in the building due to the movements of the ATMD are avoided.

The rest of this paper is organized as follows. Section \ref{sec:model_edif} introduces the mathematical model of a building equipped with an ATMD. The SMC designed with the Ackermann's formula is presented in Section \ref{sec:smc}. The desired transient and frequency responses of the closed-loop structure are described in Section \ref{sec:FT}. The proposed algorithm for tuning the sliding variable and switching gain of the SMC is explained in Section \ref{sec:tuning_alg}. Section \ref{sec:sim_exp} demonstrates the effectiveness of the proposed tuning algorithm in both simulations and experiments. Finally, Section \ref{sec:conclusions} gives the conclusions of this article.

\section{Mathematical model of a building with an ATMD}
\label{sec:model_edif}
\begin{figure}[ht]
\begin{center}
	\includegraphics[height=6cm]{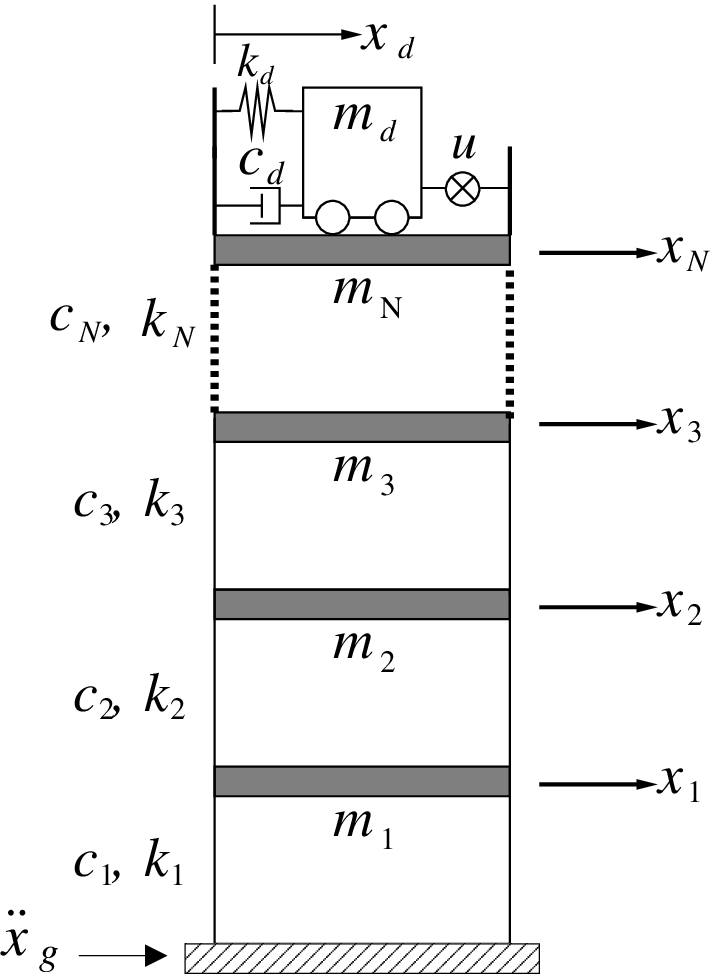}
	\caption{Building structure equipped with an ATMD mounted on its top floor.}
	\label{fig:1}
\end{center}
\end{figure}

Consider a $N$-story building that has an ATMD installed at its top floor, as shown in Figure \ref{fig:1}. The behavior of this system is described by \citep{Chopra:2001,Yu:2016}:
\begin{eqnarray}
\mathbf{M}(\ddot{\mathbf{x}}(t)+\mathbf{l} \ddot{x}_{g}(t))+\mathbf{C}\dot{\mathbf{x}}(t)+\mathbf{Kx}(t)&=&-\mathbf{\Gamma}F(t) \label{e:1} \\
m_{d}(\ddot{x}_{n}(t)+\ddot{x}_{g}(t)+\ddot{x}_{d}(t))&=&F(t)  \label{e:3} \\
F(t)&=&u(t)-k_{d}x_{d}(t)-c_{d}\dot{x}_{d}(t)-f(\dot{x}_{d}(t)) \label{e:fd}
\end{eqnarray}
where $\mathbf{M},\mathbf{C},\mathbf{K}\in \mathbb{R}^{N\times N}$ are the mass, stiffness, and damping matrices, respectively. Term $\mathbf{M}$ is a diagonal matrix composed by the floor masses $m_{i}$, $i=1,2,\ldots,N$. Moreover, $\mathbf{C}$ and $\mathbf{K}$ are tridiagonal matrices that contain the damping $c_{i}$ and stiffness $k_{i}$ coefficients between the $i$th and the $(i-1)$th floors, as depicted in Figure \ref{fig:1}. Furthermore, the term $\ddot{x}_{g}$ is the earthquake acceleration and vector $\mathbf{x}$ is given by $\mathbf{x}=[x_{1},\ x_{2}, \ldots, x_{N}]^{\mathrm{T}}$, where $x_{i}$ represents the displacement of the $i$th floor relative to the ground. Variable $x_{d}$ is the relative displacement of the ATMD with respect to the top story, and the terms $m_{d}$, $k_{d}$, $c_{d}$, and $f(\dot{x}_{d}(t))$ are the mass, stiffness, damping, and non-linear friction of the ATMD, respectively. Variable $F$ is the net force acting upon the ATMD. In addition, signal $u(t)$ is the control force applied to the damper, term $\mathbf{l}\in \mathbb{R}^{N\times 1}$ is an unity vector, and $\mathbf{\Gamma} \in \mathbb{R}^{N \times 1}$ defines the localization of the ATMD and is represented as:
\begin{equation}
\mathbf{\Gamma}=[0,0,\ldots,0,1]^{\mathrm{T}}
\end{equation} 

Since the first mode of vibration is dominant during an earthquake, the building model (\ref{e:1}) can be approximated as \cite{Chang:1995}:
\begin{equation}\label{e:2}
\begin{split}
m_{0}\ddot{x}_{0}(t)+c_{0}\dot{x}_{0}(t)+k_{0}x_{0}(t)&=-\beta_{0} m_{0}\ddot{x}_{g}(t)-F(t) \\
m_{d}(\ddot{x}_{0}(t)+\ddot{x}_{g}(t)+\ddot{x}_{d}(t))&=F(t)
\end{split}
\end{equation}
Signal $x_{0}$ represents the displacement of the dominant mode, which is given by
\begin{equation}
x_{0}(t)=\dfrac{\mathbf{\phi_{0}}^{\mathrm{T}}\mathbf{M}\mathbf{x}(t)}{\mathbf{\phi_{0}}^{\mathrm{T}}\mathbf{M}\mathbf{\phi_{0}}}
\end{equation}
where $\mathbf{\phi_{0}}\in \mathbb{R}^{N\times 1}$ is the the first mode, that is scaled to satisfy
\begin{equation}\label{e:ev}
\mathbf{\phi_{0}}^{\mathrm{T}}\mathbf{\Gamma}=1
\end{equation}
This equality, in turn, produces the approximation \cite{wu:1998}
\begin{equation}\label{e:aprox}
x_{N}(t)\approx x_{0}(t)
\end{equation}
Parameters $m_{0}, c_{0}, k_{0}$, and $\beta_{0}$ are the mass, damping, stiffness, and participation factor of the dominant mode, respectively. They are defined as \citep{Chopra:2001}:
\begin{equation}\label{e:pm}
m_{0}=\mathbf{\phi_{0}}^{\mathrm{T}}\mathbf{M}\mathbf{\phi_{0}}, \quad c_{0}=\mathbf{\phi_{0}}^{\mathrm{T}}\mathbf{C}\mathbf{\phi_{0}}, \quad 
k_{0}=\mathbf{\phi_{0}}^{\mathrm{T}}\mathbf{K}\mathbf{\phi_{0}}, \quad 
\beta_{0}=\dfrac{\mathbf{\phi_{0}}^{\mathrm{T}}\mathbf{M}\mathbf{l}}{\mathbf{\phi_{0}}^{\mathrm{T}}\mathbf{M}\mathbf{\phi_{0}}}
\end{equation}
Moreover, the natural frequency of the first mode is given by
\begin{equation}\label{e:w0}
\omega_{0}=\sqrt{\dfrac{k_{0}}{m_{0}}}
\end{equation}

Using the approximation (\ref{e:aprox}) and substituting $F(t)$ of (\ref{e:fd}) into (\ref{e:2}) yields:
\begin{equation}\label{e:4}
\begin{split}
\ddot{x}_{d}=&\left(\dfrac{m_{0}+m_{d}}{m_{0}m_{d}}\right)\left( u-c_{d}\dot{x}_{d} -f(\dot{x}_{d})-k_{d}x_{d}\right) +\dfrac{k_{0}}{m_{0}}x_{N}+\dfrac{c_{0}}{m_{0}}\dot{x}_{N}+\ddot{x}_{g}(\beta_{0}-1)\\
\ddot{x}_{N}=&-\dfrac{c_{0}}{m_{0}}\dot{x}_{N}-\dfrac{k_{0}}{m_{0}}x_{N}-\beta_{0}\ddot{x}_{g}+\dfrac{c_{d}}{m_{0}}\dot{x}_{d}+\dfrac{f(\dot{x}_{d})}{m_{0}}+\dfrac{k_{d}}{m_{0}}x_{d}-\dfrac{1}{m_{0}}u
\end{split}
\end{equation}
where argument $t$ has been omitted in the time dependent signals. 

Defining the following state variables
\begin{equation}\label{e:5}
z_{1}=x_{d}, \quad z_{2}=x_{N}, \quad z_{3}=\dot{x}_{d}, \quad z_{4}=\dot{x}_{N}
\end{equation}
allows rewriting system (\ref{e:4}) in the following state-space representation:
\begin{equation}\label{e:6}
\begin{split}
\overbrace{\left[\begin{array}{c} \dot{z}_{1} \\ \dot{z}_{2} \\ \dot{z}_{3} \\ \dot{z}_{4} \end{array}\right]}^{\mathbf{\dot{z}}}=&
\overbrace{\left[\begin{array}{cccc}
0 & 0 & 1 & 0 \\
0 & 0 & 0 & 1  \\
-\dfrac{k_{d}(m_{0}+m_{d})}{m_{0}m_{d}}  & \dfrac{k_{0}}{m_{0}} & -\dfrac{c_{d}(m_{0}+m_{d})}{m_{0}m_{d}} & \dfrac{c_{0}}{m_{0}} \\
\dfrac{k_{d}}{m_{0}}  & -\dfrac{k_{0}}{m_{0}} & \dfrac{c_{d}}{m_{0}} & -\dfrac{c_{0}}{m_{0}}  
\end{array} \right]}^{\mathbf{A}}
\overbrace{\left[\begin{array}{c} z_{1} \\ z_{2} \\ z_{3} \\ z_{4} \end{array}\right]}^{\mathbf{z}} \\ 
+&
\overbrace{\left[\begin{array}{c} 0 \\ 0 \\ \dfrac{(m_{0}+m_{d})}{m_{0}m_{d}} \\ -\dfrac{1}{m_{0}} \end{array}\right]}^{\mathbf{B}}(u-f(z_{3}))+\overbrace{\left[\begin{array}{c} 0 \\ 0 \\ \beta_{0}-1 \\ -\beta_{0} \end{array}\right]}^{\mathbf{D}}\ddot{x}_{g} 
\end{split}
\end{equation}

\section{Sliding mode control of the structure}
\label{sec:smc}
This section presents a SMC design based on the Ackermann's formula for vibration attenuation of seismically excited buildings. For this purpose, let us first define a full-state feedback controller $u_{a}$ for system (\ref{e:6}), such that the closed-loop eigenvalues $\lambda_{1}$, $\lambda_{2}$, $\lambda_{3}$, and $\lambda_{4}$ are placed in a desired location in the $s$-plane. The control law $u_{a}$ is defined as
\begin{equation}
    u_{a}=-\mathbf{k}^{\mathrm{T}}\mathbf{z} \label{e:ua}
\end{equation}
According to the Ackermann's formula \cite{ackermann:1985}, the feedback gain vector $\mathbf{k}^{\mathrm{T}} \in \mathbb{R}^{1 \times 4}$ can be computed as
\begin{equation}
\mathbf{k}^{\mathrm{T}}=\mathbf{e}^{\mathrm{T}}P(\mathbf{A}) 
\label{e:8}
\end{equation}
with
\begin{align}  
\mathbf{e}^{\mathrm{T}}=&[0,0,0,1][\mathbf{B,AB,A^{2}B,A^{3}B}]^{-1} \in \mathbb{R}^{1 \times 4} \label{e:9}\\
P(\lambda)=&(\lambda-\lambda_{1})(\lambda-\lambda_{2})(\lambda-\lambda_{3})(\lambda-\lambda_{4})  \label{e:Pc}
\end{align}  

In order to design the SMC for the system (\ref{e:6}), let us assume the following.
\begin{assumption}
\label{as2}
The eigenvalue $\lambda_{4}$ in (\ref{e:Pc}) is real and negative. 
\end{assumption}

\begin{assumption}
\label{as1}
Bounds $\delta$ and $\varpi$, corresponding to the earthquake acceleration $\ddot{x}_{g}$ and the non-linear friction $f(z_{3})$, respectively, are known and satisfy     
\begin{align}
|\ddot{x}_{g}|\leq & \delta \label{e:boe}\\ |f(z_{3})|\leq &  \varpi \label{e:bo}
\end{align} 
where the bound $\delta$ can be determined from the historical records of the ground acceleration in the region where the building is located, and the bound $\varpi$ can be determined using friction estimation techniques, such as the ones based on the Linear Extended State Observer (LESO) or the Least Squares Method (LSM). Through the LESO, the friction $f(z_{3})$ is considered as a new state of a high gain observer that is estimated using the input $u$ and the displacement $z_{1}$ of the ATMD \cite{wang2016}. On the other side, with the Least Squares method, the friction $f(z_{3})$ is contained in linear regression model, that is identified using the signals $u$ and $z_{1}$ \cite{garrido2013}. 
\end{assumption}

The following theorem presents the SMC based on the Ackermann's formula (\ref{e:8}) and analyses the behavior of the resulting closed-loop system in the sliding mode at the plane $\sigma=\boldsymbol{\eta}^{\mathrm{T}}\mathbf{z}=0$, where $\boldsymbol{\eta}^{\mathrm{T}}\in \mathbb{R}^{1\times 4}$ is a constant vector that will be automatically calculated through the proposed methodology outlined in Section \ref{sec:tuning_alg}.

\begin{theorem}
Let us consider the building structure in (\ref{e:6}), that is equipped with an ATMD, whose control force $u$ is provided by the following SMC
\begin{equation}\label{e:cmd}
u=-M_{0}\mathrm{sign} (\sigma)
\end{equation}
where $\sigma=\boldsymbol{\eta}^{\mathrm{T}}\mathbf{z}$ and $M_{0}>0$ are called sliding variable and switching gain, respectively. The vector $\boldsymbol{\eta}^{\mathrm{T}}$ is given by  
\begin{equation}\label{e:11}
\boldsymbol{\eta}^{\mathrm{T}}=[\eta_{1},\ \eta_{2}, \ \eta_{3}, \ \eta_{4}]=\mathbf{e}^{\mathrm{T}}P_{1}(\mathbf{A}), \qquad 
P_{1}(\lambda)=(\lambda-\lambda_{1})(\lambda-\lambda_{2})(\lambda-\lambda_{3}) 
\end{equation}
where $\lambda_{1}$, $\lambda_{2}$, and $\lambda_{3}$ are the desired closed-loop eigenvalues; moreover, parameter $M_{0}$ satisfies 
\begin{equation}\label{ineq3}
M_{0} > \varpi+h_{0}
\end{equation}
where $h_{0}$ is a constant such that 
\begin{equation}\label{e:ineqs}
 \qquad |u_{a}+\alpha_{1}\ddot{x}_{g}| \leq h_{0}
\end{equation}
with $u_{a}$ given in (\ref{e:ua}) and $\alpha_{1}$ defined as
\begin{equation}
\alpha_{1}=\beta_{0}(\eta_{4}-\eta_{3})+\eta_{3}
\end{equation}
Then, the trajectories of closed-loop system reach the plane $\sigma=\boldsymbol{\eta}^{\mathrm{T}}\mathbf{z}=0$ in a finite time $t_{\sigma}$, and they are confined in this plane for $t\geq t_{\sigma}$, where $t_{\sigma}\leq \sigma(0)/(M_{0}-[\varpi+h_{0}])$. Furthermore, when this plane is reached, the fourth-order dynamic system (\ref{e:6}) is reduced to the following third-order system 
\begin{equation}\label{e:12}
\mathbf{\dot{z}^{*}}=\mathbf{A}_{1}\mathbf{z}^{*}+\mathbf{B}_{1}\ddot{x}_{g}
\end{equation}
where $\mathbf{A}_{1}\in \mathbb{R}^{3\times3}$ is a matrix containing the eigenvalues $\lambda_{1}$, $\lambda_{2}$, and $\lambda_{3}$; vector $\mathbf{z}^{*}\in \mathbb{R}^{3\times1}$ is given by
\begin{equation}\label{e:14}
\mathbf{z^{*}}=[z_{1},z_{2},z_{3}]^{\mathrm{T}}
\end{equation}
and 
\begin{equation}\label{e:12a}
\mathbf{B}_{1}=\left(\begin{array}{c} 0 \\ 0 \\ \alpha_{2}  \end{array}\right) \quad \hbox{with} \quad \alpha_{2} =(\beta_{0}-1)+\dfrac{\alpha_{1}(m_{0}+m_{d})}{m_{0}m_{d}} 
\end{equation}
\end{theorem}
\begin{proof}
First, note that vector $\boldsymbol{\eta}$ in (\ref{e:11}) satisfies the following equalities \cite{Ackermann:1998}:
\begin{equation}\label{e:18}
\boldsymbol{\eta}^{\mathrm{T}}\mathbf{B}=1, \qquad \boldsymbol{\eta}^{\mathrm{T}}\mathbf{A}^{*}=\lambda_{4}\boldsymbol{\eta}^{\mathrm{T}}
\end{equation} 
where $\mathbf{A}^{*}=\mathbf{A}-\mathbf{B}\mathbf{k}^{\mathrm{T}}$. 
Adding and subtracting the term $\mathbf{B}u_{a}$ to (\ref{e:6}) yields 
\begin{equation}\label{e:13}
\dot{\mathbf{z}}=(\mathbf{A}-\mathbf{B}\mathbf{k^{\mathrm{T}}})\mathbf{z}+\mathbf{B}\left[u-u_{a}-f(z_{3})\right]+\mathbf{D}\ddot{x}_{g}
\end{equation}
The system (\ref{e:13}) is transformed into a new set of equations using the following state transformation 
\begin{equation}\label{e:15}
\mathbf{w}=\left[\begin{array}{c} \mathbf{z}^{*} \\ \sigma \end{array}\right]=\overbrace{\left[\begin{array}{cc} \mathbf{I}_{3\times3} & \mathbf{0}_{3\times1} \\ \qquad \qquad \boldsymbol{\eta}^{\mathrm{T}} \end{array}\right]}^{\mathbf{T}_{1}}\mathbf{z}=\mathbf{T}_{1}\mathbf{z}
\end{equation}
where $\mathbf{T}_{1}$ is an invertible matrix. The dynamics of the transformed system $\mathbf{\dot{w}=T}\dot{\mathbf{z}}$ is given by
\begin{align}
\mathbf{\dot{w}}=&\mathbf{T}_{1}\left( (\mathbf{A}-\mathbf{B}\mathbf{k}^{\mathrm{T}})\mathbf{z}+\mathbf{B}\left[u-u_{a}-f(z_{3})\right]+\mathbf{D}\ddot{x}_{g} \right) \nonumber \\
=&\mathbf{T}_{1}(\mathbf{A}-\mathbf{B}\mathbf{k}^{\mathrm{T}})\mathbf{T}_{1}^{-1}\mathbf{w}+\mathbf{T}_{1}\mathbf{B}[u-u_{a}-f(z_{3})]+\mathbf{TD}\ddot{x}_{g} \label{e:22}
\end{align}

Using the equalities in (\ref{e:18}) allows deducing the following structure of matrices $\mathbf{T}_{1}(\mathbf{A}-\mathbf{B}\mathbf{k}^{\mathrm{T}})\mathbf{T}_{1}^{-1}$ and  $\mathbf{T}_{1}\mathbf{B}$ in (\ref{e:22})
\begin{equation}\label{e:20}
\mathbf{T}_{1}(\mathbf{A}-\mathbf{B}\mathbf{k}^{\mathrm{T}})\mathbf{T}_{1}^{-1}=\left[\begin{array}{cc} \mathbf{A}_{1} & \mathbf{a}^{*} \\ \mathbf{0}_{1\times 3} & \lambda_{4} \end{array}\right], \qquad \mathbf{T}_{1}\mathbf{B}=\left[\begin{array}{c} \mathbf{b}^{*} \\ 1\end{array}\right] \quad \hbox{with} \quad 
\mathbf{b}^{*}=\left(\begin{array}{c} 0 \\0 \\ \dfrac{(m_{0}+m_{d})}{m_{0}m_{d}}  \end{array} \right)
\end{equation}
where $\mathbf{a}^{*} \in \mathbb{R}^{3\times 1}$, and $\mathbf{A}_{1}$ is a matrix whose eigenvalues are equal to the closed-loop eigenvalues $\lambda_{1}$, $\lambda_{2}$ and $\lambda_{3}$. Finally, vector $\mathbf{TD}$ in (\ref{e:22}) is given by 
\begin{equation}\label{e:21}
\mathbf{TD}=\left[\begin{array}{c}   \mathbf{d}_{1} \\ -\alpha_{1} \end{array}\right] \quad \hbox{with} \quad
\mathbf{d}_{1}=\left(\begin{array}{c} 0 \\0 \\ \beta_{0}-1  \end{array} \right)
\end{equation}

Substituting equations (\ref{e:20}) and (\ref{e:21}) into (\ref{e:22}) produces 
\begin{align}
\dot{\mathbf{z}}^{*}&=\mathbf{A}_{1}\mathbf{z}^{*}+\mathbf{a}^{*}\sigma+\mathbf{b}^{*}[u-u_{a}-f(z_{3})]+\mathbf{d}_{1}\ddot{x}_{g} \label{e:16} \\
\dot{\sigma}&=\lambda_{4}\sigma+u-u_{a}-f(z_{3})-\alpha_{1}\ddot{x}_{g} \label{e:17}
\end{align}

The control law $u$ is designed such that the trajectories of the structural system (\ref{e:6}) converge to the plane $\sigma=0$ in finite time, which is ensured when $\dfrac{1}{2}\dfrac{d}{dt}\sigma^{2}=\sigma \dot{\sigma} <0$. The product $\sigma \dot{\sigma}$ is given by 
\begin{equation}
\sigma\dot{\sigma}=\sigma[\lambda_{4}\sigma+u-u_{a}-f(z_{3})-\alpha_{1}\ddot{x}_{g}]=\lambda_{4}\sigma^{2}+\sigma[u-u_{a}-f(z_{3})-\alpha_{1}\ddot{x}_{g}]
\end{equation}
Since $\lambda_{4}<0$, the last expression satisfies 
\begin{equation}\label{e:inq_e}
\sigma\dot{\sigma}\leq \sigma[u-u_{a}-f(z_{3})-\alpha_{1}\ddot{x}_{g}]
\end{equation}

Substituting the control law $u=-M_{0}\mbox{sign}(\sigma)$ into (\ref{e:inq_e}) and using the inequalities (\ref{e:bo}), (\ref{ineq3}) and (\ref{e:ineqs}), we get
\begin{equation}\label{e:ineq}
\sigma\dot{\sigma}\leq \sigma\left[-M_{0}\mbox{sign}(\sigma)-u_{a}-f(z_{3})-\alpha_{1}\ddot{x}_{g}\right]\leq |\sigma|\left[-M_{0}+\varpi+h_{0} \right]\leq -|\sigma|[M_{0}-(\varpi+h_{0})]<0
\end{equation}
Therefore, the trajectories of system (\ref{e:6}) reach the surface $\sigma=\boldsymbol{\eta}^{\mathrm{T}}\mathbf{z}=0$ in a finite time $t_{\sigma}$, and remain there for $t\geq t_{\sigma}$, where $t_{\sigma}$ is computed by integrating (\ref{e:ineq}) and its value is given by  
\begin{equation}
t_{\sigma}\leq \frac{\sigma(0)}{M_{0}-(\varpi+h_{0})}
\end{equation}

Finally, in order to determine the behavior of the closed-loop system in the surface $\sigma=0$, the next solution of $\dot{\sigma}=0$ with respect to $u$ 
\begin{equation}\label{e:24}
u=u_{a}+f(z_{3})+\alpha_{1}\ddot{x}_{g}
\end{equation}
is substituted into (\ref{e:16}) to produce the sliding motion equation (\ref{e:12}). 
\end{proof}

\begin{remark}
Since the SMC (\ref{e:cmd}) and the dynamics of the closed-loop system (\ref{e:12}) in the sliding mode do not depend of $\lambda_{4}$, this parameter can take any negative value, and it is used only for the stability analysis of the SMC. 
\end{remark}

\begin{remark}
Note that the building and the ATMD are at rest or in equilibrium before an earthquake, therefore, the initial condition $z_{i}(0)$ of variables $z_{i}$, $i=1,2,3,4$ is zero, i.e., $\mathbf{z}(0)=\boldsymbol{0}$, which implies that $\sigma(0)=\boldsymbol{\eta}^{\mathrm{T}}\mathbf{z}(0)=0$, and as an consequence $t_{\sigma}=0$. 
\end{remark}

\begin{remark}
Parameter $h_{0}$ in (\ref{e:ineqs}) is required to design the SMC. In section \ref{sec:obmo}, a methodology is proposed to compute this parameter using the frequency response of the closed-loop system in the sliding mode, and the knowledge of the bound $\delta$ in (\ref{e:boe}) of the earthquake acceleration $\ddot{x}_{g}$. 
\end{remark}

\section{Analysis of the closed-loop system at the sliding mode}
\label{sec:FT}
The closed-loop system dynamics in the sliding mode is described in the Laplace domain as
\begin{equation}\label{e:n1}
(s\mathbf{I}_{3\times3}-\mathbf{A}_{1})\mathbf{Z}^{*}(s)=\mathbf{B}_{1}\mathcal{L}[\ddot{x}_{g}(t)]
\end{equation}
where $\mathcal{L}$ is the Laplace transform operator and $\mathbf{Z}^{*}(s)=[Z_{1}(s),\ Z_{2}(s), \ Z_{3}(s)]^{\mathrm{T}}$.

Using (\ref{e:n1}), the transfer function $\mathbf{Z}^{*}(s)/\mathcal{L}[\ddot{x}_{g}(t)]$ can be expressed as
\begin{equation}\label{e:n2}
\dfrac{\mathbf{Z}^{*}(s)}{\mathcal{L}[\ddot{x}_{g}(t)]}=(s\mathbf{I}_{3\times3}-\mathbf{A}_{1})^{-1}\mathbf{B}_{1}
\end{equation}

The characteristic polynomial $P_{1}(s)$ in (\ref{e:11}) corresponding to (\ref{e:n2}) can be rewritten as 
\begin{equation}\label{e:ssp}
P_{1}(s)=(s-\lambda_{1})(s-\lambda_{2})(s-\lambda_{3})=(s^{2}+2\zeta \omega_{n}s+\omega_{n}^{2})(s-\lambda_{3})
\end{equation}
where $\zeta$ and $\omega_{n}$ are positive constants, which are called damping ratio and undamped natural frequency, respectively. 

Substituting $P_{1}(s)$ into (\ref{e:n2}) leads to the following transfer functions $Z_{1}(s)/\mathcal{L}[\ddot{x}_{g}(t)]$, $Z_{2}(s)/\mathcal{L}[\ddot{x}_{g}(t)]$ and $Z_{3}(s)/\mathcal{L}[\ddot{x}_{g}(t)]$:
\begin{align}
G_{1}(s)=\dfrac{Z_{1}(s)}{\mathcal{L}[\ddot{x}_{g}(t)]}=&\mathbf{C}_{z_1}(s\mathbf{I}_{3 \times 3}-\mathbf{A}_{1})^{-1}\mathbf{B}_{1}=\dfrac{\alpha_{2}\left(s-\psi_{1}\right)}{(s-\lambda_{3})(s^{2}+2\zeta \omega_{n}s+\omega_{n}^{2})} \label{e:n4} \\
G_{2}(s)=\dfrac{Z_{2}(s)}{\mathcal{L}[\ddot{x}_{g}(t)]}=& \mathbf{C}_{z_2}(s\mathbf{I}_{3 \times 3}-\mathbf{A}_{1})^{-1}\mathbf{B}_{1}=\dfrac{- \alpha_{2}\eta_{3}\left(s-\psi_{2} \right)}{\eta_{4}(s-\lambda_{3})(s^{2}+2\zeta \omega_{n}s+\omega_{n}^{2})} \label{e:n3} \\
G_{3}(s)=\dfrac{Z_{3}(s)}{\mathcal{L}[\ddot{x}_{g}(t)]}=& \mathbf{C}_{z_3}(s\mathbf{I}_{3 \times 3}-\mathbf{A}_{1})^{-1}\mathbf{B}_{1}=\dfrac{\alpha_{2}s\left(s-\psi_{1}\right)}{(s-\lambda_{3})(s^{2}+2\zeta \omega_{n}s+\omega_{n}^{2})} \label{e:n4s} 
\end{align}
where 
\begin{equation}\label{e:zero}
\psi_{1}=-\dfrac{\eta_{2}}{\eta_{4}}, \qquad \psi_{2}=-\dfrac{\eta_{1}}{\eta_{3}},
\end{equation}
\begin{equation}
\mathbf{C}_{z_1}=[1,0,0], \qquad \mathbf{C}_{z_2}=[0,1,0], \qquad \mathbf{C}_{z_3}=[0,0,1]
\end{equation}

In order to determine the transfer function $U(s)/\mathcal{L}[\ddot{x}_{g}(t)]$, consider that non-linear function $f(z_{3})$ in (\ref{e:24}) is zero. Then, the control signal $u$ in the sliding mode is given by:
\begin{eqnarray}
u&=&u_{a}+\alpha_{1}\ddot{x}_{g}=-\mathbf{k}^{\mathrm{T}}\mathbf{z}+\alpha_{1}\ddot{x}_{g}=-\mathbf{e}^{\mathrm{T}}P(\mathbf{A})\mathbf{z}+\alpha_{1}\ddot{x}_{g} \nonumber \\ 
&=&-\mathbf{e}^{\mathrm{T}}P_{1}(\mathbf{A})[\mathbf{A}-\mathbf{I}_{4\times 4}\lambda_{4}]\mathbf{z}+\alpha_{1}\ddot{x}_{g} \nonumber \\
&=&-\boldsymbol{\eta}^{\mathrm{T}}[\mathbf{A}-\mathbf{I}_{4\times 4}\lambda_{4}]\mathbf{z}+\alpha_{1}\ddot{x}_{g}
\label{e:24e}
\end{eqnarray}

Using the state transformation equation in (\ref{e:15}), the vector $\mathbf{z}$ in (\ref{e:24e}) can be written as follows  
\begin{equation}\label{e:24c}
\mathbf{z}=\mathbf{T}_{1}^{-1}\mathbf{w}=\mathbf{T}_{1}^{-1}\left[\begin{array}{c} \mathbf{z}^{*} \\ \sigma \end{array}\right]
\end{equation}
Since $\sigma=0$ in the sliding mode, the above expression becomes
\begin{equation}\label{e:24d}
\mathbf{z}=\mathbf{T}_{1}^{-1}\left[\begin{array}{c} \mathbf{z}^{*} \\ 0 \end{array}\right]
\end{equation}
Substituting (\ref{e:24d}) into (\ref{e:24e}) yields 
\begin{equation}\label{e:24f}
u=-\boldsymbol{\eta}^{\mathrm{T}}[\mathbf{A}-\mathbf{I}_{4\times4}\lambda_{4}]\mathbf{T}_{1}^{-1}\left[\begin{array}{c} \mathbf{z}^{*} \\ 0 \end{array}\right]   +\alpha_{1}\ddot{x}_{g}
\end{equation}
Let
\begin{equation}\label{e:24g}
\boldsymbol{\nu}=-\boldsymbol{\eta}^{\mathrm{T}}[\mathbf{A}-\mathbf{I}_{4\times 4}\lambda_{4}]\mathbf{T}_{1}^{-1}=-\boldsymbol{\eta}^{\mathrm{T}}\mathbf{A}\mathbf{T}_{1}^{-1}+\lambda_{4}\boldsymbol{\eta}^{\mathrm{T}}\mathbf{T}_{1}^{-1}=-\boldsymbol{\eta}^{\mathrm{T}}\mathbf{A}\mathbf{T}_{1}^{-1}+[0,0,0,\lambda_{4}]=[\boldsymbol{\nu_{1}}, \ \nu_{2}]
\end{equation}
where $\boldsymbol{\nu} \in \mathbb{R}^{1\times4}$, $\boldsymbol{\nu_{1}} \in \mathbb{R}^{1\times3}$, and $\nu_{2} \in \mathbb{R}$. Using this definition allows rewriting (\ref{e:24f}) as
\begin{equation}\label{e:26}
u= [\boldsymbol{\nu_{1}}, \ \nu_{2}]  \left[\begin{array}{c} \mathbf{z}^{*} \\ 0 \end{array}\right]   +\alpha_{1}\ddot{x}_{g}=\boldsymbol{\nu_{1}}\mathbf{z}^{*}+\alpha_{1}\ddot{x}_{g}
\end{equation}
Note that $\boldsymbol{\nu}_1$ is a constant vector and does not depend on $\lambda_{4}$. Applying the Laplace transform to (\ref{e:26}) and using (\ref{e:n2}) produces:
\begin{equation}\label{e:27}
U(s)=\boldsymbol{\nu_{1}}(s\mathbf{I}_{3\times3}-\mathbf{A}_{1})^{-1}\mathbf{B}_{1}\mathcal{L}[\ddot{x}_{g}(t)]+\alpha_{1}\mathcal{L}[\ddot{x}_{g}(t)]
\end{equation}
which can be rewritten as the following transfer function $U(s)/\mathcal{L}[\ddot{x}_{g}(t)$ 
\begin{equation}\label{e:g4}
G_{u}(s)=\dfrac{U(s)}{\mathcal{L}[\ddot{x}_{g}(t)]}=\boldsymbol{\nu_{1}}(s\mathbf{I}_{3\times3}-\mathbf{A}_{1})^{-1}\mathbf{B}_{1}+\alpha_{1}
\end{equation}

\subsection{Transient response for $z_{1}(t)$ and $z_{2}(t)$}
\label{sec:rtrf}

In practice, it is important to impose constraints to the transient behavior of the system trajectories, such that they are within a specified limit \cite{song2016finite}, otherwise the system can have faults or can even be damaged \cite{song2018parameter}. This section analyzes the transient responses of the damper and the top floor displacements $z_{1}(t)$ and $z_{2}(t)$, respectively. Unlike references \cite{song2016finite,song2018parameter}, that consider the transient response of a closed-loop system before its trajectories reach the sliding surface $\sigma=0$, this section presents the transient analysis of the closed-loop system in the sliding mode $\sigma=0$. This analysis will permit tuning the SMC to produce sufficiently fast and damped transient responses of $z_{1}(t)$ and $z_{2}(t)$ under sudden changes in the input excitation $\ddot{x}_{g}$. For this purpose, assume that this input is a step function, and consider the following assumption:
\begin{assumption}
\label{a1}
The damping ratio $\zeta$ of the characteristic polynomial $P_{1}(s)$ in (\ref{e:ssp}) satisfies $\zeta<1$, which implies that $P_{1}(s)$ has a real pole $\lambda_{3}$ and two complex conjugate poles $\lambda_{1}$ and $\lambda_{2}$, i.e., 
\begin{equation}\label{e:sspe}
P_{1}(s)=(s^{2}+2\zeta \omega_{n}s+\omega_{n}^{2})(s-\lambda_{3})=(s+\zeta \omega_{n}+j\omega_{d})(s+\zeta \omega_{n}-j\omega_{d})(s-\lambda_{3})
\end{equation} 
\begin{equation}\label{e:lamda}
\lambda_{1}=-\zeta \omega_{n}-j\omega_{d}, \qquad \lambda_{2}=-\zeta \omega_{n}+j\omega_{d}
\end{equation}
where $\omega_{d}=\omega_{n}\sqrt{1-\zeta^{2}}$ is the damped natural frequency.
\end{assumption}

\begin{center}
\begin{figure}[ht]
\centering
\subfigure[Effect of the zero $\psi_{i}$.]
	            {\includegraphics[height=3.7cm]{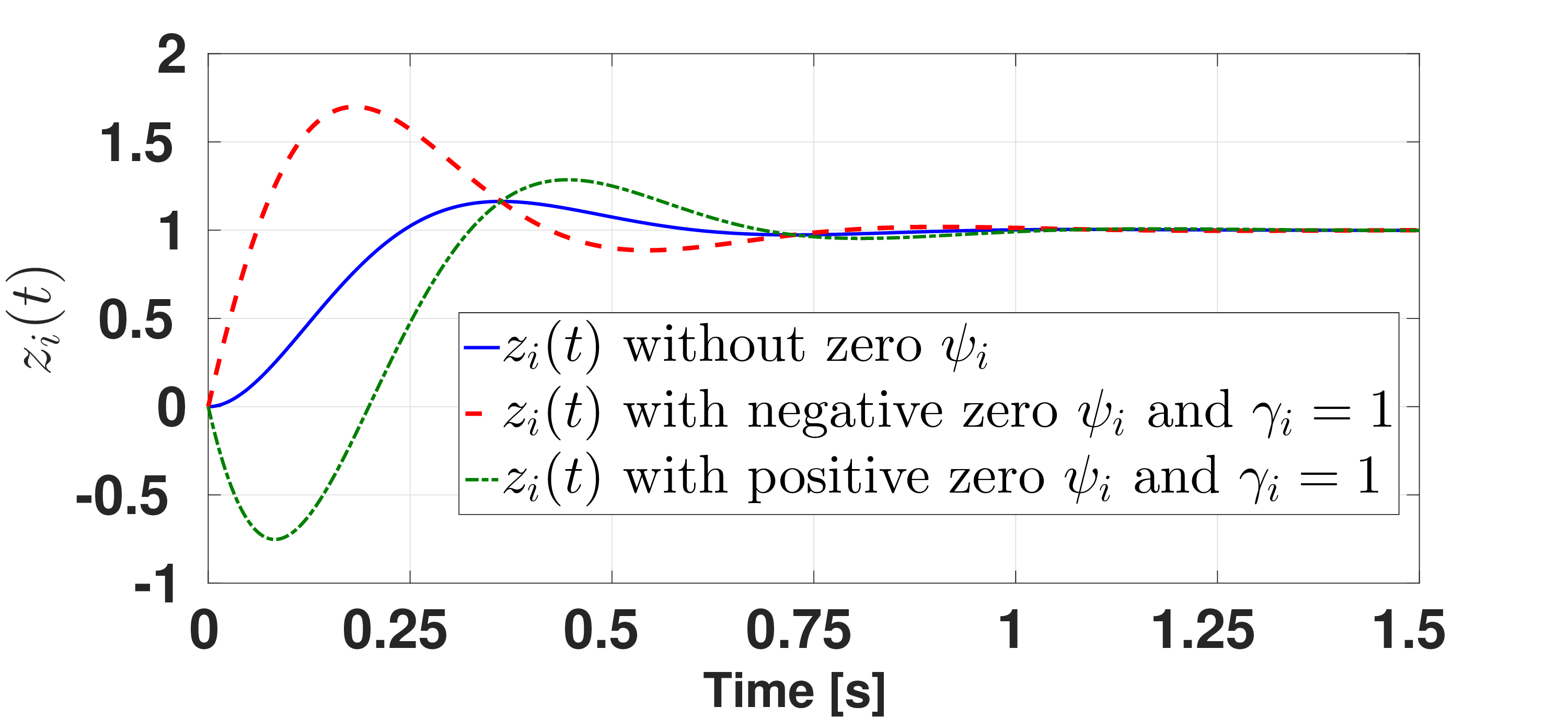}}
\subfigure[Effect of the pole $\lambda_{3}$.]
	                {\includegraphics[height=3.7cm]{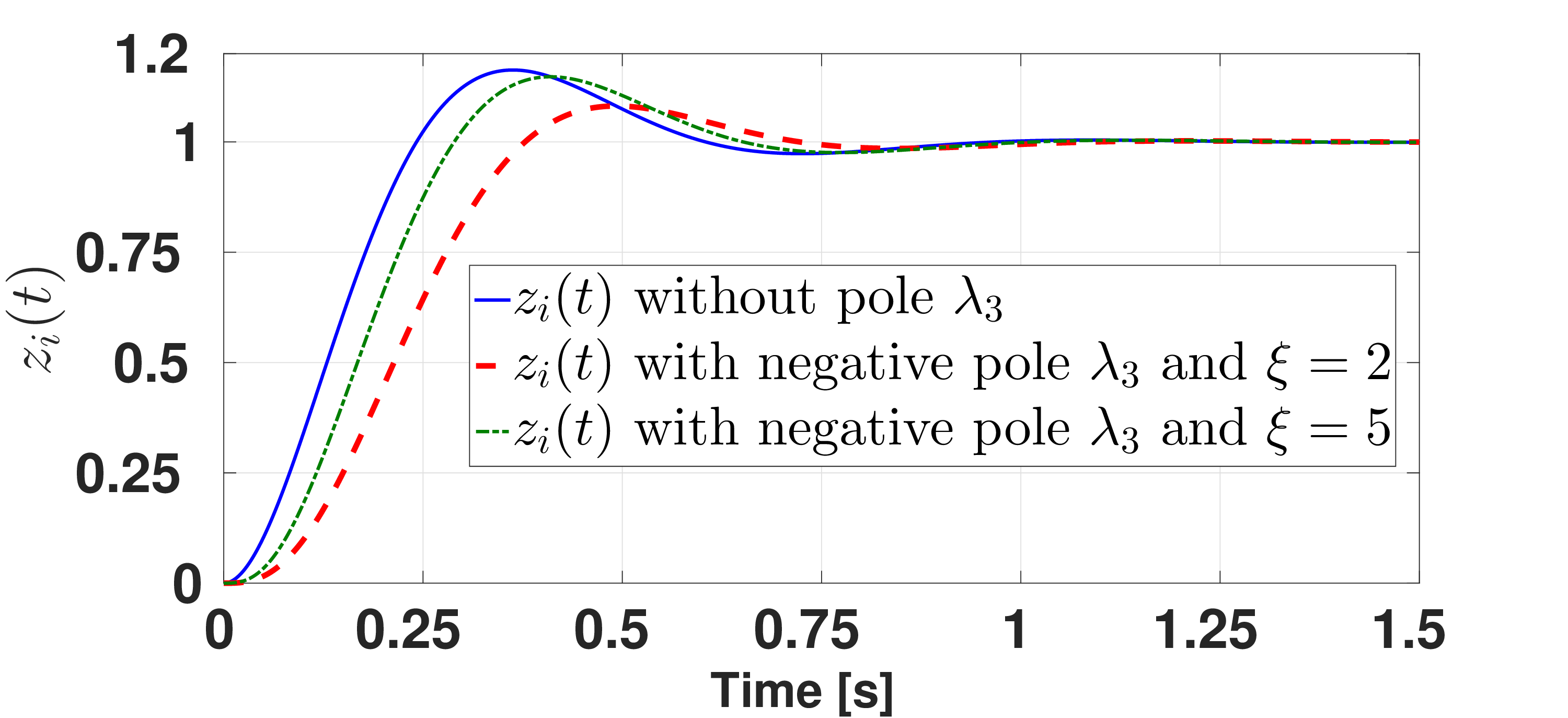}}	
\caption{Effect of the additional zero and pole in the transient response of a standard second-order system with $\omega_{n}=10$ rad/s and $\zeta=0.5$.}
\label{fig:2}
\end{figure}          
\end{center}

Taking into account Assumption \ref{a1}, the transient response of $z_{i}(t)$, $i=1,2$ can be considered as the response of a standard second-order system, that is affected by the additional pole $\lambda_{3}$ and zero $\psi_{i}$ of the transfer function $G_{i}(s)$. Thus, the response of $z_{i}(t)$, $i=1,2$ can be specified by means of its rise time $T_{r}$ and maximum overshoot $M_{P}$. The response $z_{i}(t)$ is affected by the pole $\lambda_{3}$ and zero $\psi_{i}$ \cite{Franklin:2015}, as described below:
\begin{itemize} 
\renewcommand{\labelitemi}{$\bullet$}
\item \textit{Effect of the additional zero}: The zero $\psi_{i}$ of $G_{i}(s)$ can either be positive or negative, and its effect on $z_{i}(t)$ is the following:
\begin{enumerate}
    \item If the zero $\psi_{i}$ is negative, then it has the effect of decreasing $T_{r}$ and increasing $M_{p}$ of the step response $z_{i}(t)$, as shown in Figure \ref{fig:2} (a). The increase of $M_{p}$ depends on the relation $\gamma_{i}=|\psi_{i}|/(\zeta \omega_{n})$. The smaller the relation $\gamma_{i}$, the larger the increment of $M_{p}$.
    \item If the zero $\psi_{i}$ is positive, then the transfer function $G_{i}(s)$ is non-minimum phase. In this case, the zero $\psi_{i}$ slightly increases $T_{r}$ and $M_{p}$ of the step response $z_{i}(t)$, but it produces an initial drop that appears at the beginning of the response, see Figure \ref{fig:2} (a). The peak of this drop depends on the relation $\gamma_{i}=\psi_{i}/(\zeta \omega_{n})$. The smaller the relation $\gamma_{i}$, the larger this peak will be.
\end{enumerate}

In conclusion, the zero $\psi_{i}$, either positive or negative, has little effect on the transient response of $z_{i}(t)$ for $\gamma_{i} \geq 3$, but as $\gamma_{i}$ decreases below 3, it has an increasing effect, especially when $\gamma_{i} < 1$. 
\item \textit{Effect of the additional pole}: The additional pole $\lambda_{3}<0$ of $G_{i}(s)$ tends to increase $T_{r}$ and to decrease $M_{p}$ of the step response $z_{i}(t)$, see Figure \ref{fig:2} (b). The percentage of $M_{p}$ is a function of $\xi=|\lambda_{3}|/(\zeta \omega_{n})$; the larger this relation, the smaller the percentage of $M_{p}$. The pole $\lambda_{3}$ has little effect for $\xi \geq 3$, otherwise it has an increasing effect.  
\end{itemize}

From the above analysis, it is clear that for certain values of the additional pole and zero, they have less effect on the transient response. In that case, the parameters $\zeta$ and $\omega_n$ of the dominant underdamped second-order system are tuned such that the transient responses of $z_{1}(t)$ and $z_{2}(t)$ are sufficiently fast and damped, and they do not have excessive overshoot. To this end, the tuning of the parameters $\zeta,\omega_n, \lambda_3, \psi_{1}$, and $\psi_{2}$ is carried out as follows.

\begin{itemize}
\renewcommand{\labelitemi}{$\bullet$}
    \item The damping ratio $\zeta$ of $P_{1}(s)$ will be tuned between the next interval 
\begin{equation}
\label{e:lz}
\zeta_{l} < \zeta < \zeta_{u}
\end{equation}
where the recommended values for the limits of the interval are $\zeta_{l}=$0.5 and $\zeta_{u}=$0.9. 

\item The natural frequency $\omega_n$ will be tuned based on the dominant mode frequency of the structure $\omega_{0}$. If $\omega_{n}$ is close to $\omega_{0}$, it could provoke abrupt movements of the ATMD that could excite the structure instead of protect it. On the other hand, if $\omega_{n}$ is smaller than $0.5 \omega_{0}$ could result in a slower transient response of the ATMD movements, hence an insignificant vibration attenuation. To avoid these problems, $\omega_{n}$ will be tuned between the following limits 
\begin{equation}
\label{e:lw}
\omega_{nl} < \omega_{n} < \omega_{nu}
\end{equation}
where $\omega_{nl}=0.5\omega_{0}$ and $\omega_{nu}=0.8\omega_{0}$. Moreover, the above choice helps to minimize the effect of the measurement noise and unmodeled dynamics in the closed-loop system. 

\item To neglect the effect of the pole $\lambda_{3}$ on the transient responses of $z_{1}(t)$ and $z_{2}(t)$, it is fixed to:
\begin{equation}
\label{e:lr}
\lambda_{3}=-3 \zeta \omega_{n}
\end{equation}
\item The zeros $\psi_{i}$, $i=1,2$ of $G_{i}(s)$ are selected based on the following conditions, such that they have the least possible impact on the responses $z_{1}$ and $z_{2}$. 
\begin{align}
\label{e:ineq1}
|\psi_{1}| & \geq \gamma_{1}\zeta \omega_{n} \\
|\psi_{2}| & \geq \gamma_{2}  \zeta \omega_{n}  \label{e:ineq2}
\end{align}
It is recommended to select $\gamma_{1}=5$ to guarantee that the effect of the zero $\psi_{1}$ on the ATMD displacement $z_{1}$ is insignificant. Moreover, the parameter $\gamma_{2}$ should be selected as large as possible, and we recommend $\gamma_{2}\geq 1$. 
\end{itemize}

\subsection{Frequency responses of $z_{1}(t)$, $z_{2}(t)$, $z_{3}(t)$ and $u(t)$}
The frequency response of signals $z_{1}(t)$, $z_{2}(t)$, $z_{3}(t)$ and $u(t)$ will be computed to guarantee that these signals are between acceptable limits within the frequency band of $\ddot{x}_{g}(t)$. For that, assume that excitation signal $\ddot{x}_{g}$ is a sinusoidal input, denoted as $\ddot{x}_{g}=\delta \sin (\omega t)$, where the amplitude $\delta$ represents the upper bound of $\ddot{x}_{g}$, as indicated in (\ref{e:boe}). Moreover, the frequency $\omega$ corresponds to the earthquake frequency with a bandwidth $\omega \in [\omega_{BWl}, \ \omega_{BWu}]$. Then, the steady-state of responses $z_{1}(t)$, $z_{2}(t)$, $z_{3}(t)$ and $u(t)$ are also sinusoids. By varying the frequency $\omega$ of $\ddot{x}_{g}$ between its bandwidth permits to determine if the amplitudes of $z_{1}(t)$, $z_{2}(t)$, $z_{3}(t)$ and $u(t)$ are between tolerable limits within this band.   

Let the transfer functions 
\begin{equation}
H_{i}(s)=\delta G_{i}(s), \qquad i=1,2,3,u
\end{equation}
where $G_{1}(s)$, $G_{2}(s)$, $G_{3}(s)$ and $G_{u}(s)$ are presented in section \ref{sec:FT}. The frequency responses of $z_{1}(t)$, $z_{2}(t)$, $z_{3}(t)$ and $u(t)$ are directly computed from the transfer functions $H_{1}(s)$, $H_{2}(s)$, $H_{3}(s)$ and $H_{u}(s)$ by substituting variable $s$ by $j\omega$, where $\omega \in [\omega_{BWl}, \ \omega_{BWu}]$. The root mean square (RMS) value $\kappa_{i}$ of $H_{i}(j\omega)$ is also calculated in the frequency band  of $\ddot{x}_{g}(t)$. This value is defined as  
\begin{equation}
\label{e:sig}
\kappa_{i}=\mbox{RMS}\underset{\omega \in [\omega_{BWl}, \ \omega_{BWu}]}{(|H_{i}(j\omega)|)}, \qquad i=1,2,3,u
\end{equation}
Since the predominant spectral content of earthquakes is between 1 to \SI{20}{\Hz} \cite{kayal:2008}, the parameters $\omega_{BWl}$ and $\omega_{BWu}$ will be set as $\omega_{BWl}=2\pi$~\si{\radian/\s} and $\omega_{BWu}=40\pi$~\si{\radian/\s}. 

\section{Tuning algorithm for the SMC}
\label{sec:tuning_alg}
This section presents the proposed tuning algorithm to compute: 1) the vector $\boldsymbol{\eta}$ using the transient and frequency responses of the closed-loop system; and 2) the switching gain $M_0$ by analyzing the frequency response $H_{u}(j\omega)$.

\subsection{Procedure to compute vector $\boldsymbol{\eta}$}
\label{sec:obc}
According to equation (\ref{e:11}), vector $\boldsymbol{\eta}$ depends on the parameters $\zeta$ and $\omega_{n}$ of the characteristic polynomial $P_{1}(s)$ in (\ref{e:ssp}). Let us define $\Upsilon$ as the possible set of vectors $\boldsymbol{\eta}$, with which the closed-loop system (\ref{e:12}) in the sliding mode satisfies the following three conditions:
\begin{enumerate}
\item The limits for $\zeta$ in (\ref{e:lz}) and for $\omega_{n}$ in (\ref{e:lw}), the value $\lambda_{3}=-3\zeta\omega_{n}$ for the non-dominant pole, as well as the inequalities in (\ref{e:ineq1}) and (\ref{e:ineq2}) corresponding to the zeros $\psi_{1}$ and $\psi_{2}$, respectively. 
\item The next upper limits $\bar{\kappa}_{i}$ for $\kappa_{i}$ $i=1,2,3$ in (\ref{e:sig}) given by   
\begin{equation}\label{e:bs}
\kappa_{i}\leq \bar{\kappa}_{i}, \qquad i=1,2,3
\end{equation}
where $\bar{\kappa}_{i}$, $i=1,2,3$ are positive constants, which constrain the maximum value of $|H_{i}(j\omega)|$ under the bandwidth $\omega \in [\omega_{BWl}, \ \omega_{BWu}]$, or in time-domain, the maximum permitted values of signals $z_{1}$, $z_{2}$ and $z_{3}$ in the bandwidth of $\ddot{x}_{g}$.
\item The inequality 
\begin{equation}\label{e:bs2}
\kappa_{u}+\varpi \leq \bar{\kappa}_{u}
\end{equation}
deduced from (\ref{e:24}), which indicates that the sum of the RMS value $\kappa_{u}$ of frequency response $|H_{u}(j\omega)|$ and the bound $\varpi$ of the non-linear friction $f(z_{3})$ should be less than or equal to the specified limit $\bar{\kappa}_{u}$.  
\end{enumerate}
Then, the vector $\boldsymbol{\eta} \in \Upsilon$ used by the SMC, denoted as $\boldsymbol{\eta}_{*}$, is obtained by minimizing either of the following two performance indexes (PIs)
\begin{align}\label{e:pin}
J_{z_{2}}=&\min \limits_{\boldsymbol{\eta} \in \Upsilon}\kappa_{2}(\boldsymbol{\eta}) \\
J_{u}=&\min \limits_{\boldsymbol{\eta} \in \Upsilon}\kappa_{u}(\boldsymbol{\eta})  \label{e:pin2}
\end{align}
where parameters $\kappa_{2}$ and $\kappa_{u}$ show their dependency on $\boldsymbol{\eta}$. Note that $J_{z_{2}}$ and $J_{u}$ are related to the ability of the SMC to minimize the top floor displacement $z_{2}$ and the control force $u$, respectively. Hence the sliding variable $\sigma=\boldsymbol{\eta}_{*}^{\mathrm{T}}\mathbf{z} $ guarantees a minimal of $z_{2}$ or $u$, while ensuring that the RMS values of the closed-loop signals are within acceptable limits and their transient responses are sufficiently fast and damped.

\subsection{Procedure to compute the switching gain $M_{0}$}
\label{sec:obmo}
According to inequality (\ref{ineq3}), the value of $M_{0}$ should satisfy $M_{0}>\varpi+h_{0}$. Among these two terms of the sum, only parameter $\varpi$ is assumed to be known according to Assumption \ref{as1}. This section presents a methodology for estimating parameter $h_{0}$, that is subsequently employed to compute $M_{0}$. Note that from (\ref{e:ineqs}), the parameter $h_{0}$ must satisfy the condition $h_{0} \geq |u_{a}+\alpha_{1}\ddot{x}_{g}|$, where the frequency response of the signal $u_{a}+\alpha_{1}\ddot{x}_{g}$ is given by $H_{u}(j\omega)$. Define $\chi$ as the maximum magnitude of $|H_{u}(j\omega)|$ between the interval $\omega \in [\omega_{BWl}, \ \omega_{BWu}]$, where $\chi$ is computed using the vector $\boldsymbol{\eta}_{*}$ that minimizes either of the performance indexes (\ref{e:pin}) and (\ref{e:pin2}), i.e., 
\begin{equation}\label{e:chi}
\chi=\max\underset{\omega \in [\omega_{BWl}, \ \omega_{BWu}]}{(|H_{u}(j\omega,\boldsymbol{\eta}_{*})|)}
\end{equation}
Since $\chi\geq |u_{a}+\alpha_{1}\ddot{x}_{g}|$ in the spectrum of $\ddot{x}_{g}$, we will select $h_{0}=\chi$. Thus, the switching gain $M_{0}$ of the SMC is computed as:
\begin{equation}\label{e:mo}
M_{0}=\varpi+\chi+\varsigma
\end{equation}
where $\varsigma$ is a small positive constant that guarantees the inequality (\ref{ineq3}), and that in this work is fixed to 0.5.

\begin{figure} 
\centering
\begin{tikzpicture}[node distance = 3mm, auto, scale=0.65, every node/.style={transform shape}]
\node (Start) [startstop] {Start};
\node (Step02) [block, align=left, text width = 10cm, below = 8mm of Start] { {\centerline {DEFINE}}\\[2mm]
\begin{itemize}
\item System parameters: $\omega_{0}$ in (\ref{e:w0}); {\bf A, B, D} in (\ref{e:6}); and Bounds $\delta$ (\ref{e:boe}) and $\varpi$ (\ref{e:bo})
\item Damping ratio limits $\zeta_{l}$ and $\zeta{_u}$ in (\ref{e:lz}), and increment $\Delta_{\zeta}$
\item Natural frequency limits $\omega_{nl}$ and $\omega_{nu}$ in (\ref{e:lw}), and increment $\Delta_{\omega_{n}}$\\
\item Lower bounds $\gamma_{1}$ (\ref{e:ineq1}) and $\gamma_{2}$ (\ref{e:ineq2}) for the zeros $\psi_{1}$ and $\psi_{2}$, respectively\\ 
\item Maximum signal limits: $\bar{\kappa}_{i}$, $i=1,2,3,u$ in (\ref{e:bs}) and (\ref{e:bs2})\\
\item The performance index $J_{z_{2}}$ (\ref{e:pin}) or $J_{u}$ (\ref{e:pin2}) to minimize  
\end{itemize}
};

\node (Step03) [block, align=center, text width = 7cm, below = of Step02] { {\centerline {SET}}\\[2mm]
$\zeta=\zeta_{l}$ and $j=0$};
\node (Step05) [decision, below = 15mm of Step03, aspect=2] {Is $\zeta \leq \zeta{_u}$?};
\node (Step03a) [block, align=center, text width = 5cm, below = 5mm of Step05] { {\centerline {SET}}\\[2mm]
$\omega_{n}=\omega_{nl}$};
\node (Step06) [decision, below = 5mm of Step03a, aspect=2] {Is $\omega_n \leq \omega_{nu}$?};
\node (Step09) [block, align=center, text width = 10cm, below = 5mm of Step06] {Compute $\lambda_{1}$ and $\lambda_{2}$ in (\ref{e:lamda}), $\lambda_{3}=-3\zeta\omega_{n}$, $\boldsymbol{\eta}$ (\ref{e:11}), and zeros $\psi_{1}$ and $\psi_{2}$ in (\ref{e:zero})};
\node (Step10) [decision, below = of Step09, aspect=5, align=center] {Is  $|\psi_{1}|/(\zeta \omega_{n}) \geq \gamma_{1}$ and $|\psi_{2}|/(\zeta \omega_{n}) \geq \gamma_{2}$?};
\node (Step11) [block, align=center, text width = 10cm, below = 5mm of Step10] {Calculate vector $\mathbf{k}$ (\ref{e:8}); matrices $\mathbf{B}_{1}$ (\ref{e:12a}), $\mathbf{T}_{1}$ (\ref{e:15}) and $\mathbf{A}_{1}$ (\ref{e:20}); Transfer functions $G_{1}(s)$ (\ref{e:n4}), $G_{2}(s)$ (\ref{e:n3}), $G_{3}(s)$ (\ref{e:n4s}) and $G_{u}(s)$ (\ref{e:g4}); and frequency response RMS values $\kappa_{i}$, $i=1,2,3,u$ in (\ref{e:sig})};
\node (Step12) [decision, below = 5mm of Step11, aspect=1.25, align=center] {Is \\ ($\kappa_{1}\leq \bar{\kappa}_{1}$) \& ($\kappa_{2}\leq \bar{\kappa}_{2}$) \\ \& ($\kappa_{3}\leq \bar{\kappa}_{3}$) \& ($\kappa_{u}+\varpi)\leq \bar{\kappa}_{u}$) \\ ?};  
\node (Step13a) [block, align=center, text width = 2.5cm, below = 5mm of Step12] {$j=j+1$};
\node (Step13) [block, align=center, text width = 10cm, below = 5mm of Step13a] {Save the $j^{\mbox{th}}$ tuple as\\ 
$\Gamma_{j}=\{\zeta,\omega_{n}, \kappa_{1},\kappa_{2},\kappa_{3},\kappa_{u},\lambda_{1},\lambda_{2},\lambda_{3},\psi_{1},\psi_{2},\boldsymbol{\eta}\}$, where $\boldsymbol{\eta} \in \Upsilon$};
\node (Step14) [block, align=center, text width = 2.5cm, right = 7mm of Step12] {$\omega_n=\omega_n+\Delta_{\omega_{n}}$};
\node (Step07) [block, align=center, text width = 2cm, right = 7mm of Step06] {$\zeta=\zeta+\Delta_\zeta$};
\node (Step15) [decision, aspect=2, align=center,left = 70mm of Step06,shift={(0,-0.5)}] {Is $j\geq 1$ ?};
\node (Stepx) [blockx, blue, align=center, text width = 3cm, below = 0.7cm of Start, shift={(1.1,0.2)}] {};
\node (Stepy) [blocky, blue, align=center, text width = 3cm, below = -1.8cm of Step05, shift={(1.1,0.2)}] {};
\node (Stepz) [blockz, blue, align=center, text width = 3cm, below = -1.8cm of Step15, shift={(-1.55,1)}] {};
\node [above=-0.0cm of Stepx.south,shift={(3.9,0)},align=right,blue,text width = 5cm] {\textcircled{1} INITIALIZE \\PARAMETERS AND \\DEFINE SIGNAL LIMITS AND PI};
\node [below=-0.0cm of Stepy.north,shift={(3.9,0)},align=right, blue, text width = 5cm] {\textcircled{2} CALCULATE \\ VECTORS $\boldsymbol{\eta} \in \Upsilon$ \\ \& SAVE TUPLES $\Gamma_{j}$};
\node [below=-0.0cm of Stepz.north,shift={(-2.4,0)},align=left, blue, text width = 5cm] {
\textcircled{3} CALCULATE $\boldsymbol{\eta}_{*}$ \& $M_0$ FROM SAVED TUPLES $\Gamma_{j}$};
\node [nn, left of=Step13,node distance=6.1cm] (nx) {};

\node (Step16) [block, align=center, text width = 3cm, left = 1.5cm of Step15] {It is not possible to find $\boldsymbol{\eta}_{*}$ and $M_{0}$ with this initialization};
\node (Step17) [block, align=center, text width = 6cm, below = of Step15, shift={(0,-0.5)}] {Using all the tuples $\Gamma_{j}$, $j=1,\ldots,\mathcal{N}$, display three-dimensional plots of $(\zeta,\omega_{n},\kappa_{i})$, $i=1,2,3,u$};
\node (Step18) [block, align=center, text width = 6cm, below = of Step17] {Among these $\mathcal{N}$ tuples, select the one that contains the vector $\boldsymbol{\eta}_{*}$ that minimizes the performance index $J_{z_{2}}$ (\ref{e:pin}) or $J_{u}$ (\ref{e:pin2}). This tuple is denoted as $\Gamma_{*}$};
\node (Step19) [block, align=center, text width = 6cm, below = of Step18] {Compute $\chi$ (\ref{e:chi}), which is the maximum value of $H_{u}(j\omega)$ obtained using $\boldsymbol{\eta}_{*}$};
\node (Step20) [block, align=center, text width = 6cm, below = of Step19] {Compute the switching gain $M_{0}$ using (\ref{e:mo})};
\node (Step21) [block, align=center, text width = 6cm, below = of Step20] {Return the tuple $\Gamma_{*}$ and $M_{0}$};
\node (Stop) [startstop, align=center, below = 14mm of Step21] {Stop};

\draw [arrow] (Start) -- (Step02);
\draw [arrow] (Step02) -- (Step03);
\draw [arrow] (Step03) -- (Step05);
\draw [arrow] (Step05.south) node[anchor=north east]{Yes} -- (Step03a);
\draw [arrow] (Step03a) -- (Step06);
\draw [arrow] (Step06.south) node[anchor=north west]{Yes} -- (Step09);
\draw [arrow] (Step09) -- (Step10);
\draw [arrow] (Step10.south) node[anchor=north west]{Yes} -- (Step11);
\draw [arrow] (Step11) -- (Step12);
\draw [arrow] (Step12.south) node[anchor=north west]{Yes} -- (Step13a);
\draw [arrow] (Step12.east) node[anchor=south west]{No} -- (Step14);
\draw [arrow] (Step13a) -- (Step13);
\draw [arrow] (Step13) -| (Step14);
\draw [arrow] (Step10.east) node[anchor=south west]{No} -| (Step14);
\draw [arrow] (Step14.east) -- node[pos=0.75, inner sep=2pt]{}++(0.25,0) |- ($ (Step06.north) + (0mm,2.5mm) $);
\draw [arrow] (Step06.east) node[anchor=south west]{No} -- (Step07);
\draw [arrow] (Step07) |- ($ (Step05.north) + (0mm,2.5mm) $);

\draw [arrow] (Step05.west) node[anchor=south east] {No} -| (Step15);
\draw [arrow] (Step15.west) node[anchor=south east]{No} -- (Step16);
\draw [arrow] (Step15.south) node[anchor=north west]{Yes} -- (Step17);
\draw [arrow] (Step17) -- (Step18);
\draw [arrow] (Step18) -- (Step19);
\draw [arrow] (Step19) -- (Step20);
\draw [arrow] (Step20) -- (Step21);
\draw [arrow] (Step21) -- (Stop);
\draw [arrow] (Step16) |- ($ (Stop.north) + (0mm,10mm) $);
\draw [very thick, dashed] (Step13) -- (nx);
\draw [arrow,very thick, dashed] (nx) |- (Step17);
\draw [arrow,very thick, dashed] (nx) |- (Step18);

|\end{tikzpicture}
\caption{Proposed automatic tuning algorithm.}
\label{Fig:Temp1}
\end{figure}
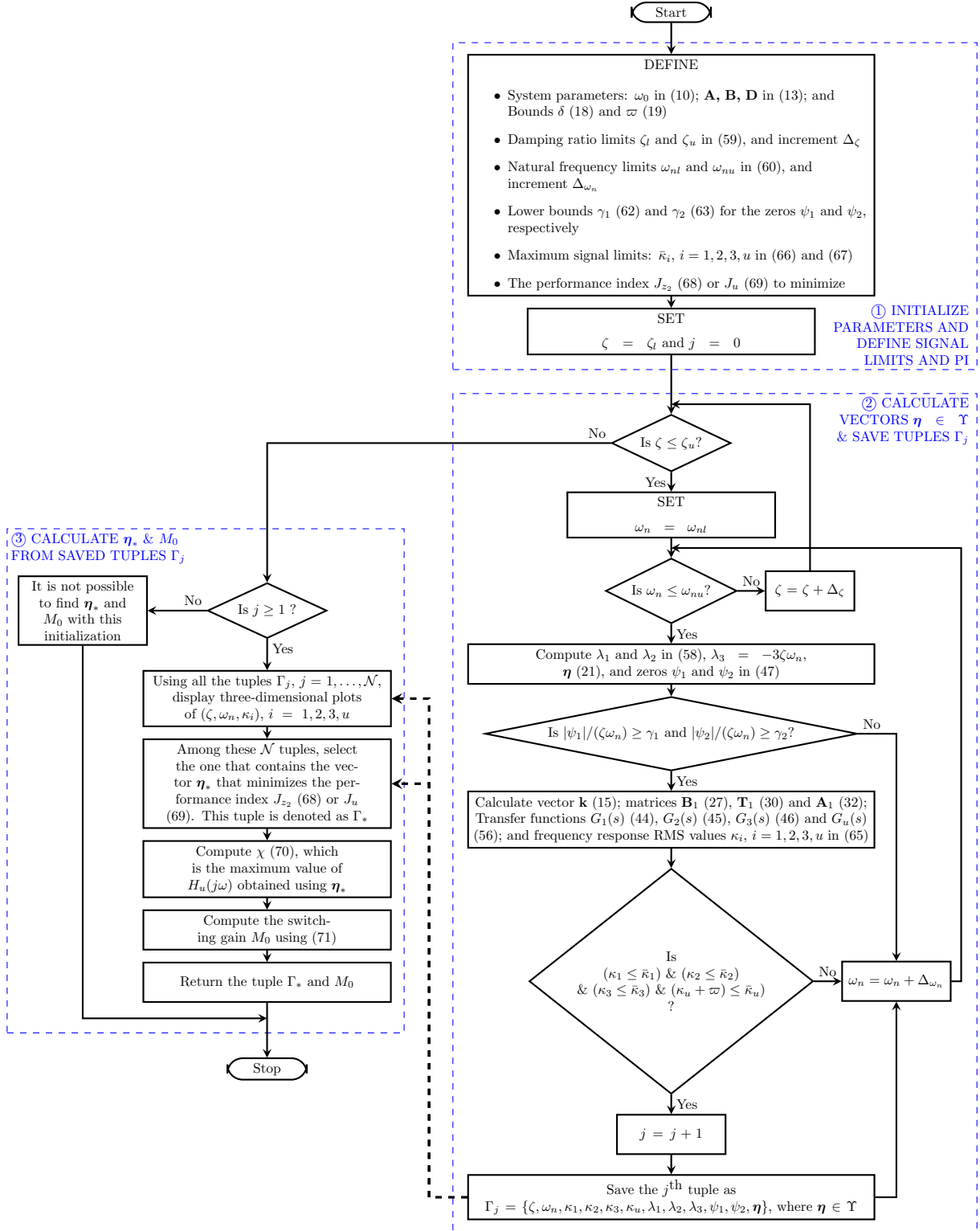

Finally, Figure \ref{Fig:Temp1} shows the proposed tuning algorithm that computes the vector $\boldsymbol{\eta}_{*}$ and gain $M_{0}$ of the SMC. This algorithm implements the procedure outlined in the present section, and it is programmed in the Matlab software. It has three main phases: 1) initialization of system parameters, definition of limits for signals and parameters, and selection of the PI to be minimized; 2) calculation of feasible vectors $\boldsymbol{\eta}$ for the SMC, which are saved in tuples $\Gamma_j=\{\zeta,\omega_{n}, \kappa_{1},\kappa_{2},\kappa_{3},\kappa_{u},\lambda_{1},\lambda_{2}, \lambda_{3},\psi_{1},\psi_{2},\boldsymbol{\eta}\}$, where  $j=1,\ldots,\mathcal{N}$; and 3) searching in these $\mathcal{N}$ tuples to determine the vector $\boldsymbol{\eta}_{*}$ that minimizes the PI $J_{z_{2}}$ (\ref{e:pin}) or $J_{u}$ (\ref{e:pin2}), and its corresponding switching gain $M_{0}$. The terms $\Delta_{\zeta}$ and $\Delta_{\omega_{n}}$ in this algorithm represent the increments for $\zeta$ and $\omega_{n}$ from their lower to upper limits.  

\section{Results and discussion}
\label{sec:sim_exp}
The performance of the proposed tuning algorithm for the SMC, based on the Ackermann's formula, is evaluated by means of numerical and experimental studies. In both studies the SMC is programed in the Matlab/Simuink software and the data is sampled at \SI{1}{\ms}. The SMC in (\ref{e:cmd}) causes the chattering effect, that is a discontinuous force at $\sigma=0$ that cannot be applied to the ATMD in practice. There exists several chattering reduction techniques such as the Boundary layer solution \cite{Slotine:1991}, Dynamic Terminal SMC \cite{fei2020dynamic}, Super-twisting SMC \cite{fei2020fractional}, and Higher-order SMC \cite{levant2003higher}, just to mention a few. In this paper, we use the Boundary layer solution, where the sign function of the SMC is approximated by means of following continuous saturation function
\begin{equation}\label{e:aprox2}
u=-M_{0}\mbox{sign} (\sigma)\approx -M_{0}\mbox{sat}(\sigma)=\left\{
               \begin{array}{rl}
                 -M_{0}      & \mathrm{if\ } \sigma > \epsilon \\
                  {\sigma}/{\epsilon}& \mathrm{if\ } -\epsilon \leq \sigma \leq \epsilon \\
                 M_{0}     & \mathrm{if\ } \sigma <-\epsilon
               \end{array} 
             \right.
\end{equation}
where $\epsilon$ is a positive constant that is fixed to $\epsilon=0.05$. 

\begin{figure}[ht]
\centering
\resizebox{0.8\textwidth}{!}{
\begin{tikzpicture}[auto, node distance=1.2cm,>=latex']

\node [nn, name=input] {};
\node [build, right of=input,node distance=3.5cm] (building) {$\dot{\mathbf{z}}=\mathbf{A}\mathbf{z}+\mathbf{B}\left[u-f(z_{3})\right]+\mathbf{D}\ddot{x}_{g}$};
\node [nn, right of=building,node distance=6cm] (output) {};
\node [nn, right of=building,node distance=4.25cm] (n1) {};
\node [blk, above of=n1,node distance=1.5cm] (sgain) {$\boldsymbol{\eta}^{\mathrm{T}}$};
\node [blk, left of=sgain,node distance=2.4cm] (sign) {$-M_0 \mathrm{sat}(.)$};
\node [blka, text width = 2cm, align=center, left of=sign, node distance=3.2cm,shift={(0,1.3)}] (algo) {Tuning algorithm in Figure \ref{Fig:Temp1}};
\node [blka, text width = 5.8cm, align=center, left of=algo, node distance=5.2cm] (param) {Define: ${\bf A},{\bf B},{\bf D},\omega_{0}, \delta, \varpi$, $\zeta_{l}, \zeta{_u}, \Delta_{\zeta}, \omega_{nl}, \omega_{nu}, \Delta_{\omega_{n}}$, $\gamma_{1}, \gamma_{2}$, $\bar{\kappa}_{i}$, $i=1,2,3,u$, and PI $J_{z_{2}}$ or $J_{u}$.};

\draw [-latex] (input) -- node  {$\ddot{x}_g$} (building);
\draw [-latex,very thick] (building) -- node [below] {$\mathbf{z}=[x_{d}, x_{N}, \dot{x}_{d}, \dot{x}_{N}]^{\mathrm{T}}$} (output);
\draw [-latex,very thick] (n1) -- node [above, very near end] {} (sgain);
\draw [-latex] (sgain) -- node [above] {$\sigma$} (sign);
\draw [-latex] (sign) -| node [anchor=south, near start] {$u$} (building);
\draw [-latex,very thick, dashed] (algo) -| node [right, near end] {$\boldsymbol{\eta}=\boldsymbol{\eta}^*$} (sgain);
\draw [-latex,very thick, dashed] (algo) -| node [right, near end] {$M_0$} (sign);
\draw [-latex,very thick] (param) -- node [above, very near end] {} (algo);
\end{tikzpicture} }
\caption{Block diagram showing the implementation of the proposed control strategy.}
\label{block_con}
\end{figure}
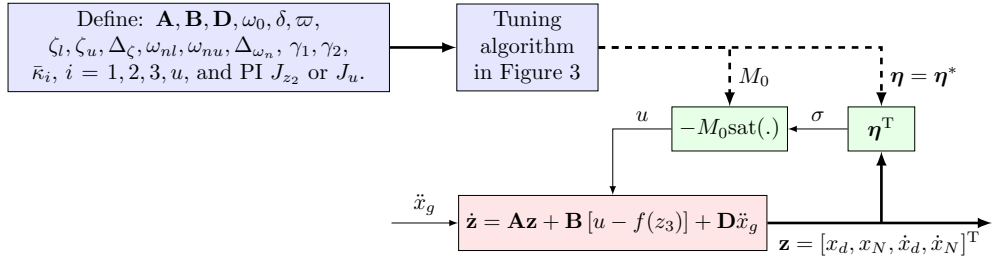

For the implementation of the proposed scheme, first the parameters of the SMC in (\ref{e:aprox2}), i.e., the vector $\boldsymbol{\eta}_{*}$ of the sliding variable and the switching gain $M_0$, that guarantees that the top floor and the ATMD displacements and velocities remain within some specified limits, are calculated by the proposed tuning algorithm outlined in Section \ref{sec:tuning_alg}. Then, these parameters are used in the SMC (\ref{e:aprox2}) to attenuate the vibrations of the seismically excited building (\ref{e:6}) via the ATMD. The block diagram representation of the control implementation is presented in Figure \ref{block_con}.

\subsection{Numerical simulation}
\label{sec:simulations}
A reduced scale five-story building with an ATMD mounted on its top floor is simulated. The structure is excited through the North-South component of the El Centro (California, 1940) earthquake, whose magnitude has been scaled. The mass and stiffness of each floor are equal to \SI{10}{\kg} and $1.21\times10^{4}\si{\N / \m}$, respectively. The mass $m_{0}$ and stiffness $k_{0}$ of the dominant first mode are \SI{28.07}{\kg} and $2.75\times10^{3}\si{\N / \m}$, respectively. Therefore, the natural frequency of this mode is $\omega_{0}=\SI{9.9}{\radian / \s}$ or $f_{0}=\SI{1.58}{\Hz}$. It is assumed that the structure has Rayleigh damping, where the damping ratios of the first and second modes are equal to 0.01. The mass $m_{d}$ of the ATMD is \SI{1.4}{\kg}, that is 5\% of the modal mass $m_{0}$. Moreover, the tuned parameters $c_{d}$ and $k_{d}$ are selected as $c_{d}=\SI{3.54}{\N\s/\m}$ and $k_{d}=\SI{121.66}{\N/\m}$, which are the optimum parameters according to \cite{rana1998}. Using these parameters, the natural frequency $\omega_{d}=\sqrt{k_{d}/m_{d}}$ and damping ratio $\zeta_{d}=c_{d}/(2m_{d}\omega_{d})$ of the ATMD are equal to $0.94\omega_{0}$ and 0.135, respectively. Finally, the function $f(z_{3})$ in (\ref{e:6}) is modeled as Coulomb friction acting between the ATMD and the floor
on which it is attached. It is given by:
\begin{equation}\label{e:fc}
f(z_{3})=\mu_{d} \mbox{sign} (z_{3})
\end{equation}
where $\mu_{d}$ is the Coulomb friction coefficient with a value of \SI{0.35}{\N}. 

\begin{center}
\begin{figure}[ht]
\centering
\subfigure[Mesh plot of $(\zeta,\omega_{n},\kappa_{1}$).]
	                {\includegraphics[height=3.7cm]{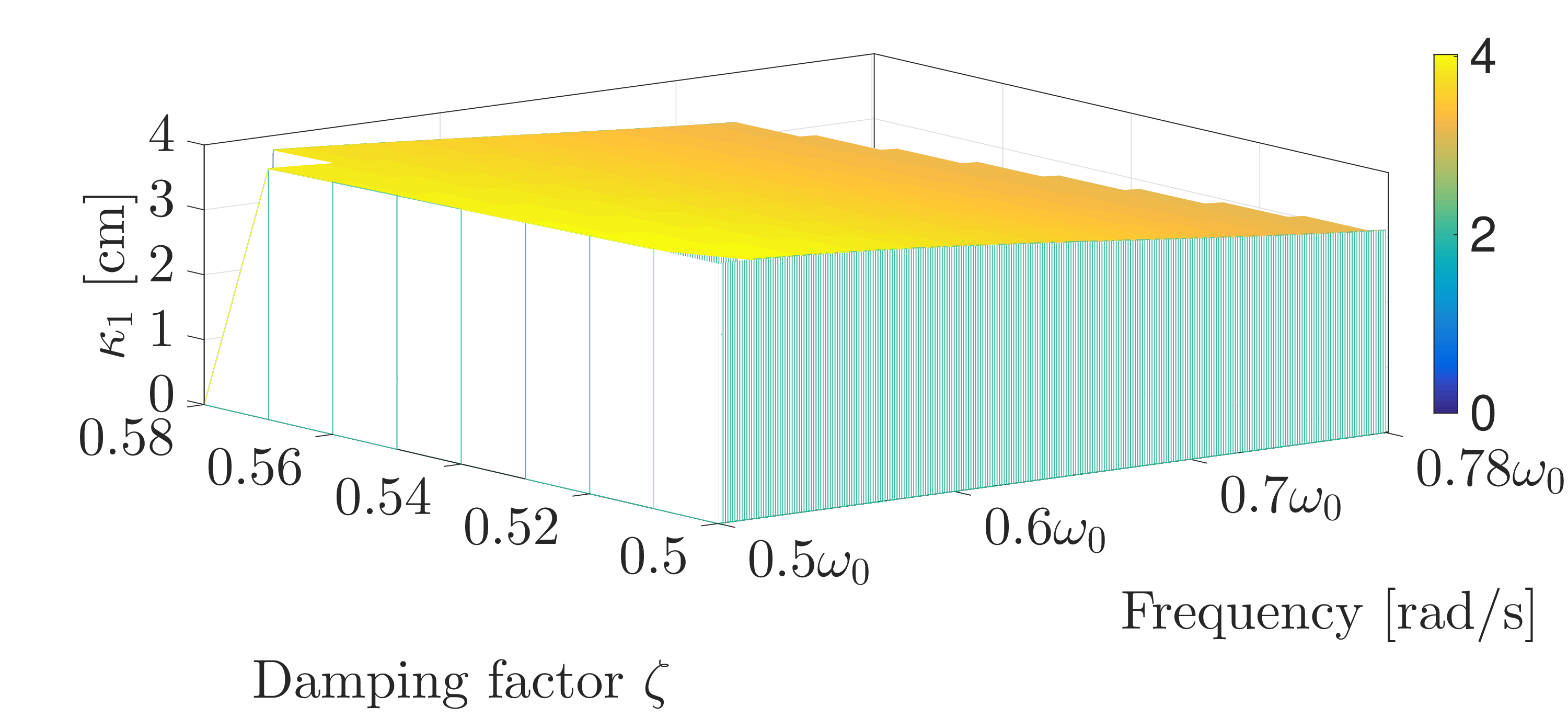}}
\subfigure[Mesh plot of $(\zeta,\omega_{n},\kappa_{2}$).]
	                {\includegraphics[height=3.7cm]{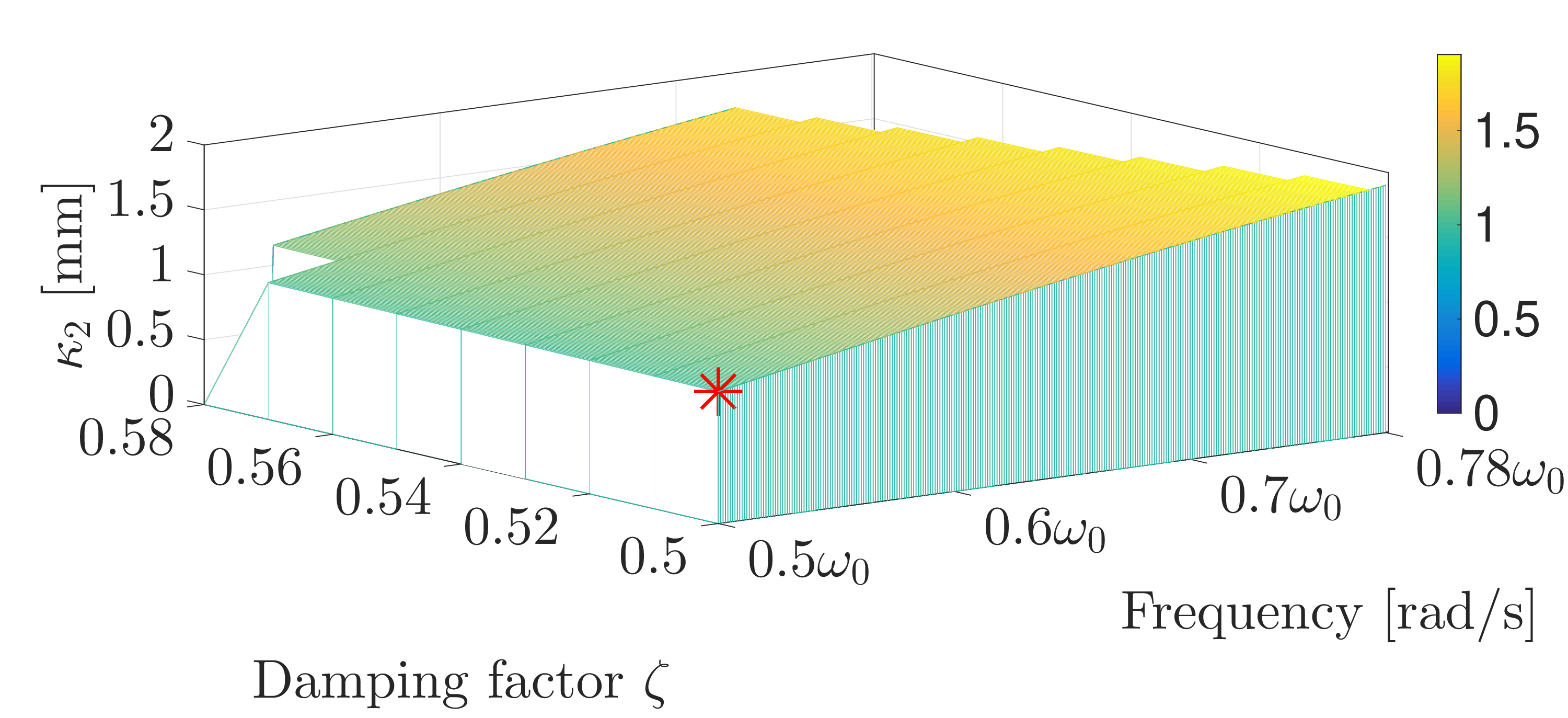}}	
\subfigure[Mesh plot of $(\zeta,\omega_{n},\kappa_{3}$).]
	                {\includegraphics[height=3.7cm]{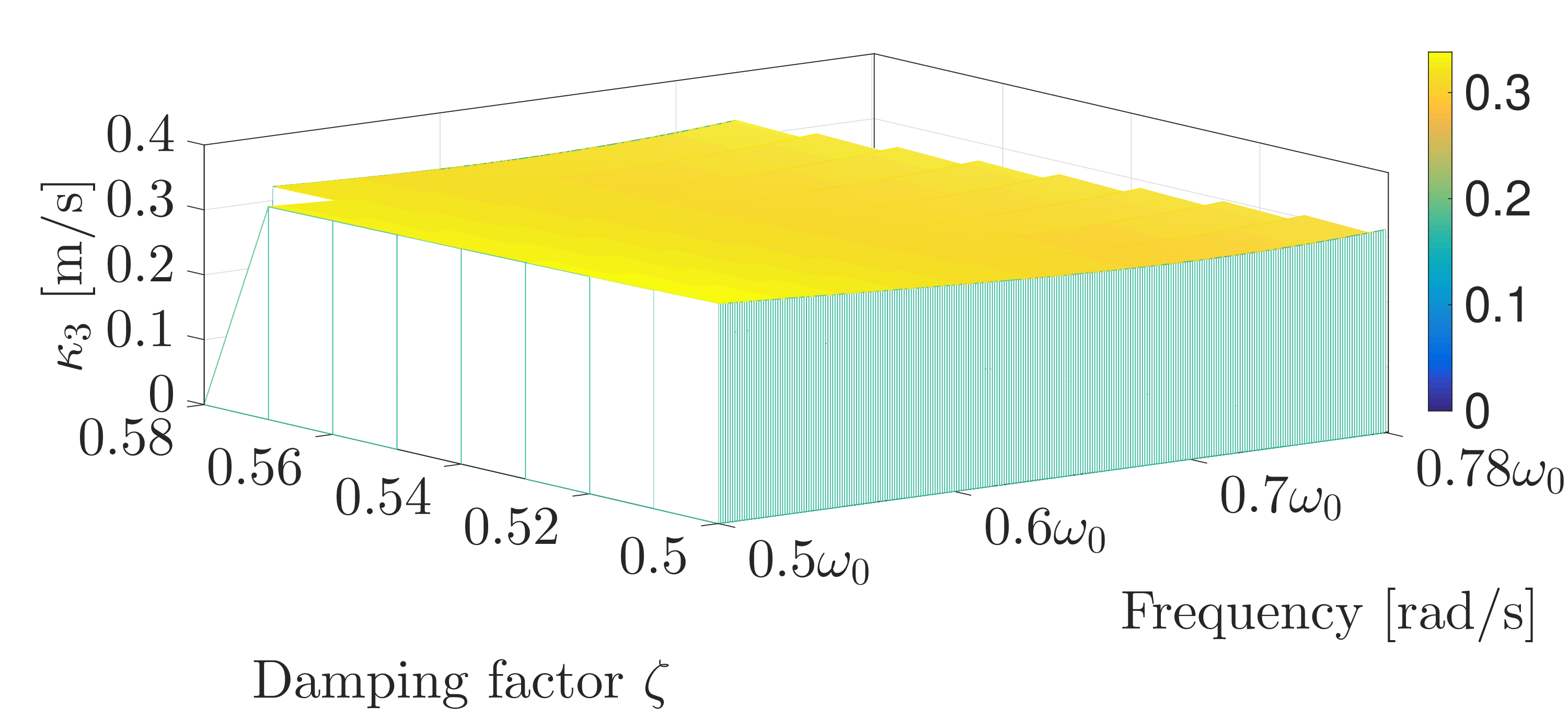}}
\subfigure[Mesh plot of $(\zeta,\omega_{n},\kappa_{u}$).]
	                {\includegraphics[height=3.7cm]{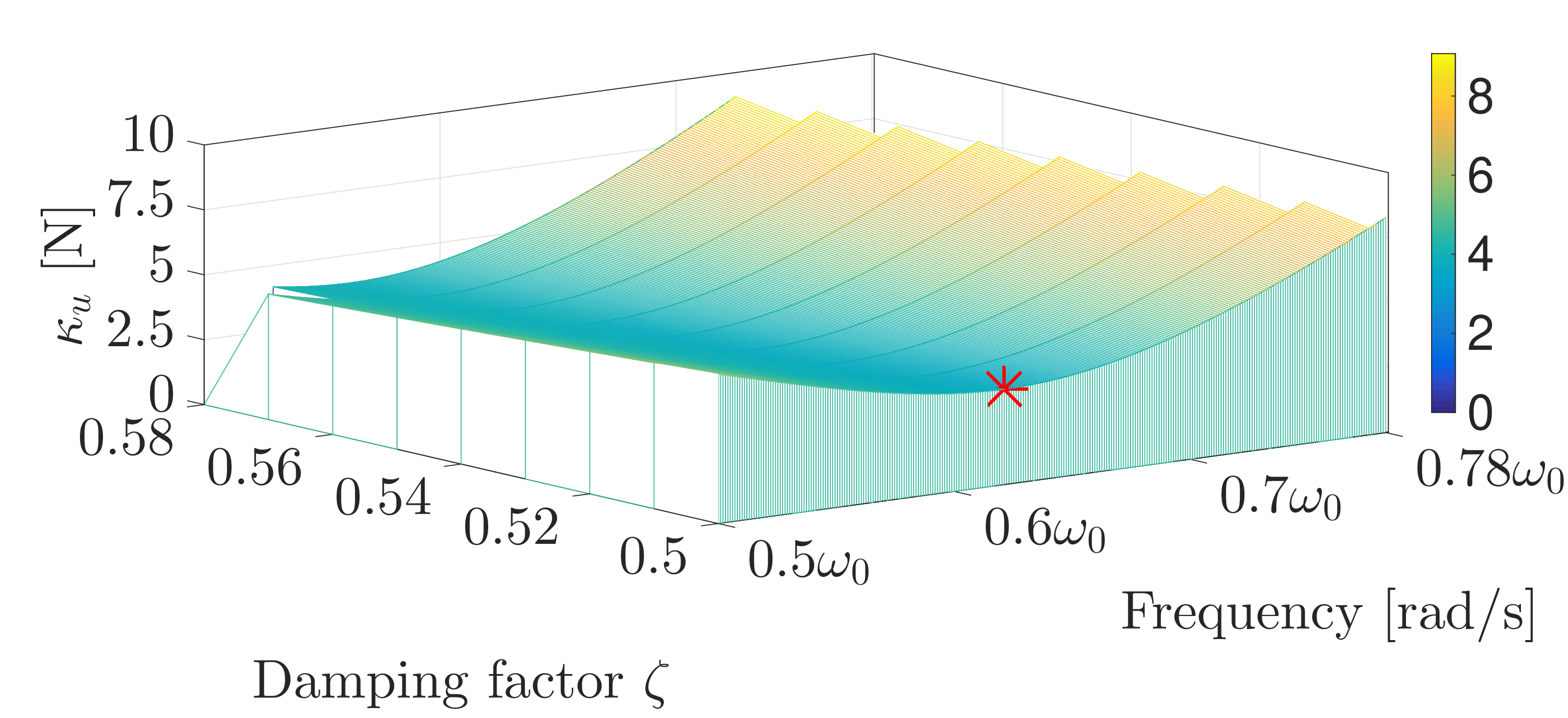}}	
\caption{Three-dimensional surfaces of $\kappa_{1}$, $\kappa_{2}$, $\kappa_{3}$ and $\kappa_{u}$.}
\label{fig:kappa}
\end{figure}          
\end{center}

In order to compute the parameters $\boldsymbol{\eta}_{*}$ and $M_{0}$ of the SMC, the proposed tuning algorithm is initialized as follows: $\delta=\SI{0.5}{\m/\s^2}$, $\varpi=\SI{0.5}{\N}$, $\gamma_{1}=5$, $\gamma_{2}=1$, $\zeta_{l}=0.5$, $\zeta{_u}=0.9$, $\Delta_{\zeta}=0.01$, $\omega_{nl}=0.5\omega_{0}$, $\omega_{nu}=0.8\omega_{0}$, and $\Delta_{\omega_{n}}=0.01$. Moreover, the signals $z_{1}$, $z_{2}$, $z_{3}$ and $u$ are constrained using the bounds $\bar{\kappa}_{1}=\SI{20}{\cm}$, $\bar{\kappa}_{2}=\SI{10}{\mm}$, $\bar{\kappa}_{3}=\SI{70}{\cm / \s}$, and $\bar{\kappa}_{u}=\SI{12}{\N}$, respectively. The proposed algorithm displays three-dimensional surfaces of $\kappa_{1}$, $\kappa_{2}$, $\kappa_{3}$ and $\kappa_{u}$ computed with the vectors $\boldsymbol{\eta} \in \Upsilon$. These plots are shown in Figure \ref{fig:kappa} and they indicate that the feasible sets of parameters $\zeta$ and $\omega_{n}$, that satisfy the desired transient responses of $z_{i}$, $i=1,2$ and frequency responses of $z_{i}$, $i=1,2,3$ and $u$, are $\zeta \in [0.5,0.58]$ and $\omega_{n} \in [0.5\omega_{0},0.78\omega_{0}]$. The minimum value of the surfaces of $\kappa_{2}$ and $\kappa_{u}$ is indicated with a red asterisk. According to Figure \ref{fig:kappa} (b), the value of $\kappa_{2}$ monotonically increases with the increment of either $\zeta$ or $\omega_ {n}$, i.e., the minimum of $\kappa_{2}$ is located at ($\zeta_{l}, \omega_{nl}$). Moreover, the mesh plot shown in Figure \ref{fig:kappa} (d) indicates that the minimum of $\kappa_{u}$ is located at $\zeta=\zeta_{l}$. Thus, using other initial parameters of $\zeta_{l}$ and $\omega_{nl}$ in the tuning algorithm changes the coordinates ($\zeta,\omega_{n}$) of the minimums of $\kappa_{2}$ and $\kappa_{u}$. Table \ref{tab:tuples} presents the tuples $\Gamma_{*}=\{ \zeta,\omega_{n},\kappa_{1},\kappa_{2},\kappa_{3},\kappa_{u},\lambda_{1},\lambda_{2},\lambda_{3},\psi_{1},\psi_{2},\boldsymbol{\eta}_{*} \}$ and the switching gains $M_{0}$ returned by proposed algorithm by minimizing the performance indexes $J_{z_{2}}$ and $J_{u}$. 

\begin{table*}[ht]
\centering
\caption{Tuples $\Gamma_{*}$ returned by the proposed algorithm for the numerical structure.}
\label{tab:tuples}
\begin{adjustbox}{width=1\textwidth}
\begin{tabular}{ccccccccccccc} \toprule
\multirow{2}{*}{PI} & \multirow{2}{*}{$\zeta$} & $\omega_{n}$ &  $\kappa_{1}$ & $\kappa_{2}$ & $\kappa_{3}$ & $\kappa_{u}$ & $\lambda_{1,2}$ & $\lambda_{3}$ & $\psi_{1}$ & $\psi_{2}$ & \multicolumn{2}{c}{SMC parameters}\\
\cmidrule(lr){4-7} \cmidrule(lr){8-9} \cmidrule(lr){10-11} \cmidrule(lr){12-13}
   &   & (rad/s)  & (cm)  & (mm) & (cm/s)&(N) & (rad/s) & (rad/s) & (rad/s) & (rad/s) & $\boldsymbol{\eta}_{*}^{\mathrm{T}}$ & $M_{0}$ (N)  \\ \midrule
{$J_{z_{2}}$} & {0.5} & {0.50$\omega_{0}$}  & {4.0} & {1.0} & {33.8} & {5.76} & -$2.48 \pm 4.29j$  & {-7.43} & {-29.64} & {-2.99} & [2.6,-289.2,0.87,-9.76] & {24.13} \\
{$J_{u}$} & {0.5} & {0.62$\omega_{0}$}  & {3.7} & {1.4} & {31.4} & {3.68} & -$3.08 \pm 5.33j$ & {-9.23} & {2114.6} & {-3.72} & [5,-312.3,1.34,0.15] & {20.35} \\
\bottomrule
\end{tabular}
\end{adjustbox}
\end{table*}

\begin{center}
\begin{figure}[ht]
\centering
\subfigure[Ground acceleration $\ddot{x}_{g}(t)$.]
	                {\includegraphics[height=3.7cm]{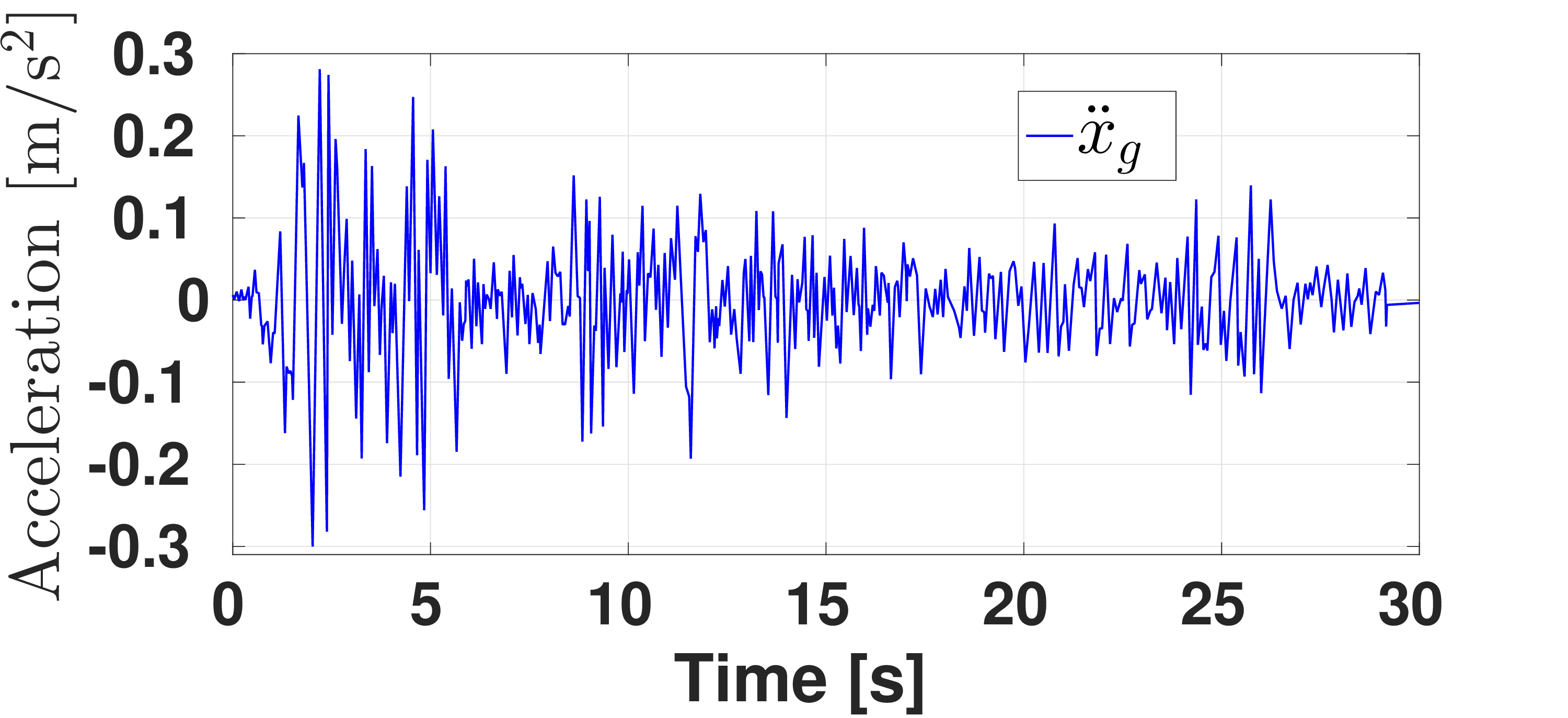}}
\subfigure[Top floor displacement $z_{2}(t)$.]
	                {\includegraphics[height=3.7cm]{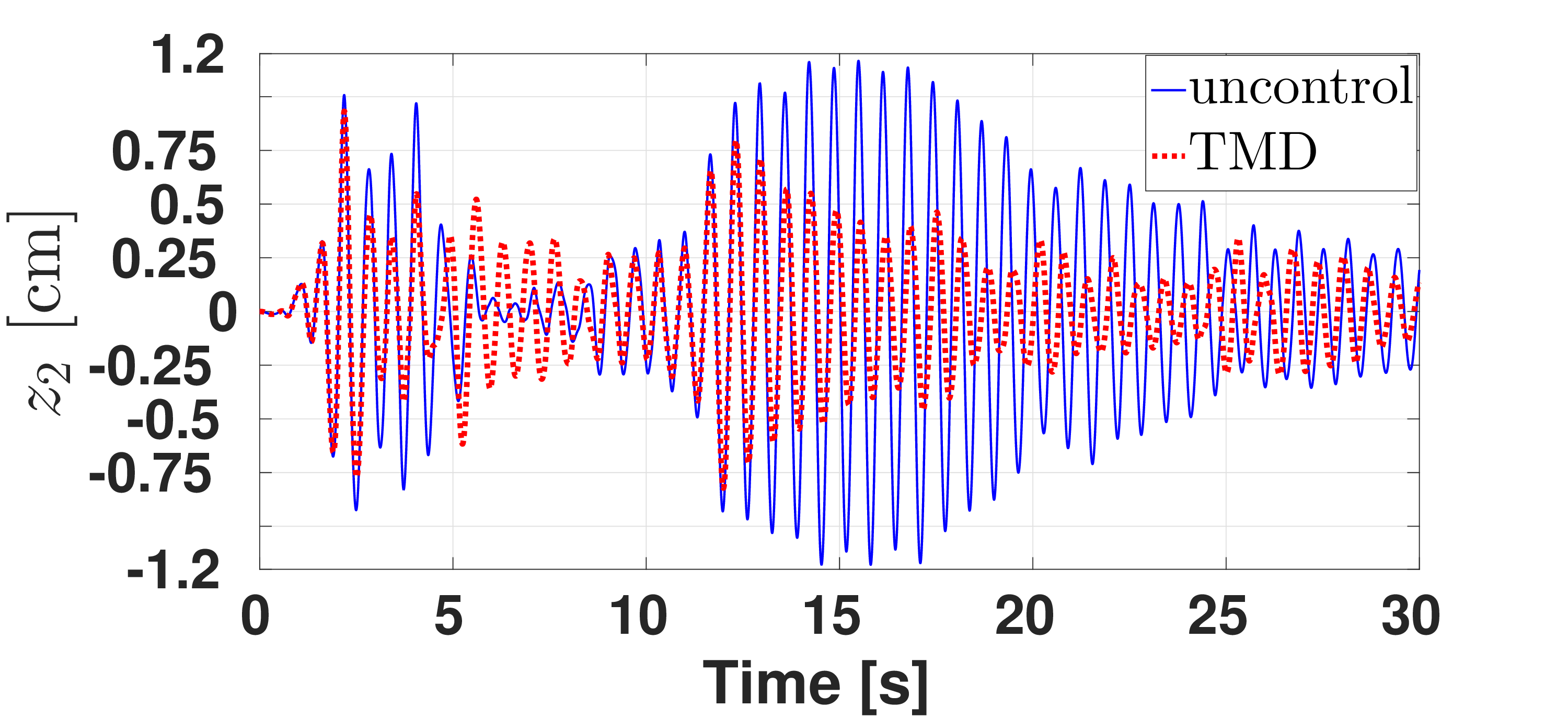}}	
\caption{Signals $\ddot{x}_{g}(t)$ and $z_{2}$.}
\label{fig:3a}
\end{figure}          
\end{center}

\begin{center}
\begin{figure}[H]
\centering
\subfigure[ATMD displacement $z_{1}(t)$.]
	                {\includegraphics[height=3.7cm]{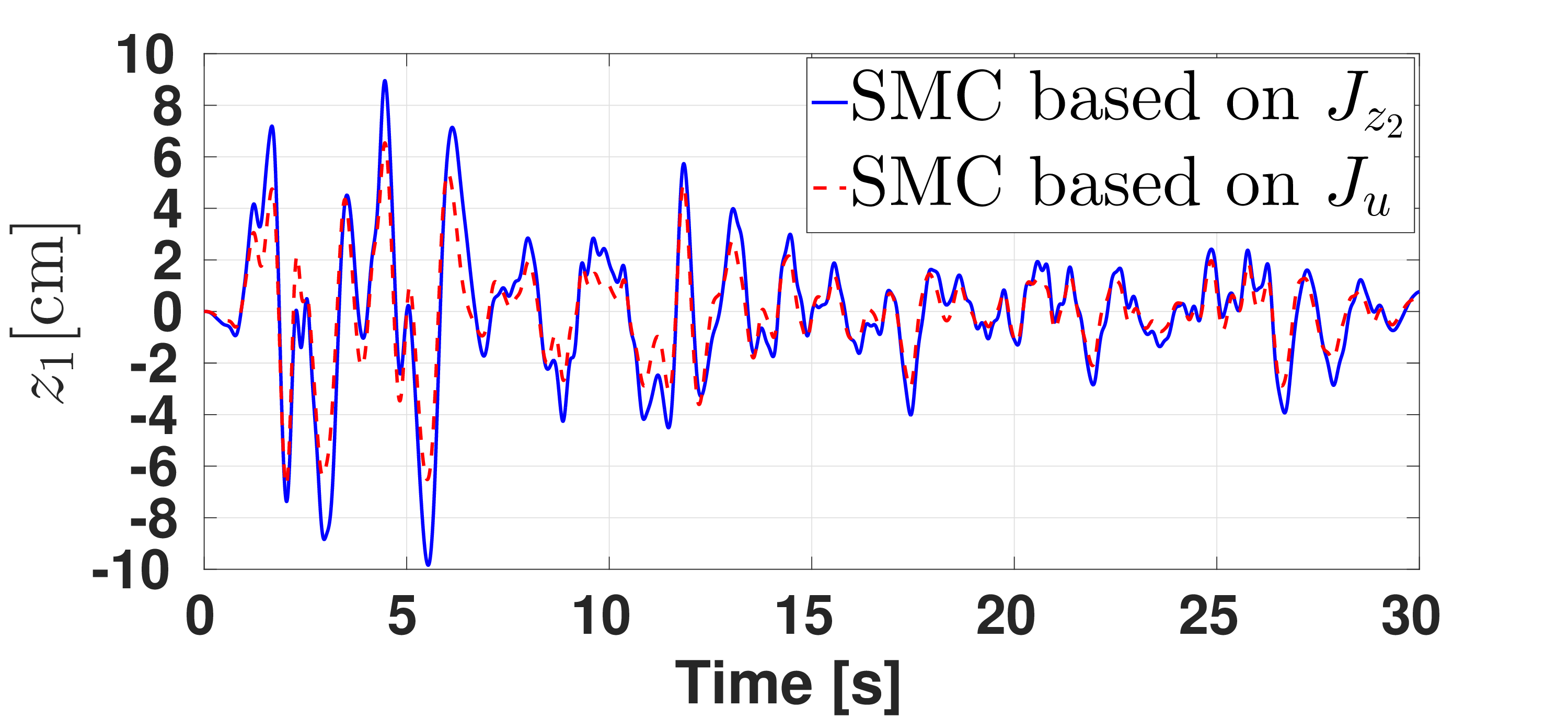}}
\subfigure[Top floor displacement $z_{2}(t)$.]
	                {\includegraphics[height=3.7cm]{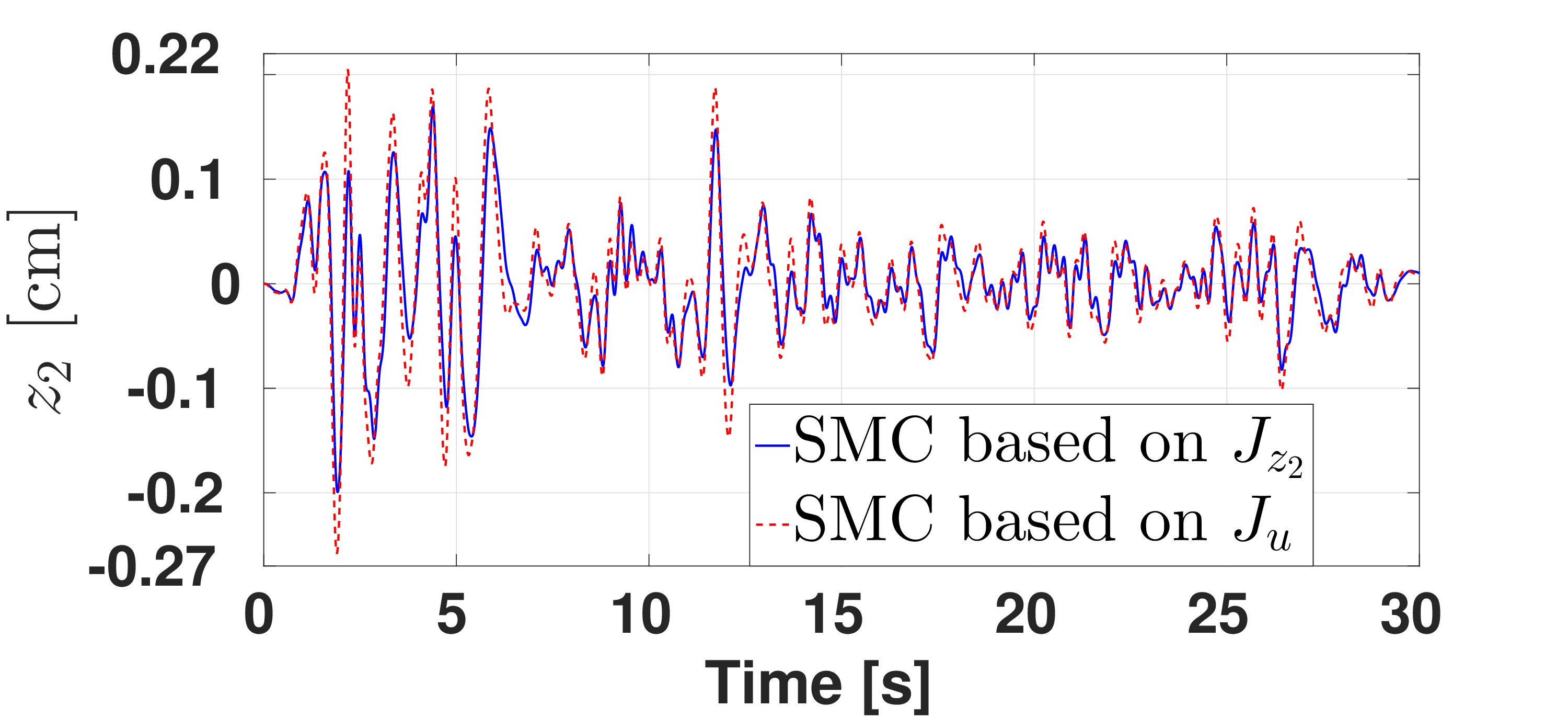}}	
\subfigure[ATMD velocity $z_{3}(t)$.]
	            {\includegraphics[height=3.7cm]{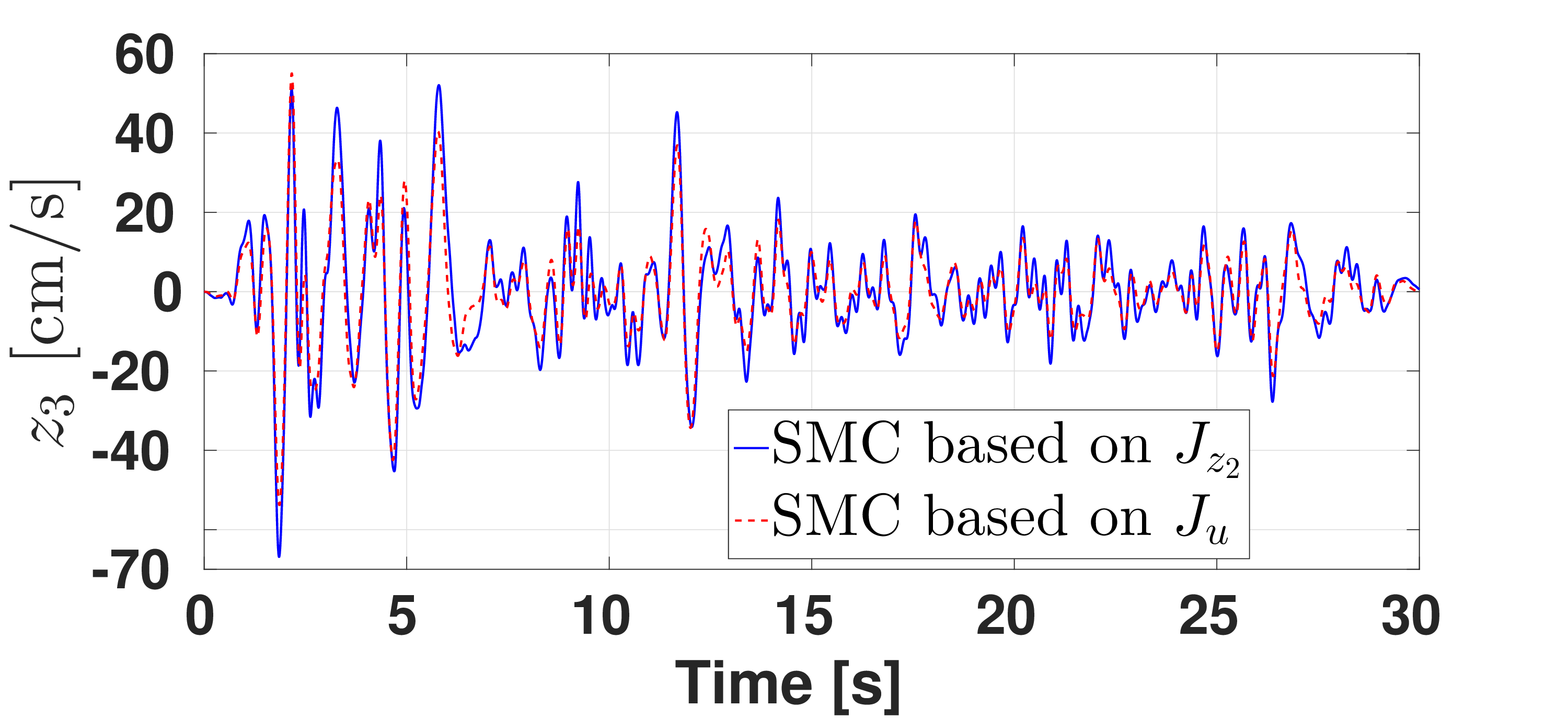}}
\subfigure[Control force $u(t)$.]
	                {\includegraphics[height=3.7cm]{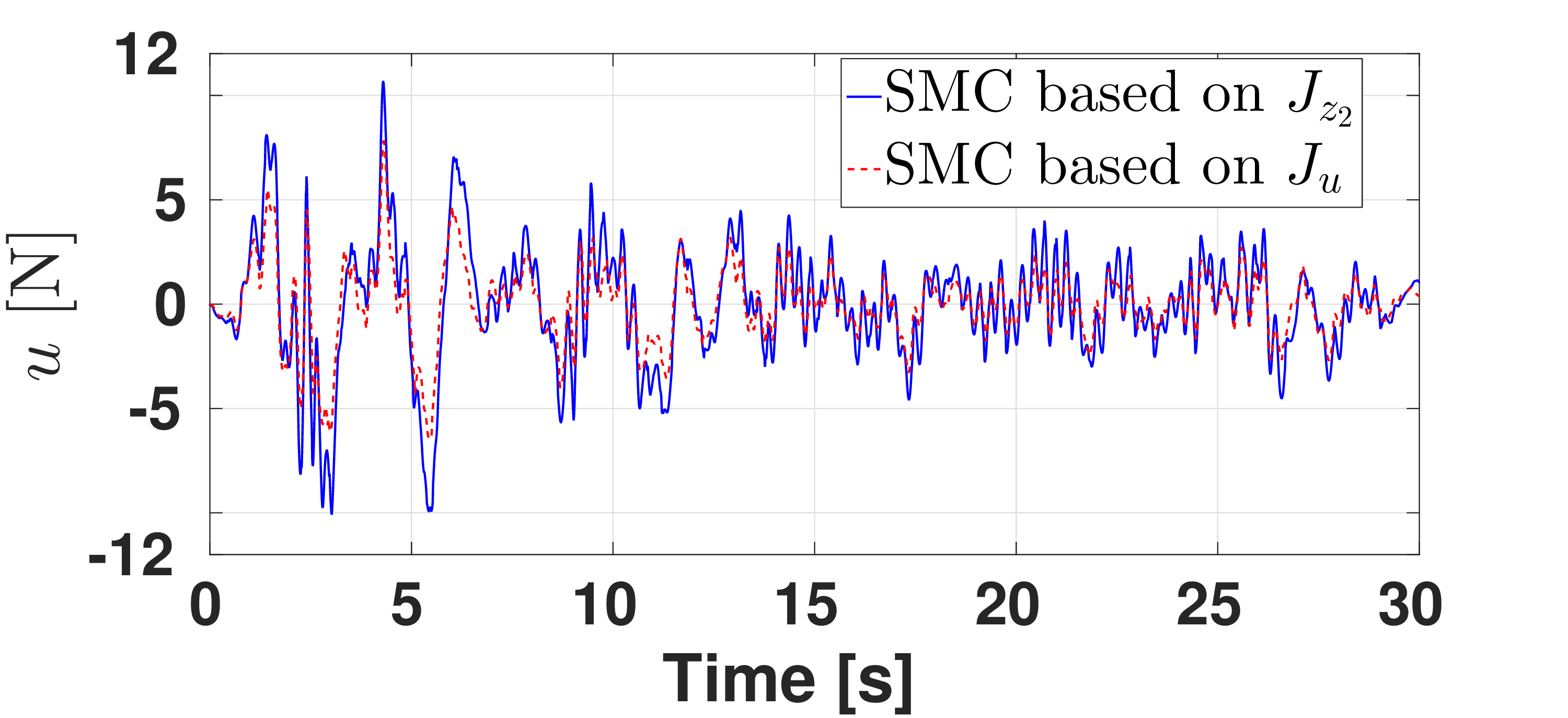}}	
\caption{Signals $z_{1}(t),z_{2}(t),z_{3}(t)$, and $u(t)$.}
\label{fig:6}
\end{figure}          
\end{center}

Figure \ref{fig:3a} (a) depicts the earthquake excitation signal $\ddot{x}_{g}$, whereas Figure \ref{fig:3a} (b) shows the top floor displacement $x_{5}=z_{2}$ of the uncontrolled system, that is compared with the displacement $z_{2}$ produced by the TMD. Furthermore, Figures \ref{fig:6} (a)-(d) present, respectively, the signals $z_{1}$, $z_{2}$, $z_{3}$ and $u$, which are produced with the SMC using the parameters $\boldsymbol{\eta}_{*}$ and $M_{0}$ that minimize the PIs $J_{z_{2}}$ and $J_{u}$. By comparing Figures \ref{fig:3a} (b) and \ref{fig:6} (b), we can conclude that the ATMD, based on the proposed SMC and either of the PIs $J_{z_{2}}$ and $J_{u}$, provides greater reduction of the building displacements when compared to the TMD. 

Now, let us define $z_{i}^{\mathrm{rms}}$, $u^{\mathrm{rms}}$ and $z_{i}^{\mathrm{peak}}$, $u^{\mathrm{peak}}$ as the RMS and peak values of signals $z_{i}$ and $u$, respectively. Moreover, let us denote $\mathcal{R}(z_{2}^{\mathrm{rms}})$ and $\mathcal{R}(z_{2}^{\mathrm{peak}})$ as the vibration attenuation percentages of $z_{2}^{\mathrm{rms}}$ and $z_{2}^{\mathrm{peak}}$, respectively, which are given by:
\begin{align}
    \mathcal{R}(z_{2}^{\mathrm{rms}})=&\left(1-\dfrac{(z_{2}^{\mathrm{rms}})_{\text{controlled}}}{(z_{2}^{\mathrm{rms}})_{\text{uncontrolled}}}   \right)\times 100  \\ 
    \mathcal{R}(z_{2}^{\mathrm{peak}})=&\left(1-\dfrac{(z_{2}^{\mathrm{peak}})_{\text{controlled}}}{(z_{2}^{\mathrm{peak}})_{\text{uncontrolled}}}   \right)\times 100
\end{align} 

The larger these percentages, the better the attenuation of the signal $z_{2}$. Table \ref{tab:rms_val_edif} presents these values corresponding to the TMD and ATMD during the period of $t=0$ to $t=\SI{30}{\s}$. This table indicates that the ATMD allows reducing more than three times the peak value of the floor displacement $z_{2}=x_{5}$ when compared to the TMD. For both SMCs, the percentages $\mathcal{R}(z_{2}^{\mathrm{rms}})$ and $\mathcal{R}(z_{2}^{\mathrm{peak}})$ of attenuation for the displacement $z_{2}$ are close to 90\% and 80\%, respectively. Note that the SMC that minimize the PI $J_{z_{2}}$ (respectively $J_{u}$) produce the smallest $z_{2}^{\mathrm{rms}}$ (respectively $u^{\mathrm{rms}}$), as expected. Finally, by comparing Tables \ref{tab:tuples} and \ref{tab:rms_val_edif}, it is possible to observe that the values $z_{i}^{\mathrm{rms}}$, $i=1,2,3,u$ can be considered as an scaled version of the parameters $\kappa_{i}$, since the relation $\kappa_{i}/z_{i}^{\mathrm{rms}}$ is between 1.5 to 2.65. 
\begin{table*}[ht] 
\centering
\caption{RMS and peak values of $z_{i}$, $i=1,2,3,4$ and $u(t)$.}
\label{tab:rms_val_edif}
\begin{adjustbox}{width=1\textwidth}
\footnotesize
\begin{tabular}{c c c c c c c c c c c c c c} \toprule
	\multicolumn{1}{c}{\multirow{2}{*}{{Controller}}} & \multicolumn{1}{c}{\multirow{2}{*}{PI}}& $z_{1}^{\mathrm{rms}}$ & $z_{1}^{\mathrm{peak}}$ & $z_{2}^{\mathrm{rms}}$& $z_{2}^{\mathrm{peak}}$ & $z_{3}^{\mathrm{rms}}$& $z_{3}^{\mathrm{peak}}$ & $z_{4}^{\mathrm{rms}}$& $z_{4}^{\mathrm{peak}}$ & $u^{\mathrm{rms}}$ & $u^{\mathrm{peak}}$ & $\mathcal{R}(z_{2}^{\mathrm{rms}})$ & $\mathcal{R}(z_{2}^{\mathrm{peak}})$\\
	\cmidrule(lr){3-4} \cmidrule(lr){5-6} \cmidrule(lr){7-8} \cmidrule(lr){9-10} \cmidrule(lr){11-12} \cmidrule(lr){13-14}
 & &\multicolumn{2}{c}{(cm)} & \multicolumn{2}{c}{(mm)}  & \multicolumn{2}{c}{(cm/s)}   &  \multicolumn{2}{c}{(mm/s)}  & \multicolumn{2}{c}{(N)} & \multicolumn{2}{c}{(\%)} \\ \midrule
      Uncontrol   &--  & --  & --   & 4.70  & 11.79  & --  & -- & 46.48   & 118.02  & 0  & 0 & 0 & 0\\ 
      TMD   &--  & 0.33  & 1.33   & 2.67  & 9.33  & 3.19  & 13.38 & 25.8   & 93.8  & 0  & 0 & 43.19 & 20.87 \\ 
	  ATMD  & $J_{z_{2}}$& 2.61 & 9.84 & 0.48  & 2.0 & 14.0 & 66.9& 3.93 & 17.61 &2.78 & 10.66 & 89.79 & 83.04\\
	  ATMD  & $J_{u}$& 1.86 & 6.74 & 0.58  & 2.58 & 11.81 & 55.04 & 5.11 & 26.49 & 1.81 & 7.81 & 87.66 & 78.12\\ \bottomrule
\end{tabular}
\end{adjustbox}
\end{table*}

\subsection{Experimental results}
\label{sec:er}
\begin{figure}[H]
\begin{center}
	\includegraphics[height=6cm]{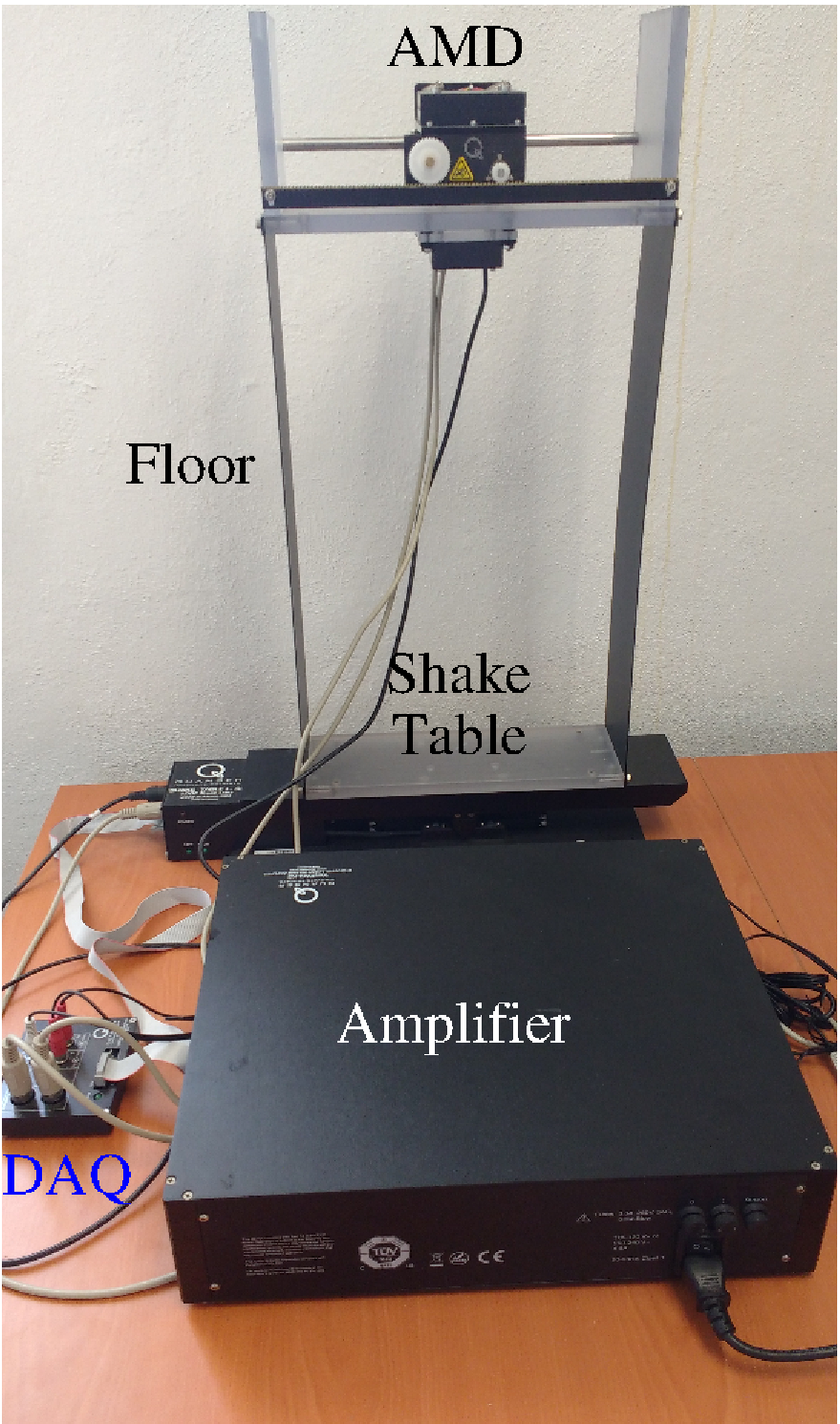}
	\caption{Experimental structure with an AMD.}
	\label{fig:5}
\end{center}
\end{figure}

Figure \ref{fig:5} shows the experimental structure, which was developed by the Quanser company. It consists of a shake table, model STI-40, that excites a reduced-scale single-story building controlled by an AMD. The shake table has the following characteristics: total travel of $\SI{40.0} {\mm}$ (i.e., $\pm \SI{20.0} {\mm})$, and it is driven by a ball-screw drive mechanism, that is coupled to a direct current motor, model Magmotor S23. The linear position of the shake table is provided by an encoder with a resolution of $\SI{1.22}{\micro\meter}$. The dimensions of the building are:  \SI{0.32}{\m} length, \SI{0.11}{\m} width, and \SI{0.50}{\m} height. The two columns of the structure are made of steel, and they have a cross-section of $1.75\times$\SI{108.1}{\mm}, and a mass of \SI{0.24}{\kg}. The structure has an accelerometer with a measurement range of $\pm$\SI{5}{\g}, that measures the floor acceleration relative to the ground. Moreover, the top of the structure contains the AMD, which is a linear cart made of aluminum, that has a maximum travel or stroke of $\pm$\SI{9.5}{\cm}, and whose position is detected through an encoder with a resolution of $\SI{22.75}{\micro\meter}$. Furthermore, the AMD is actuated by a Faulhaber Coreless DC motor, model 2338S006, which is coupled to a rack and pinion mechanism. All the sensors and motors of the experimental structure are energized with a power amplifier, model VoltPAQ-X1. Moreover, data acquisition is performed with a Q2-USB board, that communicates with Matlab/Simulink by means of the QUARC real-time control software. Further details of the shake table, building, and the AMD can be found at their user manuals \cite{st:manual}, \cite{building:manual}, and \cite{amd:manual}, respectively.

\begin{center}
\begin{figure}[ht]
\centering
    {\includegraphics[height=3.7cm]{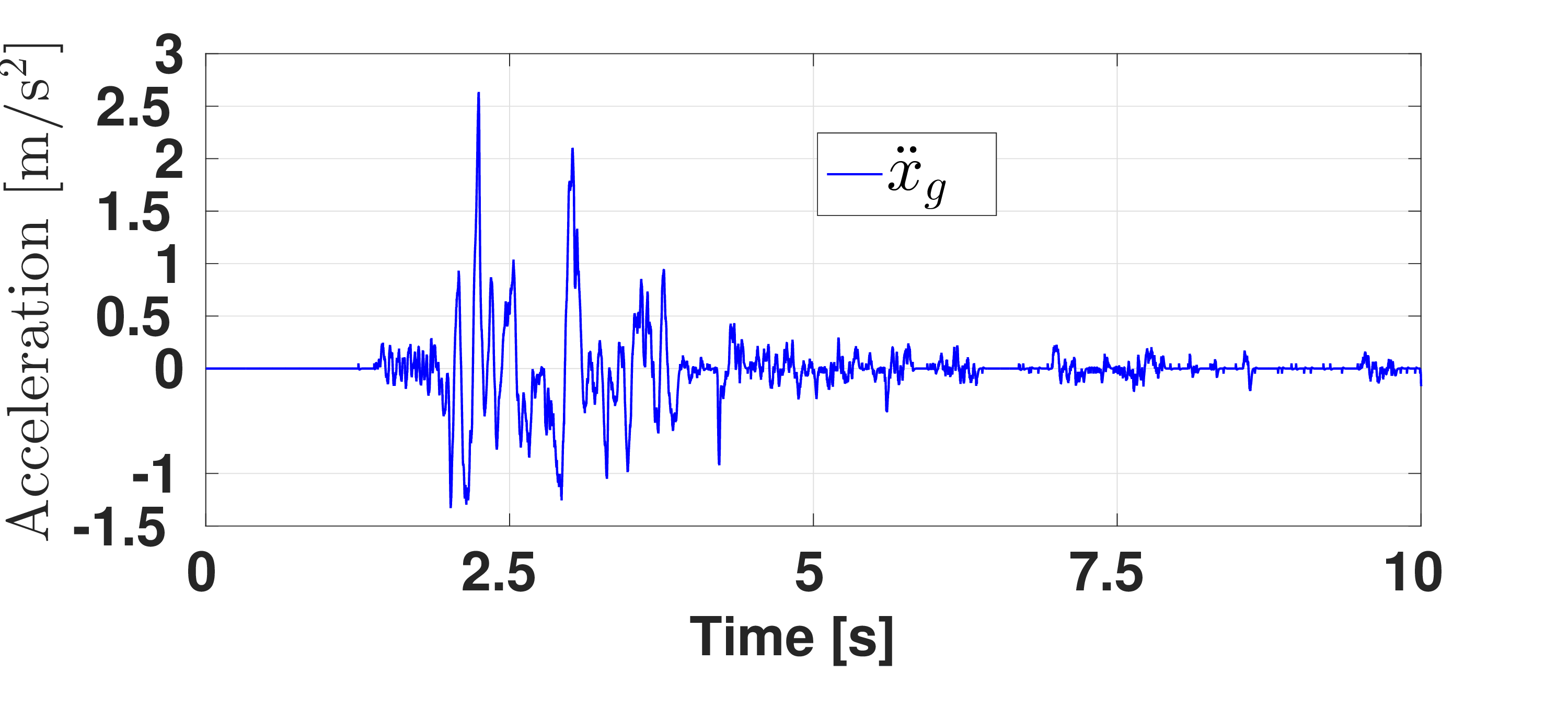}}
\caption{Shake table acceleration $\ddot{x}_{g}(t)$.}
\label{fig:7}
\end{figure}          
\end{center}

During the experiments, the structure is excited with the 1994 Northridge earthquake, which is scaled in time and amplitude, as shown in Figure \ref{fig:7}. The model of the structure is given in (\ref{e:6}), where the nonlinear friction $f(z_{3})$ is given in (\ref{e:fc}). The parameters of the experimental structure are listed in Table \ref{tab:par_est}, where $c_{d}$ represents the viscous friction between the damper and the floor. These parameters are provided by Quanser with exception of the Coulomb friction coefficient $\mu_{d}$ of $f(z_{3})$. To identify this parameter, the AMD is represented as a filtered linear regression model and the Least Squares method described in \cite{garrido2013} is applied, that uses the AMD displacement $z_{1}$ and its input force $u$. On the other hand, the displacement $z_{2}$ and velocity $z_{4}$ of the story, as well as the velocity $z_{3}$ of the AMD are not available. Hence, they are estimated by means of a Luenberger state observer \cite{Ogata:2002}, that uses the measurements from the encoders of the shake table and AMD, the acceleration of the floor, and the input force of the damper.

\begin{table}[ht]
\centering
\caption{Parameters of the experimental structure.}
\label{tab:par_est}
\small
\begin{tabular}{c c c} \toprule
	Parameter & Value & Units \\ \midrule
	            $m_{0}$& 1.84 & \si{\kg} \\ 
              $m_{d}$ & 0.79 &  \si{\kg} \\ 
              $k_{0}$ & 226.23 & \si{\N/\m} \\ 
              $k_{d}$ & 0 &   \si{\N/\m} \\ 
              $c_{0}$ & 0.16 & \si{\N\s/\m} \\ 
              $c_{d}$ &6.85 & \si{\N\s/\m} \\ 
              $\mu_{d}$ &0.43 & \si{\N} \\ 
              $\beta_{0}$ & 1 & -- \\  
              $\omega_{0}$ & 11.08 & \si{\radian/\s}\\  \bottomrule
\end{tabular}
\end{table}

\begin{center}
\begin{figure}[ht]
\centering
\subfigure[Mesh plot of $(\zeta,\omega_{n},\kappa_{1}$).]
	                {\includegraphics[height=3.7cm]{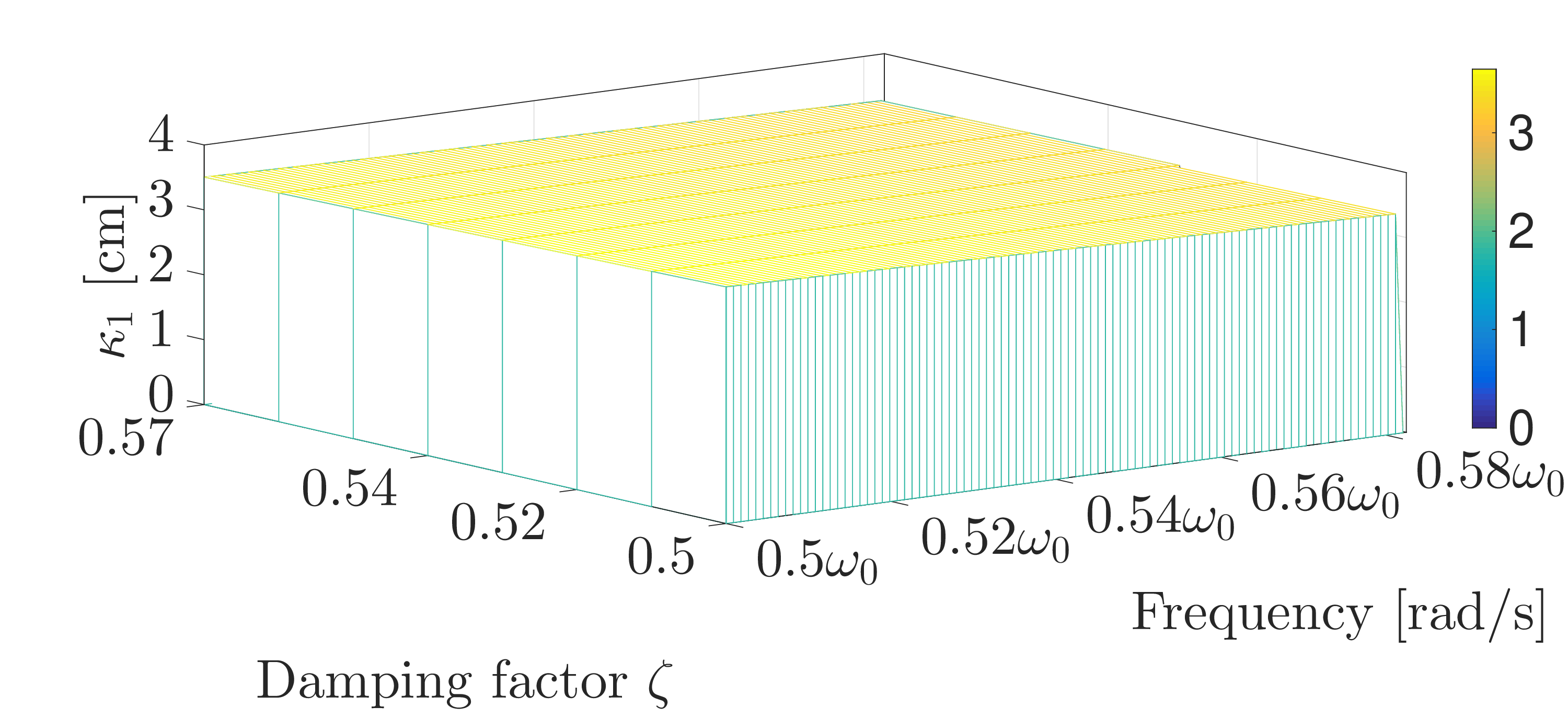}}
\subfigure[Mesh plot of $(\zeta,\omega_{n},\kappa_{2}$).]
	                {\includegraphics[height=3.7cm]{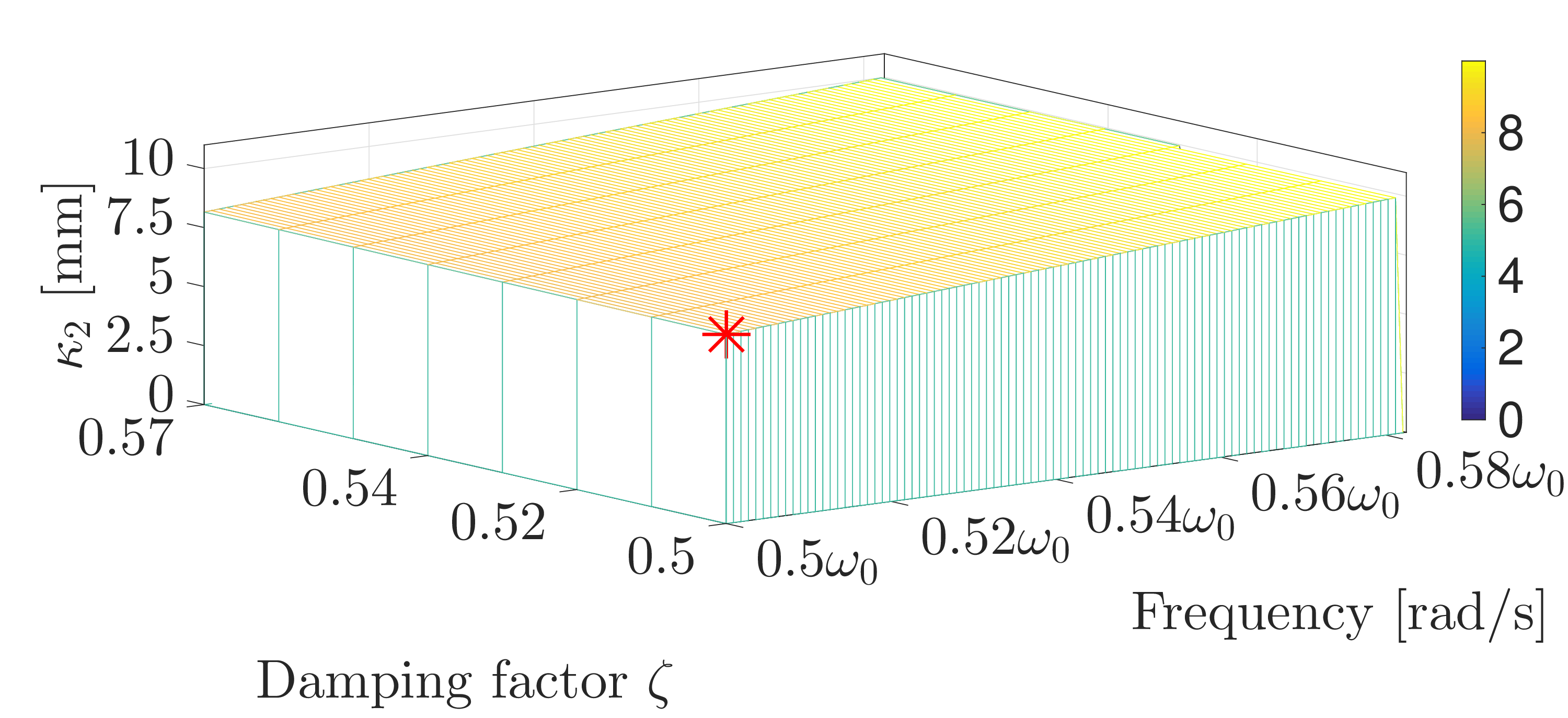}}	
\subfigure[Mesh plot of $(\zeta,\omega_{n},\kappa_{3}$).]
	                {\includegraphics[height=3.7cm]{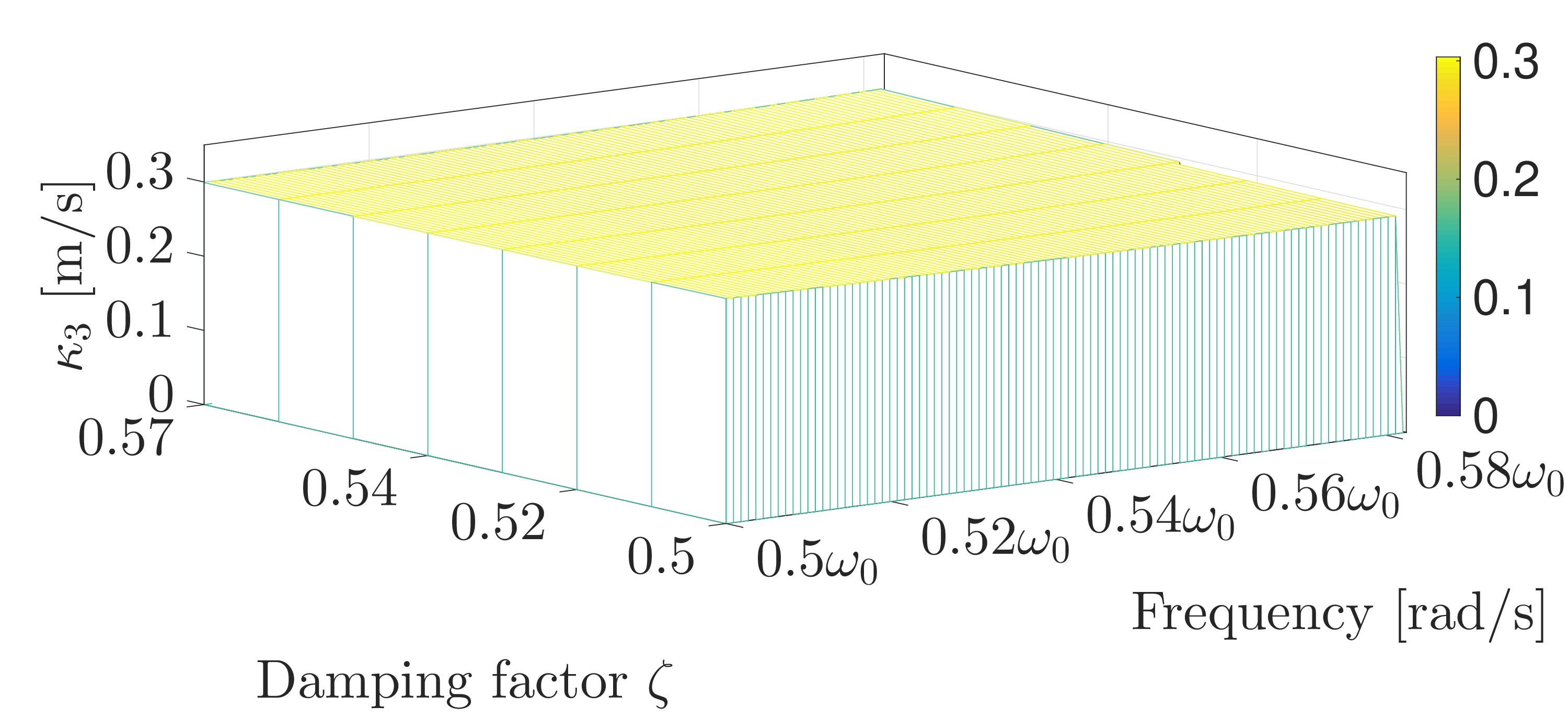}}
\subfigure[Mesh plot of $(\zeta,\omega_{n},\kappa_{u}$).]
	                {\includegraphics[height=3.7cm]{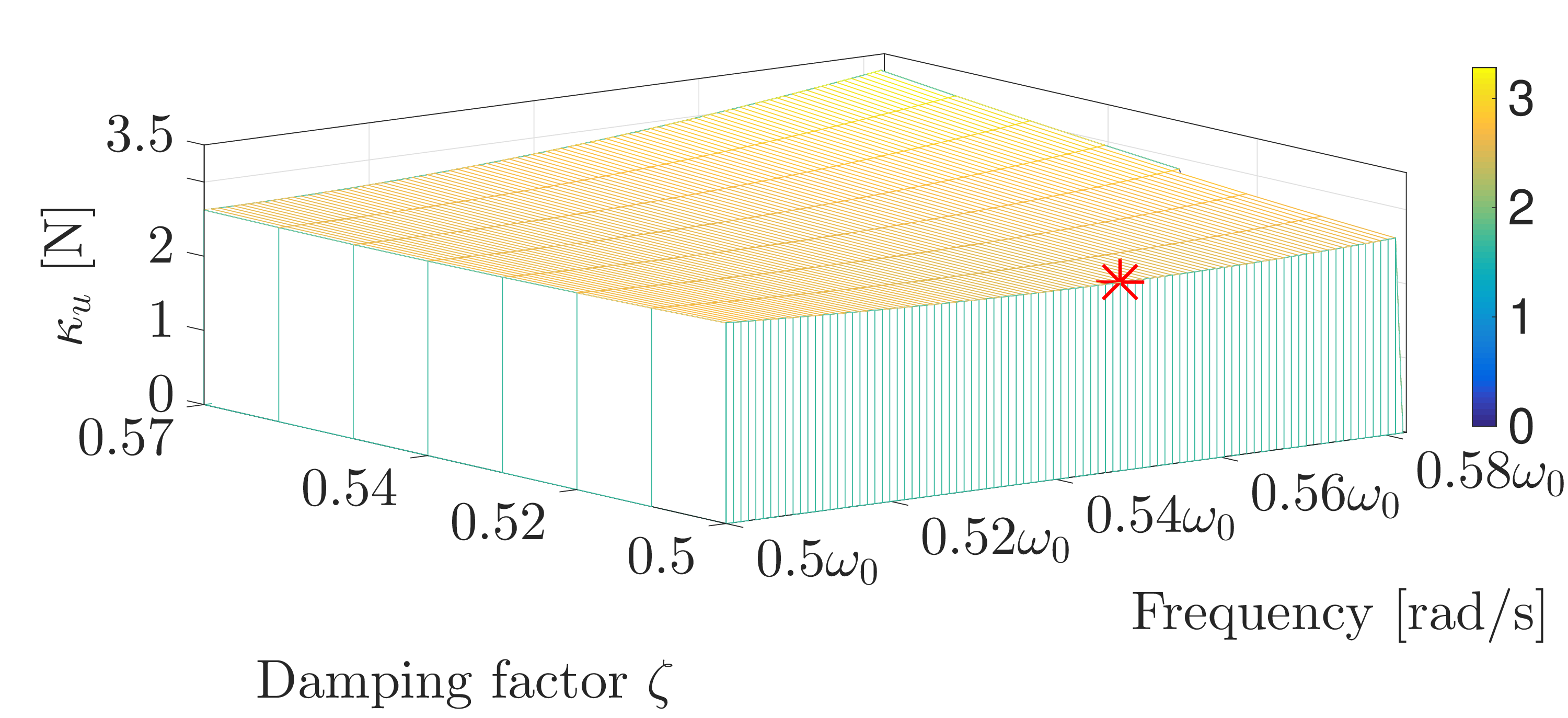}}	
\caption{Three-dimensional surfaces of $\kappa_{1}$, $\kappa_{2}$, $\kappa_{3}$ and $\kappa_{u}$.}
\label{fig:kappa_e}
\end{figure} 
\end{center}

In order to compute the parameters $M_{0}$ and $\boldsymbol{\eta}_{*}$ of the SMC, the tuning algorithm is initialized with the parameters employed in the numerical simulation described in section \ref{sec:simulations}, with the exception of parameter $\delta$, that is set as $\delta=\SI{3}{\m / \s^2}$, and the maximum allowed values of signals $z_{1}$, $z_{2}$, $z_{3}$ and $u$, which are specified by means of the upper bounds $\bar{\kappa}_{1}=\SI{5}{\cm}$, $\bar{\kappa}_{2}=\SI{10}{\mm}$, $\bar{\kappa}_{3}=\SI{32}{\cm / \s}$, and $\bar{\kappa}_{u}=\SI{10}{\N}$, respectively. The tuning algorithm returns the mesh plots of $\kappa_{1}$, $\kappa_{2}$, $\kappa_{3}$ and $\kappa_{u}$, which are displayed in Figure \ref{fig:kappa_e} and indicate that the feasible intervals for $\zeta$ and $\omega_{n}$ are $\zeta \in [0.5,0.57]$ and $\omega_{n} \in [0.5\omega_{0},0.58\omega_{0}]$. The parameters returned by the proposed tuning algorithm are presented in Table \ref{tab:tuples2}. 

\begin{table*}[ht]
\centering
\caption{Tuples $\Gamma_{*}$ returned by the proposed algorithm, and closed-loop system parameters with the LQR and OSMC.}
\label{tab:tuples2}
\begin{adjustbox}{width=1\textwidth}
\begin{tabular}{ccccccccccccc} \toprule
\multirow{2}{*}{PI} & \multirow{2}{*}{$\zeta$} & $\omega_{n}$  & $\kappa_{1}$ & $\kappa_{2}$ & $\kappa_{3}$ & $\kappa_{u}$ & $\lambda_{1,2}$ & $\lambda_{3}$ & $\psi_{1}$ & $\psi_{2}$ & \multicolumn{2}{c}{SMC parameters}\\%

\cmidrule(lr){4-7} \cmidrule(lr){8-9} \cmidrule(lr){10-11} \cmidrule(lr){12-13}
  &  & (rad/s)   & (cm)  & (mm) & (cm/s)& (N) & (rad/s) & (rad/s) & (rad/s) & (rad/s) & Sliding variable  & Gain (N)  \\ \midrule

{$J_{z_{2}}$} & {0.50} & {0.50$\omega_{0}$} &  {3.65} & {8.0} & {30.33} & {2.71} & {-$2.77 \pm 4.80j$} & {-8.31} & {-99.15} & {-3.33} &  {$\boldsymbol{\eta}_{*}^{\mathrm{T}}=$[1.64,-19.94,0.49,-0.20]}  &{$M_{0}=13.52$} \\

{$J_{u}$} & {0.50} & {0.55$\omega_{0}$} &  {3.50} & {9.2} & {29.61} & {2.54} & {-$3.04 \pm 5.26j$} & {-9.11} & {162.45} & {-3.65} & {$\boldsymbol{\eta}_{*}^{\mathrm{T}}$=[2.16,-20.64,0.59,0.13]} &{$M_{0}=12.05$} \\

{$J_{\mathrm{LQR}}$} & {0.46} & {0.69$\omega_{0}$} &  {--} & {--} & {--} & {--} & {-$3.52 \pm 6.82j$} & {-6.77} & {--} & {--} & {--} &{--} \\

{$J_{\mathrm{OSMC}}$} & {0.46} & {0.69$\omega_{0}$} &  {--} & {--} & {--} & {--} & {-$3.48 \pm 6.77j$} & {-6.86} & {--} & {--} & {$\boldsymbol{\vartheta}^{\mathrm{T}}$=[2.55,-16.81,0.68,0.41]} &{$M_{1}=13.52$} \\
\bottomrule
\end{tabular}
\end{adjustbox}
\end{table*}

\begin{center}
\begin{figure}[ht]
\centering
\subfigure[$H_{1}(j\omega,\boldsymbol{\eta}_{*})$.]
	                {\includegraphics[height=3.7cm]{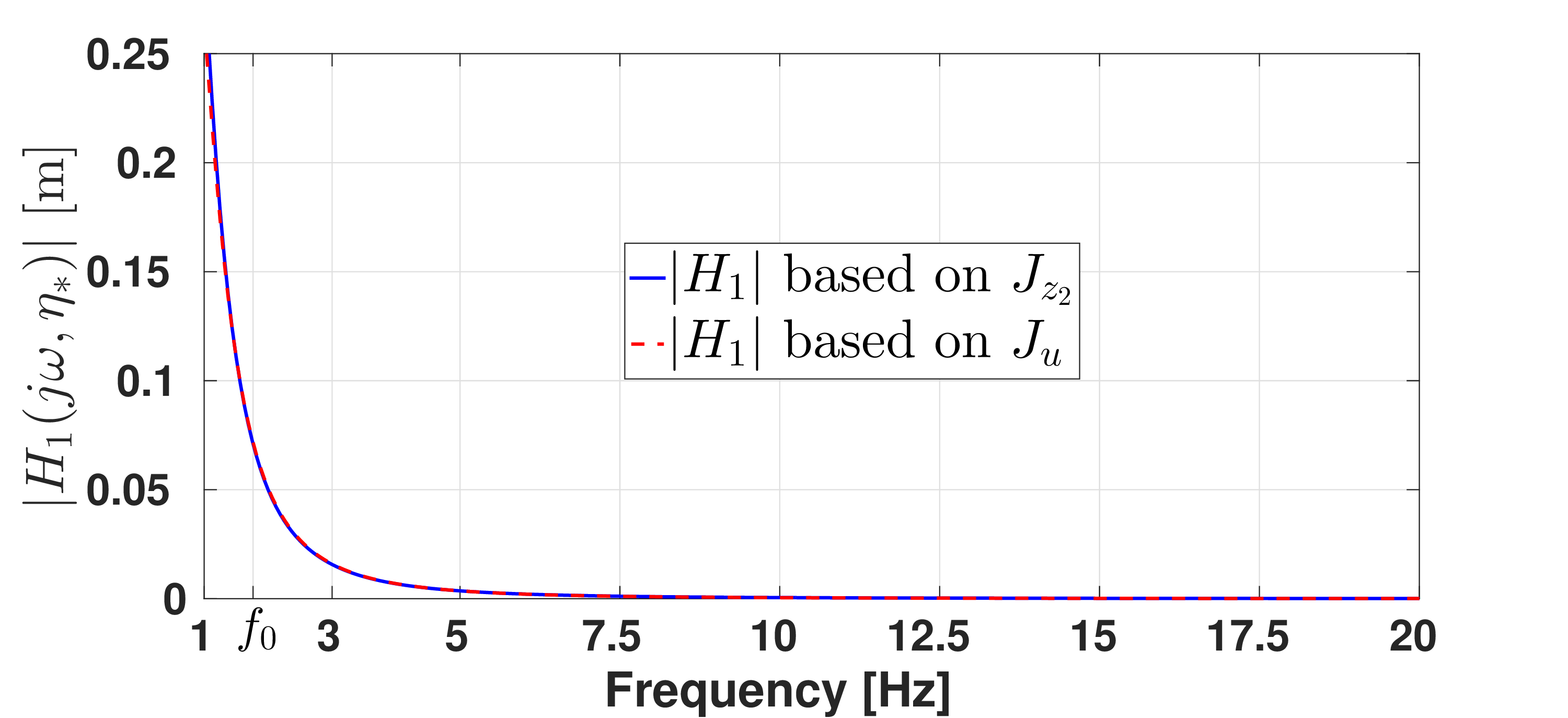}}
\subfigure[$H_{2}(j\omega,\boldsymbol{\eta}_{*})$.]
	                {\includegraphics[height=3.7cm]{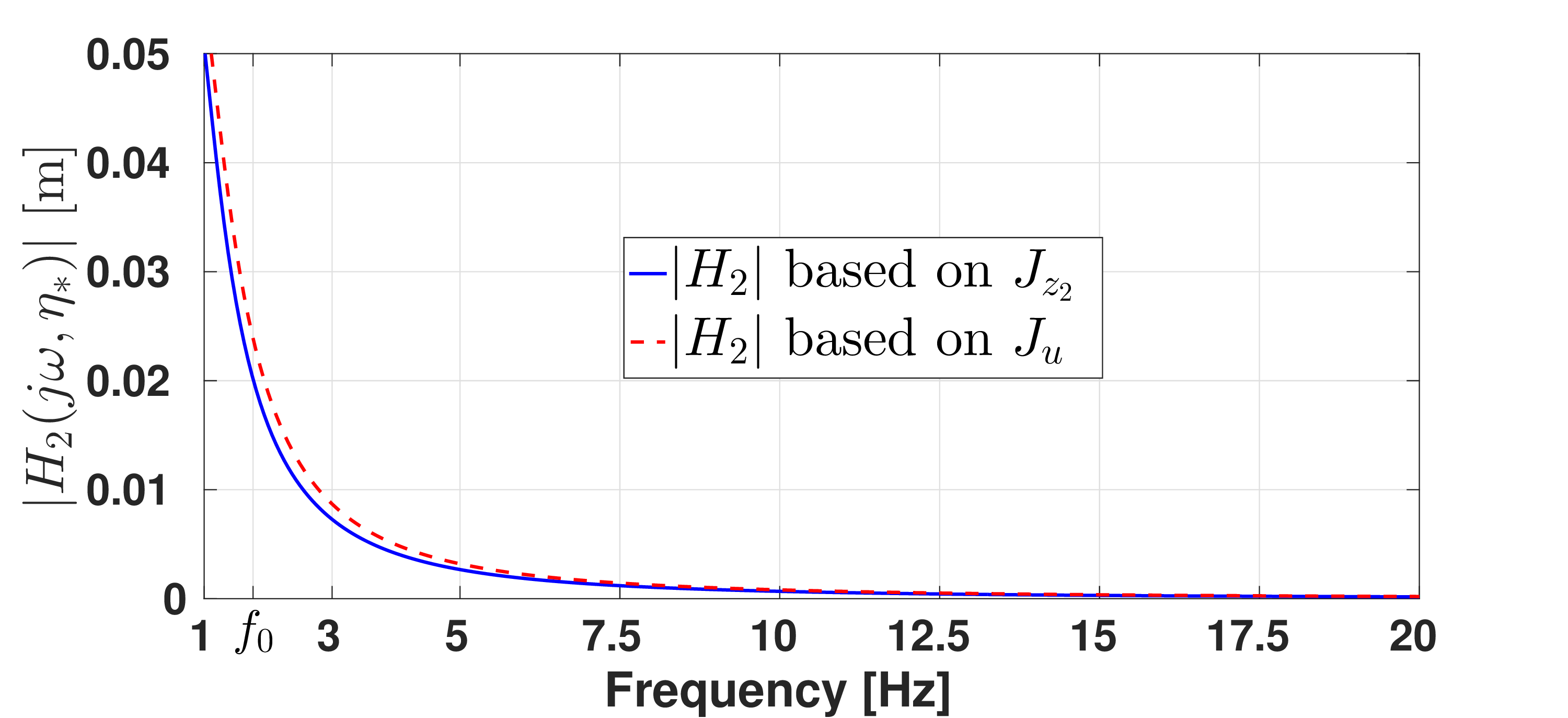}}	
\subfigure[$H_{3}(j\omega,\boldsymbol{\eta}_{*})$.]
	                {\includegraphics[height=3.7cm]{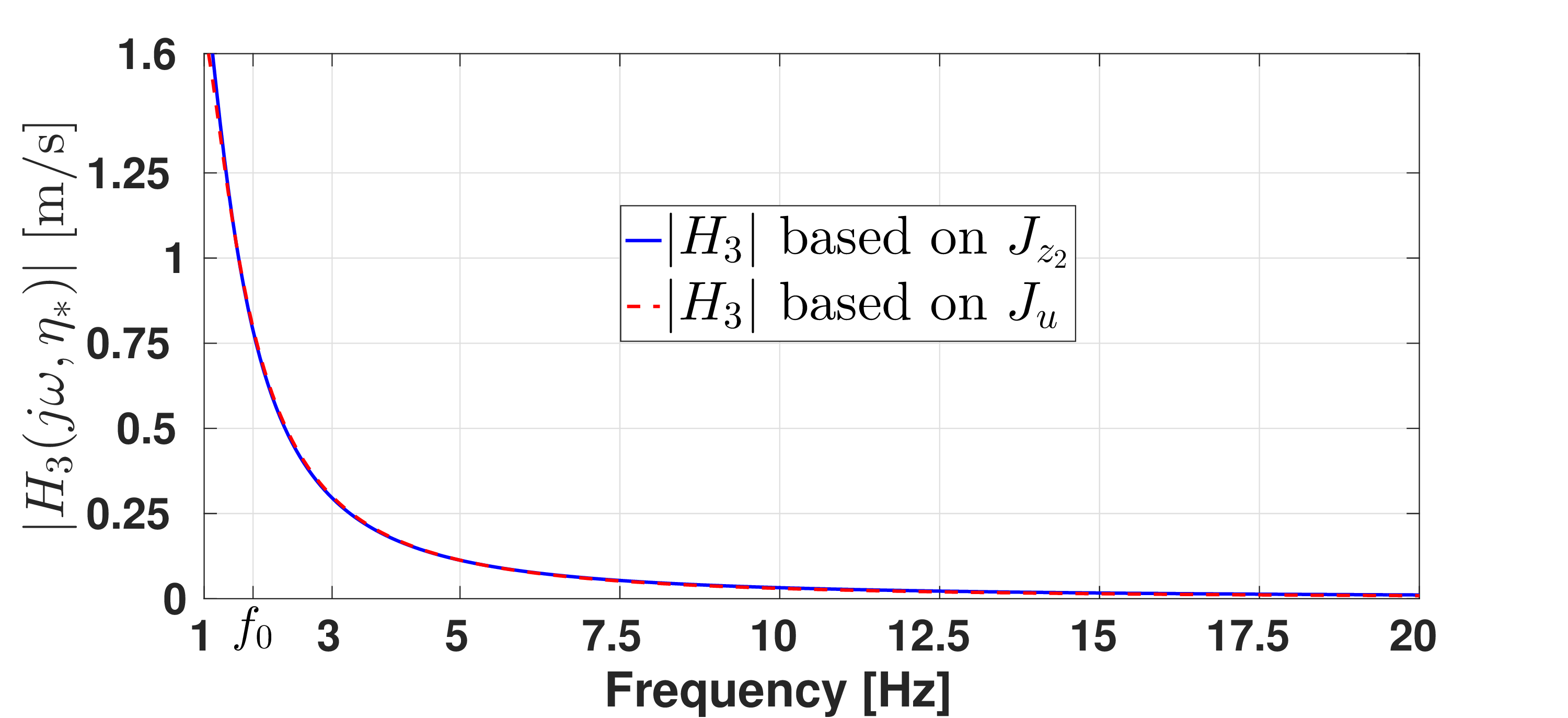}}
\subfigure[$H_{u}(j\omega,\boldsymbol{\eta}_{*})$.]
	                {\includegraphics[height=3.7cm]{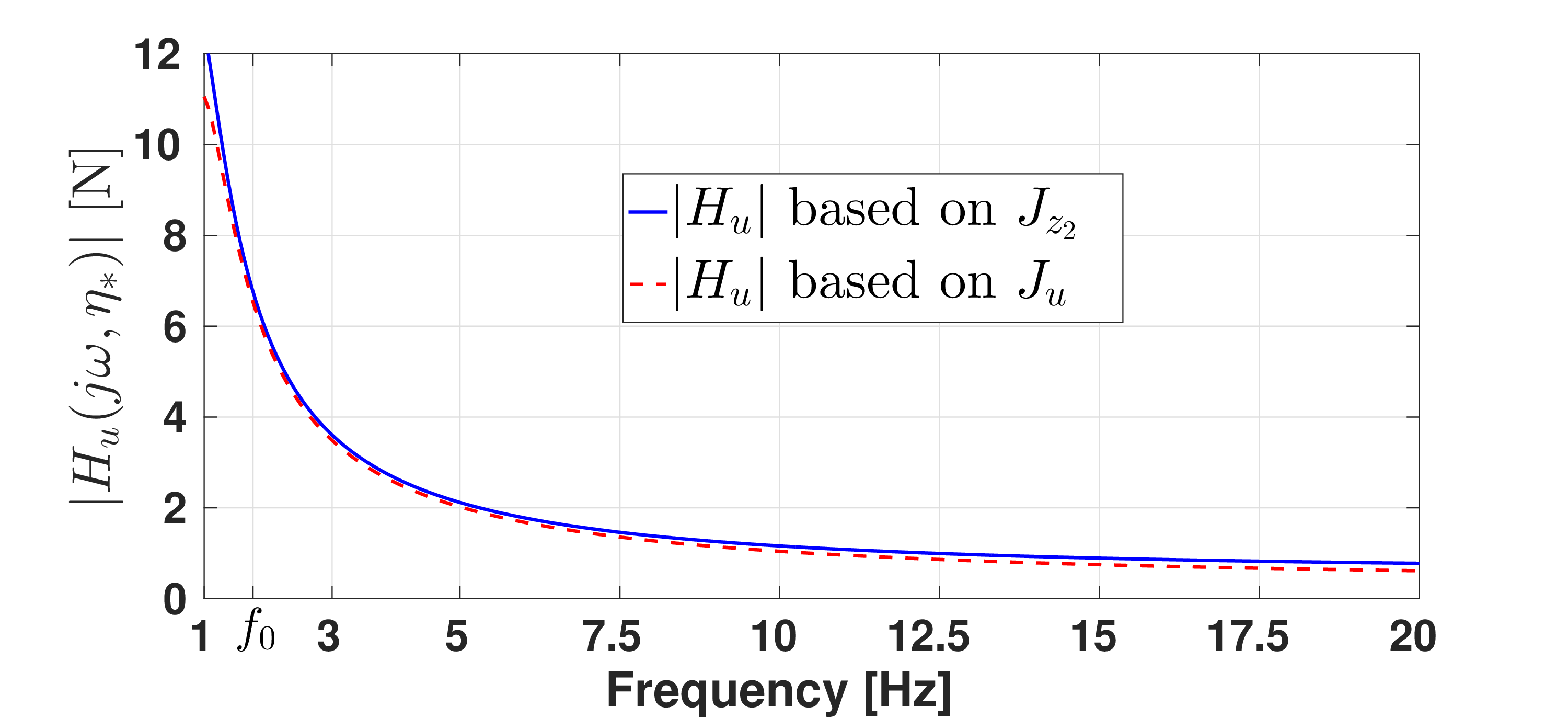}}	
\caption{Magnitude of frequency responses of $H_{i}(j\omega,\boldsymbol{\eta}_{*})$, $i=1,2,3,u$ using the vector $\boldsymbol{\eta}_{*}$ that minimizes the PI $J_{z_{2}}$.}
\label{fig:fr}
\end{figure}          
\end{center}

Figure \ref{fig:fr} presents the frequency responses of the filters $H_{i}(j\omega,\boldsymbol{\eta}_{*})$, $i=1,2,3,u$ corresponding to the vectors $\boldsymbol{\eta}_{*}$ that minimize the PIs $J_{z_{2}}$ and $J_{u}$. This figure also indicates the natural frequency of the structure $f_{0}=\omega_{0}/(2\pi)$ in Hz. It shows that at low frequencies, the magnitude of $H_{i}(j\omega,\boldsymbol{\eta}_{*})$ is larger than $\bar{\kappa}_{i}$; however, the RMS value $\kappa_{i}$ of $H_{i}(j\omega,\boldsymbol{\eta}_{*})$ is smaller than $\bar{\kappa}_{i}$ in the bandwidth of the earthquake $\ddot{x}_{g}$. 

\subsubsection{Comparison with the LQR}
The proposed SMC is compared with the LQR \cite{Franklin:2015}, which minimizes the PI  
\begin{equation}\label{e:id}
J_{\mathrm{LQR}}=\int_{0}^{\infty}(\mathbf{z^{\mathrm{T}}Qz}+ru^{2})dt
\end{equation}
where $\mathbf{Q}\in \mathbb{R}^{4 \times 4}$ is a symmetric matrix such that $\mathbf{Q}\geq 0$; moreover $r$ is a positive constant. Both $\mathbf{Q}$ and $r$ are designed according to the literature \cite{Bryson:1975,Franklin:2015}, where the entries of $\mathbf{Q}=\mbox{diag}[q_{11},q_{22},q_{33},q_{44}]$ satisfy
\begin{equation}\label{e:id5}
q_{ii}=\dfrac{1}{\mbox{maximum acceptable value of} \ z_{i}^{2}}, \qquad i=1,2,3,4
\end{equation}
and the parameter $r$ is given by
\begin{equation}\label{e:id6}
r=\dfrac{1}{\mbox{maximum acceptable value of} \ u^{2}}
\end{equation}

The maximum acceptable values for $z_{i}$, $i=1,2,3$ and $u$ are, respectively, specified by the constants $\bar{\kappa}_{j}$, $j=1,2,3,u$, which were mentioned in Section \ref{sec:er}. The remaining maximum acceptable value for $z_{4}$ is chosen as 0.1 m/s. Thus, coefficients $q_{ii}$, $i=1,2,3,4$ in (\ref{e:id5}) and $r$ in (\ref{e:id6}) are given by:
\begin{equation}
q_{11}=400, \quad q_{22}=10000, \quad q_{33}=9.77, \quad q_{44}=100, \quad r=0.01
\end{equation}
The control law $u$ that minimizes the performance index $J_{\mathrm{LQR}}$ in (\ref{e:id}) is computed as 
\begin{equation}
    u=-\mathbf{k}_{\mathrm{J}}\mathbf{z}
\end{equation}
where $\mathbf{k}_{\mathrm{J}}=r^{-1}\mathbf{B}^{\mathrm{T}}\mathbf{P}_{1}$, and matrix $\mathbf{P}_{1}=\mathbf{P}_{1}^{\mathrm{T}}>0$ is the solution of the Riccati equation 
\begin{equation}\label{e:id3}
\mathbf{A}^{\mathrm{T}}\mathbf{P}_{1}+\mathbf{P}_{1}\mathbf{A}-\mathbf{P}_{1}\mathbf{B}r^{-1}\mathbf{B}^{\mathrm{T}}\mathbf{P}_{1}=-\mathbf{Q}
\end{equation}

Solving (\ref{e:id3}) yields $\mathbf{k}_{\mathrm{J}}=[200,\ -1276.5,\ 49, \ 17.32]$, and the closed-loop eigenvalues $\lambda_{1}$, $\lambda_{2}$ and $\lambda_{3}$ given in Table \ref{tab:tuples2}, as well as $\lambda_{4}=-77.98$. Using the dominant poles $\lambda_{1}$ and $\lambda_{2}$ permits computing the parameters $\zeta$ and $\omega_{n}$ shown in this Table. Note that these LQR parameters do not belong to the feasible intervals of $\zeta$ and $\omega_{n}$ corresponding to the SMC. 

\subsubsection{Comparison with the OSMC}

The designed SMC is also compared with the OSMC \cite{utkin2013}, that is based on the minimization of the following PI: 
\begin{equation}\label{e:pi_osmc}
J_{\mathrm{OSMC}}=\int_{0}^{\infty}\mathbf{z^{\mathrm{T}}Qz} dt
\end{equation}

Note that the PI {(\ref{e:pi_osmc})} is quite different from the PI {(\ref{e:id})} corresponding to the LQR, since the former does not impose a penalty cost on the control effort $u$. Meanwhile, both of these PIs use the same matrix $\mathbf{Q}$.

To determine the OSMC, the structure model (\ref{e:6}) needs to be written in its regular form \cite{alwi2011}. For that, the vector $\mathbf{B}$ of this model is partitioned as $\mathbf{B}=[\boldsymbol{\mathcal{B}}_{1} \ \mathcal{B}_{2}]^{\mathrm{T}}$, where $\boldsymbol{\mathcal{B}}_{1}=\left[0,0,(m_{0}+m_{d})/m_{0}m_{d}\right]^{\mathrm{T}}$ and $\mathcal{B}_{2}=-1/m_{0}$. This vector is used in the following non-singular transformation matrix $\mathbf{T}_{2}$, which produces a state transformation from $\mathbf{z}$ to $\mathbf{v}\in \mathbb{R}^{4\times 1}$

\begin{equation}\label{e:ct}
\mathbf{v}=\left[\begin{array}{c} \mathbf{v}_{1} \\ \mathrm{v}_{2} \end{array}\right]=\overbrace{\left[\begin{array}{cc} \mathbf{I}_{3\times3} & -\boldsymbol{\mathcal{B}}_{1}\mathcal{B}_{2}^{-1} \\  \mathbf{0}_{3\times1}  & \mathcal{B}_{2}^{-1} \end{array}\right]}^{\mathbf{T}_{2}}\mathbf{z}=\mathbf{T}_{2}\mathbf{z}
\end{equation}

In the new state $\mathbf{v}$, the building model {(\ref{e:6})} is written in the following regular form:
\begin{align}
    \mathbf{\dot{v}}_{1}=& \mathbf{A}_{11}\mathbf{v}_{1}+ \mathbf{A}_{12}\mathrm{v}_{2}+\boldsymbol{\mathcal{D}}_{1}\ddot{x}_{g}  \label{rf1}\\
    \dot{\mathrm{v}}_{2}=&\mathbf{A}_{21}\mathbf{v}_{1}+ A_{22}\mathrm{v}_{2}+(u-f(z_{3}))+\mathcal{D}_{2}\ddot{x}_{g}
\end{align}
where $\mathbf{A}_{11} \in \mathbb{R}^{3\times 3}$, $\mathbf{A}_{12} \in \mathbb{R}^{3\times 1}$, $\mathbf{A}_{21} \in \mathbb{R}^{1\times 3}$, $A_{22} \in \mathbb{R}$, $\boldsymbol{\mathcal{D}}_{1}=\left[0,0,-(\beta_{0}m_{0}+m_{d})/m_{d}\right]^{\mathrm{T}}$, and $\mathcal{D}_{2}=\beta_{0}m_{0}$. 

For obtaining the control effort $u$ that minimizes (\ref{e:pi_osmc}), it is assumed that  $\boldsymbol{\mathcal{D}}_{1}=\mathbf{0}$ \cite{khatibinia2020optimal}. This assumption leads to the optimal sliding surface given by
\begin{equation}\label{e:ss}
    \sigma=\boldsymbol{\vartheta}^{\mathrm{T}}\mathbf{z}, \quad \mbox{with} \quad \boldsymbol{\vartheta}^{\mathrm{T}}=\left[ \boldsymbol{\mathcal{K}}, \ 1\right]\mathbf{T}_{2}
\end{equation}
where 
\begin{equation}\label{e:gain_k}
\boldsymbol{\mathcal{K}}=-Q_{22}^{-1}\left(\mathbf{A}_{12}^{\mathrm{T}} \mathbf{P}_{2}+\mathbf{Q}_{21}\right), \quad \mbox{and} \quad \left[\mathbf{T}_{2}^{-1}\right]^{\mathrm{T}}\mathbf{Q}\mathbf{T}_{2}^{-1}=\left[\begin{array}{cc}
 \mathbf{Q}_{11}    &  \mathbf{Q}_{12}\\
  \mathbf{Q}_{21}   &   Q_{22}
\end{array}      \right]
\end{equation}

Moreover, the matrix $\mathbf{P}_{2}=\mathbf{P}_{2}^{\mathrm{T}}>0$ satisfies the following Riccati equation 
\begin{equation}\label{e:re_osmc3}
\mathbf{\widehat{\mathcal{A}}}^{\mathrm{T}}\mathbf{P}_{2}+\mathbf{P}_{2}\mathbf{\widehat{\mathcal{A}}}-\mathbf{P}_{2} \mathbf{A}_{12}Q_{22}^{-1} \mathbf{A}_{12}^{\mathrm{T}}\mathbf{P}_{2}=-\mathbf{\widehat{\mathcal{Q}}}
\end{equation}
with
\begin{equation}
    \mathbf{\widehat{\mathcal{A}}}=\mathbf{A}_{11}- \mathbf{A}_{12}Q_{22}^{-1} \mathbf{Q}_{21}, \quad \mbox{and} \quad \mathbf{\widehat{\mathcal{Q}}}=\mathbf{Q}_{11}- \mathbf{Q}_{12}Q_{22}^{-1} \mathbf{Q}_{21}
\end{equation}

Finally, the control law $u$ of the OSMC that minimizes the PI (\ref{e:pi_osmc}) is given by \cite{alwi2011}:
\begin{equation}\label{e:u_osmc}
    u=-\left( \boldsymbol{\vartheta}^{\mathrm{T}}\mathbf{B} \right)^{-1}\left[\boldsymbol{\vartheta}^{\mathrm{T}} \mathbf{A}\mathbf{z}+M_{1}\mbox{sign} (\sigma)  \right]
\end{equation}
where the switching gain $M_{1}$ is selected to satisfy the reachability condition $\sigma \dot{\sigma}<0$, such that the system trajectories reach the sliding surface $\sigma=\boldsymbol{\vartheta}^{\mathrm{T}}\mathbf{z}=0$ in finite time. The product $\sigma \dot{\sigma}$ satisfies
\begin{equation}\label{e:ineqf}
\sigma \dot{\sigma}=\sigma\left[-M_{1}\mbox{sign} (\sigma) - \boldsymbol{\vartheta}^{\mathrm{T}}\mathbf{B}f(z_{3})+\boldsymbol{\vartheta}^{\mathrm{T}}\mathbf{D}\ddot{x}_{g}  \right] \leq -|\sigma|[M_{1}-(|\boldsymbol{\vartheta}^{\mathrm{T}}\mathbf{B}|\varpi+|\boldsymbol{\vartheta}^{\mathrm{T}}\mathbf{D}|\delta)]<0
\end{equation}
for
\begin{equation}\label{e:M1}
    M_{1}>|\boldsymbol{\vartheta}^{\mathrm{T}}\mathbf{B}|\varpi+|\boldsymbol{\vartheta}^{\mathrm{T}}\mathbf{D}|\delta
\end{equation}

It is worth mentioning that the closed-loop system in the sliding mode $\sigma=0$ is reduced to the following third-order system:
\begin{equation}\label{e:ros}
        \mathbf{\dot{v}}_{1}=\left[\mathbf{A}_{11}- \mathbf{A}_{12}\boldsymbol{\mathcal{K}}\right]\mathbf{v}_{1}
\end{equation}

Like in the case of the proposed SMC, the sign term of the OSMC $u$ in (\ref{e:u_osmc}) is substituted by a saturation function in order to apply it to the experimental prototype. Solving the Riccati equation (\ref{e:re_osmc3}) produces the gain $\boldsymbol{\mathcal{K}}=[2.55,\ -16.81,\ 0.68]$ in (\ref{e:gain_k}), and substituting it into (\ref{e:ss}) yields the vector $\boldsymbol{\vartheta}$ shown in Table \ref{tab:tuples2}. The switching gain should satisfy $M_{1}>\SI{2.97}{\N}$, which is fulfilled by selecting $M_{1}=\SI{13.52}{\N}$, that is equal to the value $M_{0}$ of the SMC designed with the proposed algorithm by minimizing the PI $J_{z_{2}}$. The OSMC generates the closed-loop system (\ref{e:ros}), whose eigenvalues $\lambda_{1}$, $\lambda_{2}$ and $\lambda_{3}$ are given in Table \ref{tab:tuples2}. Note that the parameters $\zeta$ and $\omega_{n}$ of this system are equal to those obtained with the LQR.

\begin{remark}
By comparing the tuned SMC based on the Ackermann's formula and the OSMC, it is possible to establish the following differences: \begin{enumerate}
    \item Tuning the SMC requires the transformation matrix $\mathbf{T}_{1}$ in (\ref{e:15}), whereas the design of the OSMC is based on the matrix $\mathbf{T}_{2}$ in (\ref{e:ct}) that converts the building model to the regular form. 
    \item To design the SMC based on the Ackermann's formula, it is assumed that the matching condition $\mathbf{D} \in \mathrm{span} (\mathbf{B})$ is not satisfied. Therefore, the proposed tuning algorithm for the SMC considers the effect of the seismic excitation signal $\ddot{x}_{g}$ on the transient and frequency responses of the structure and damper. On the other hand, the design of the OSMC assumes that this matching condition is satisfied, since the effect of $\ddot{x}_{g}$ is omitted in equation (\ref{rf1}) by considering $\boldsymbol{\mathcal{D}}_{1}=\mathbf{0}$. 
     \item The design of the OSMC only considers the minimization of the system states $\mathbf{z}$ and does not consider the control effort $u$. On the other side, the proposed tuning algorithm can minimize either the control effort $u$ of the SMC or the top floor displacement $z_2$, ensuring that the displacements and velocities of the building and damper are within the specified limits.
\end{enumerate}
\end{remark}

\subsubsection{Responses produced by the LQR, OSMC, and the SMC tuned with the proposed algorithm}
\begin{center}
\begin{figure}[H]
\centering
	  {\includegraphics[height=3.7cm]{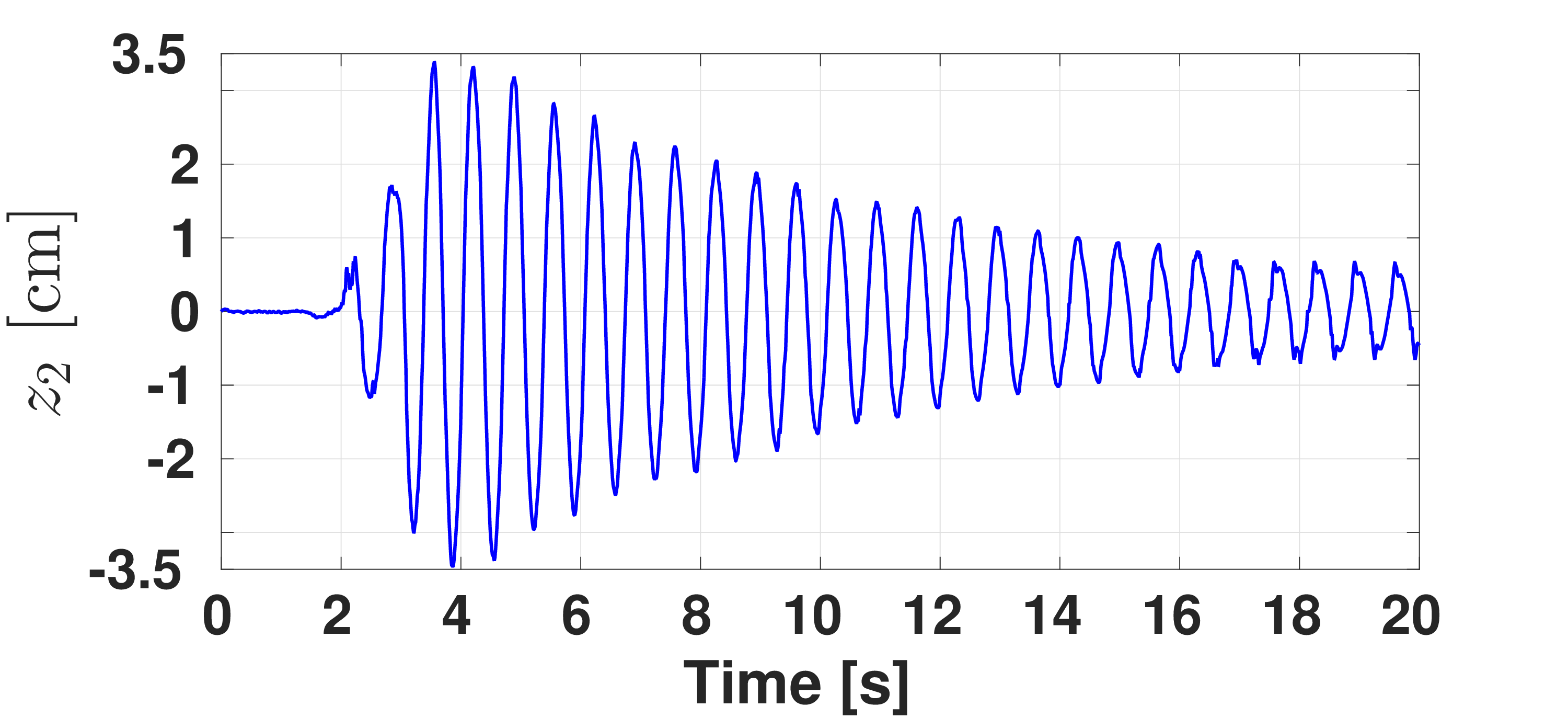}}	
\caption{Uncontrolled response of $z_{2}(t)$.}
\label{fig:7a}
\end{figure}          
\end{center}

\begin{center}
\begin{figure}[H]
\centering
\subfigure[Signals $z_{1}(t)$.]
	            {\includegraphics[height=3.7cm]{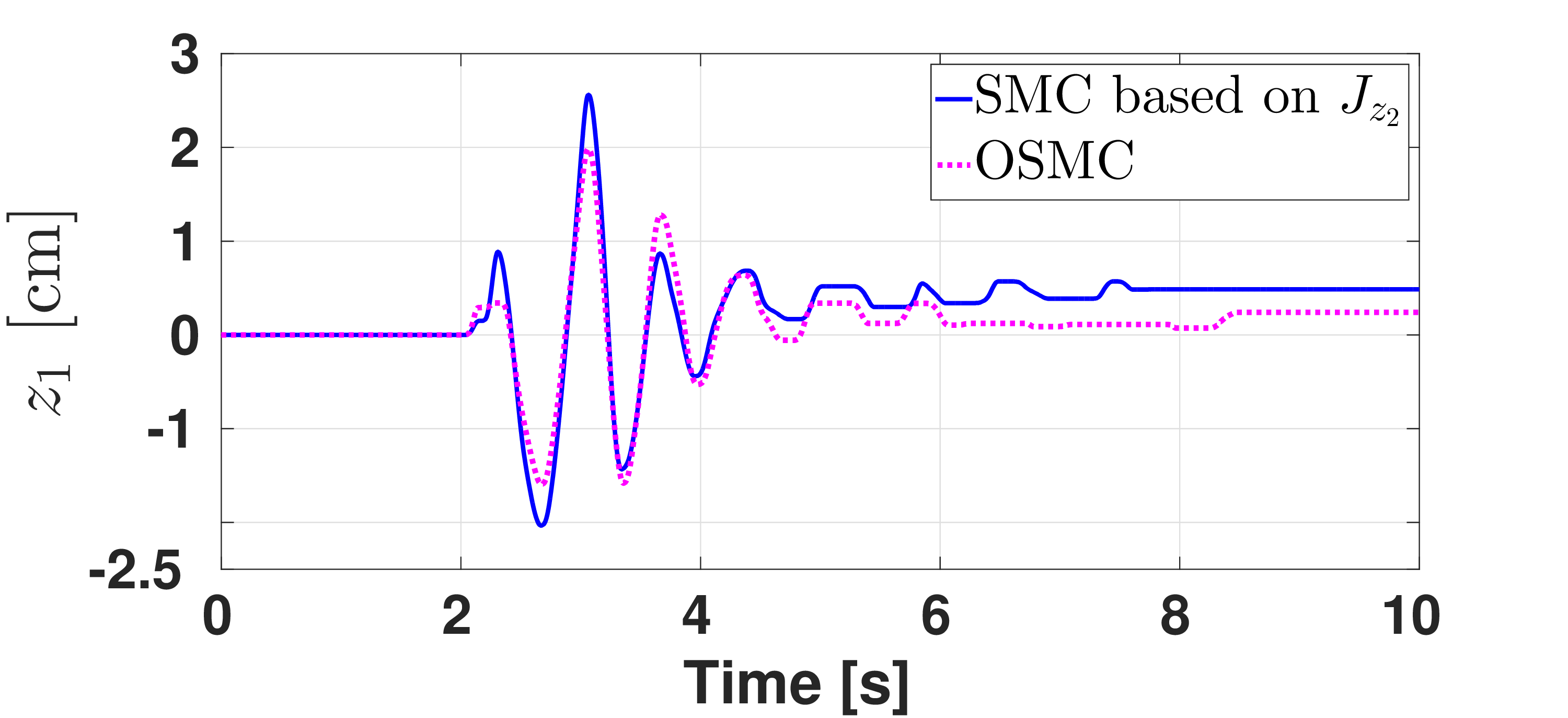}}
\subfigure[Signals $z_{1}(t)$.]
	            {\includegraphics[height=3.7cm]{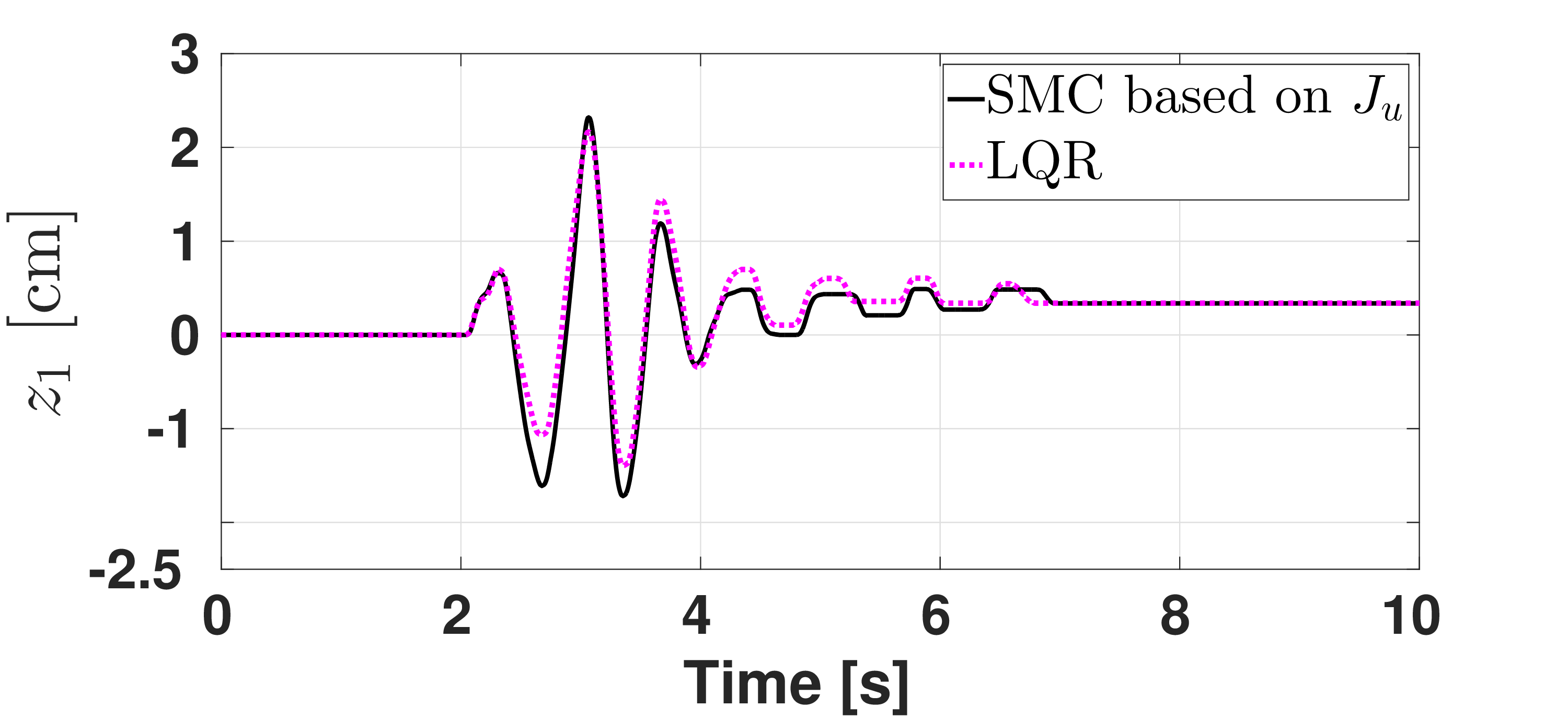}}	
\caption{Displacements of the AMD.}
\label{fig:9}
\end{figure}          
\end{center}

\begin{center}
\begin{figure}[H]
\centering
\subfigure[Signals $z_{2}(t)$.]
	            {\includegraphics[height=3.7cm]{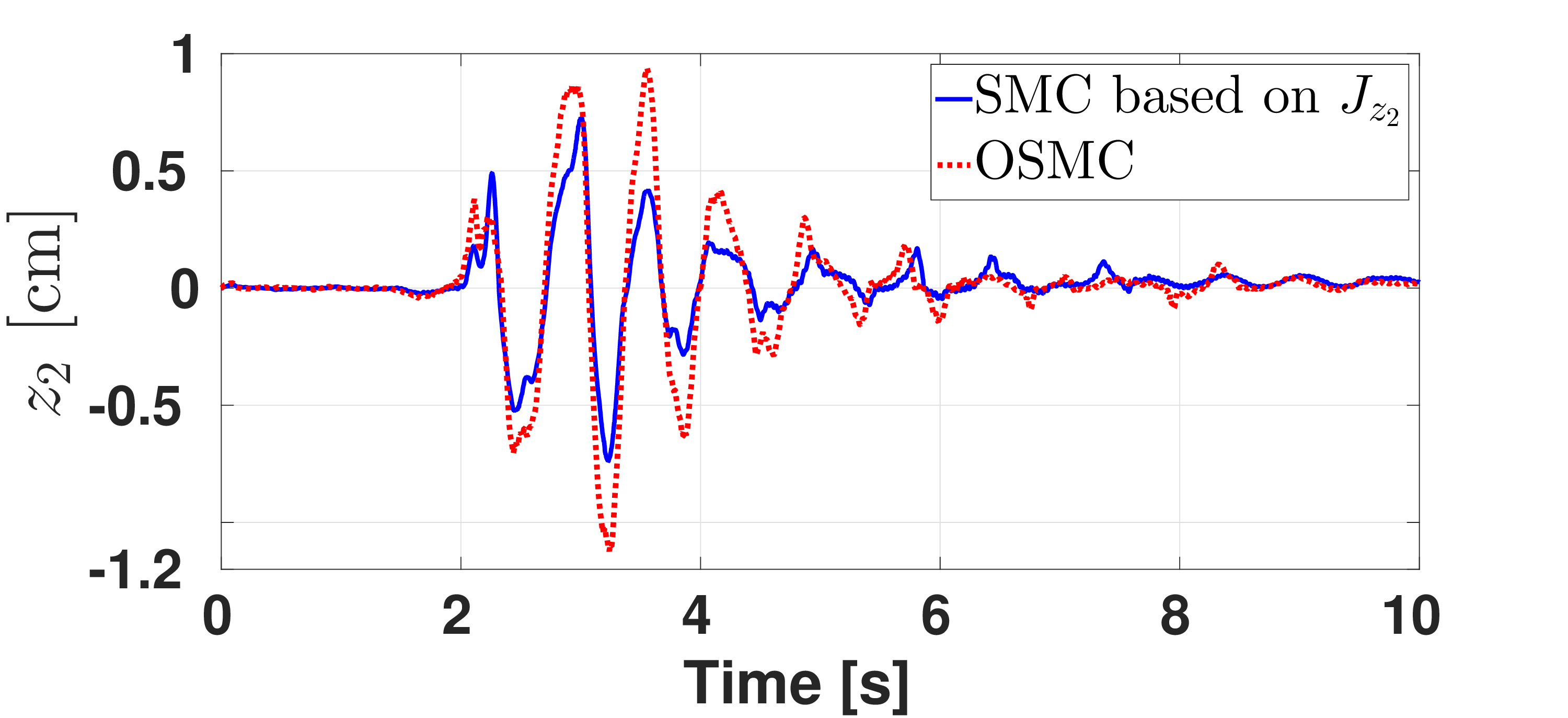}}
\subfigure[Signals $z_{2}(t)$.]
	            {\includegraphics[height=3.7cm]{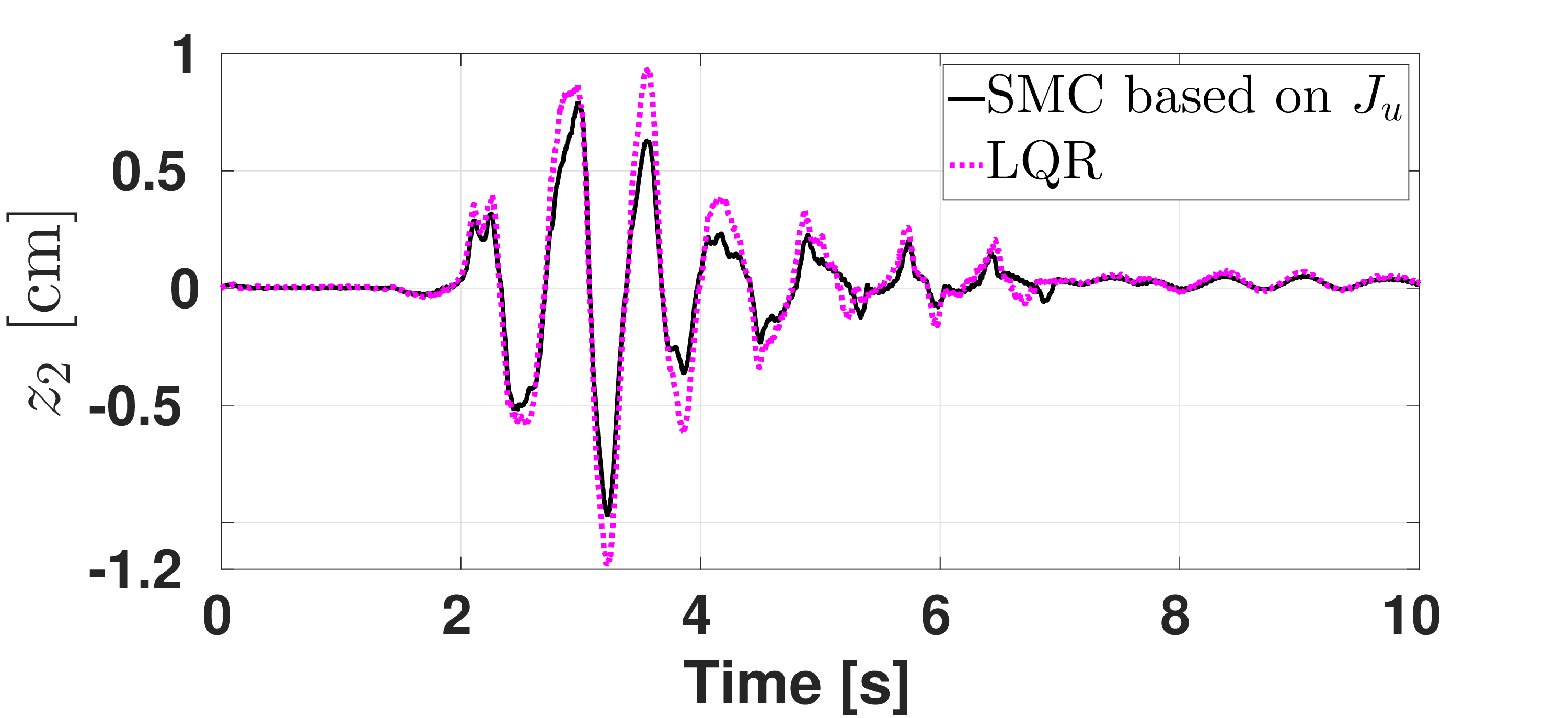}}	
\caption{Displacements of the floor.}
\label{fig:10}
\end{figure}          
\end{center}

\begin{center}
\begin{figure}[H]
\centering
\subfigure[Signals $z_{3}(t)$.]
	            {\includegraphics[height=3.7cm]{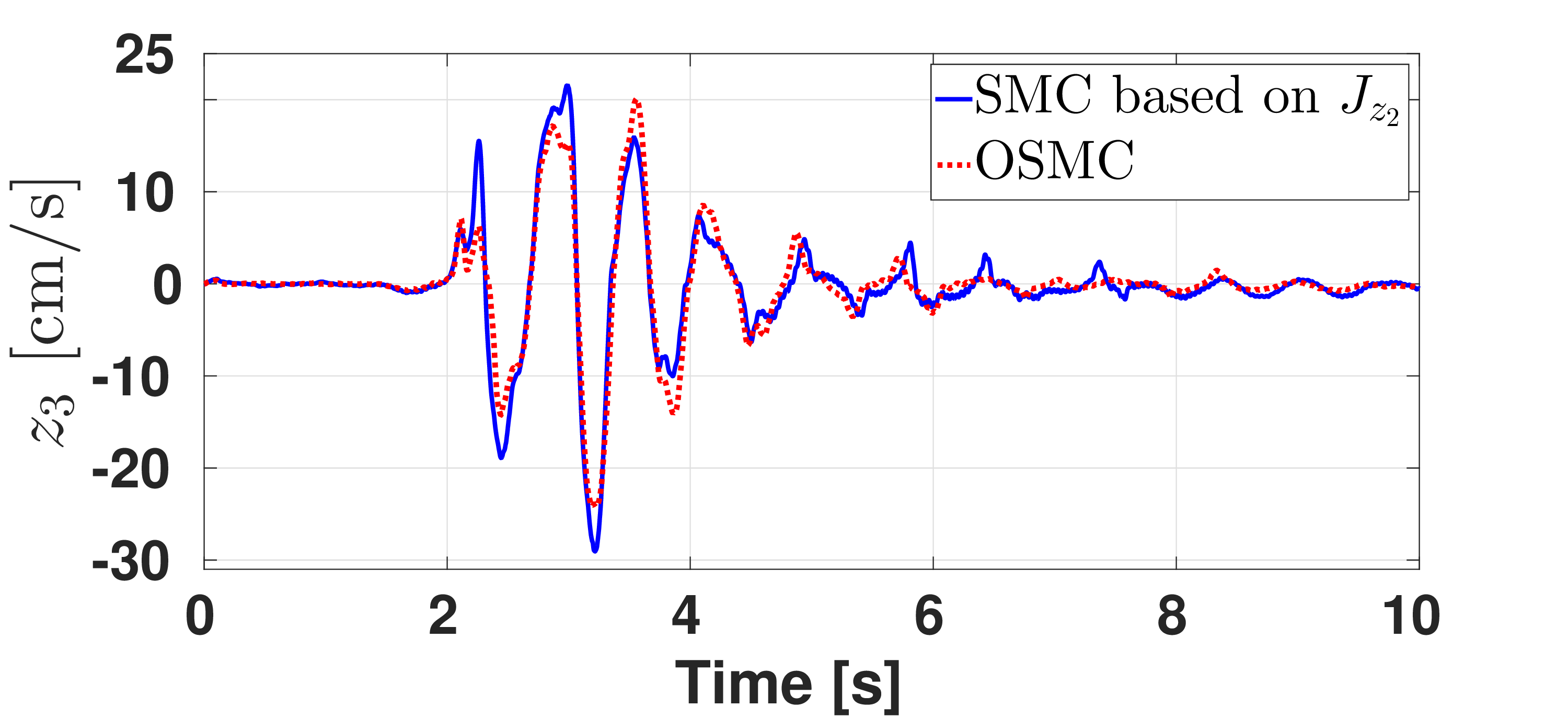}}
\subfigure[Signals $z_{3}(t)$.]
	            {\includegraphics[height=3.7cm]{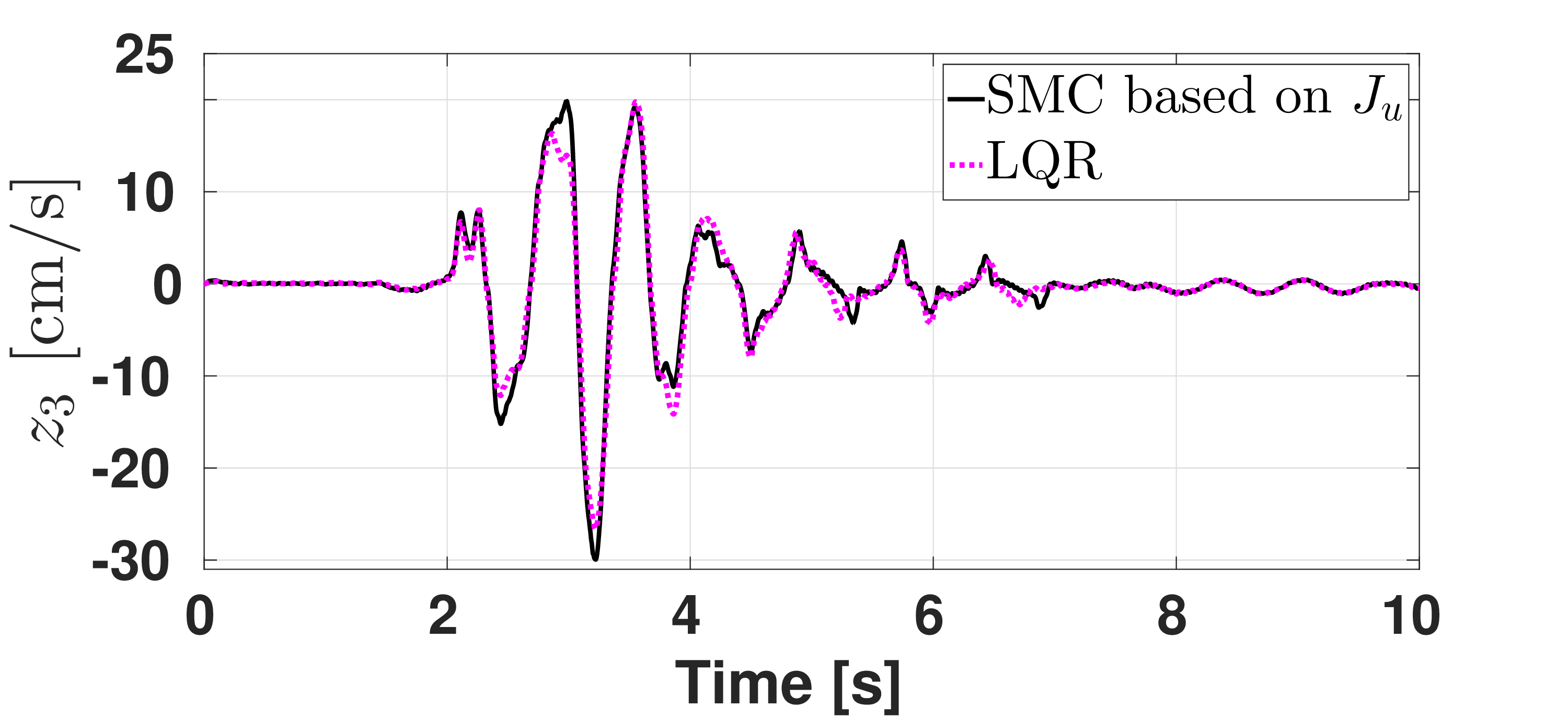}}	
\caption{Velocities of the AMD.}
\label{fig:11}
\end{figure}          
\end{center}

\begin{center}
\begin{figure}[H]
\centering
\subfigure[Signals $z_{4}(t)$.]
	            {\includegraphics[height=3.7cm]{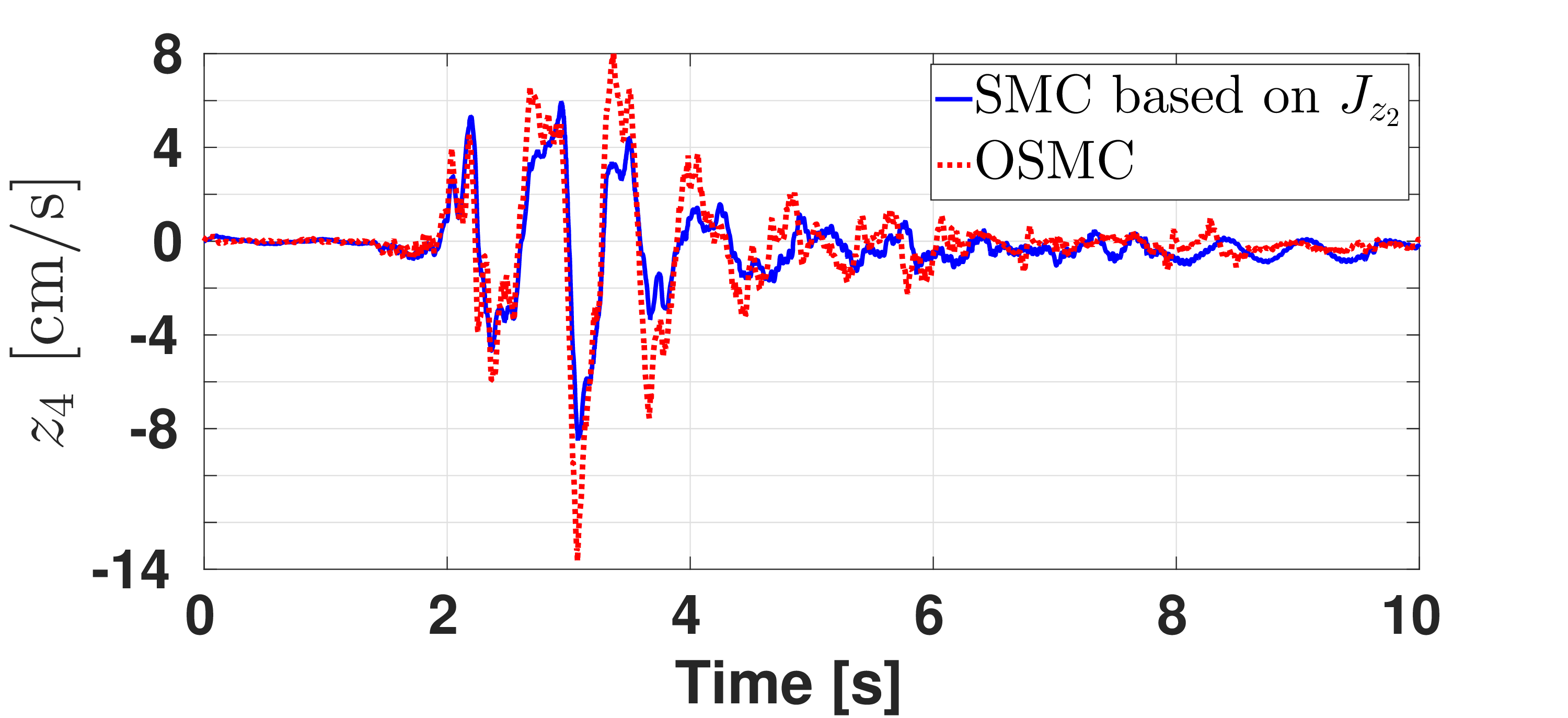}}
\subfigure[Signals $z_{4}(t)$.]
	            {\includegraphics[height=3.7cm]{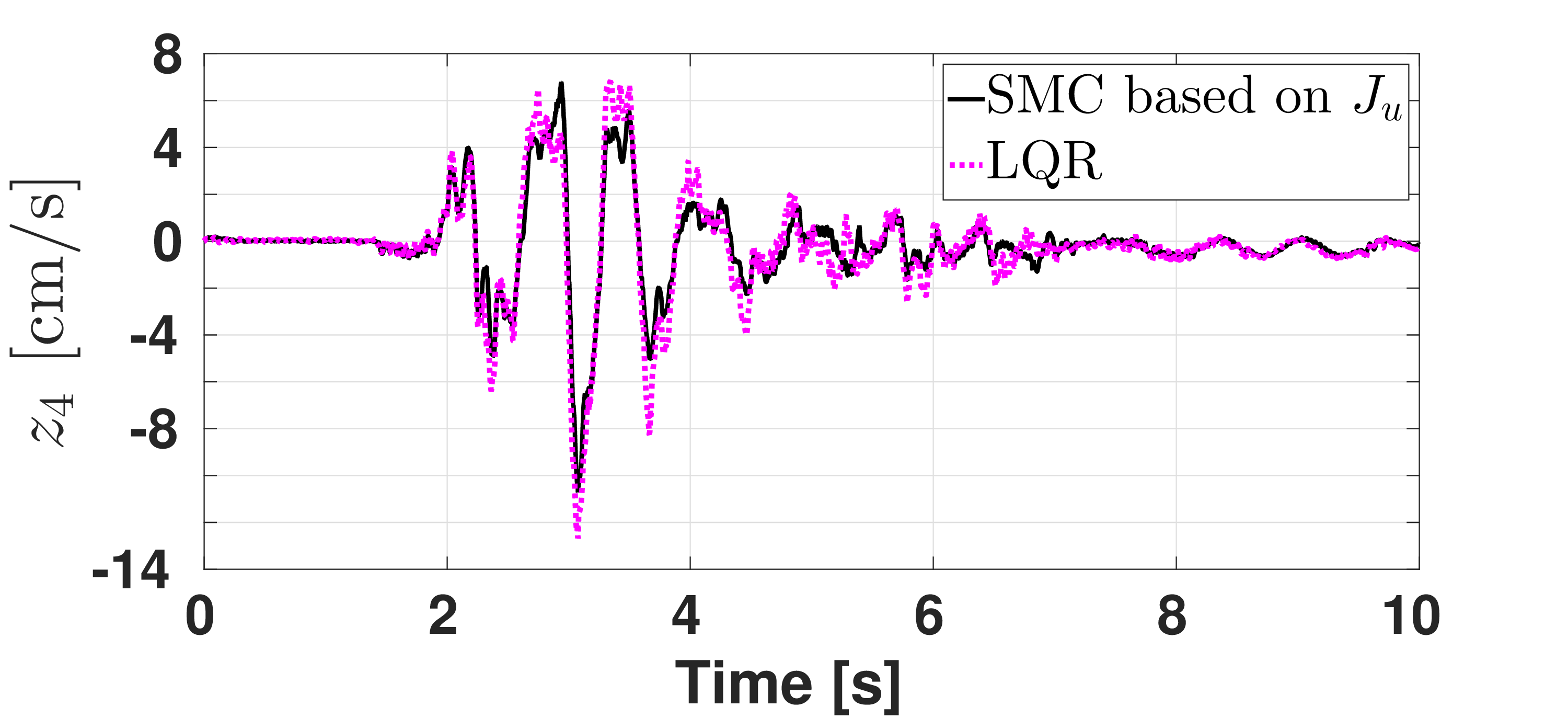}}	
\caption{Velocities of the floor.}
\label{fig:12}
\end{figure}          
\end{center}

\begin{center}
\begin{figure}[H]
\centering
\subfigure[Signals $u(t)$.]
	            {\includegraphics[height=3.7cm]{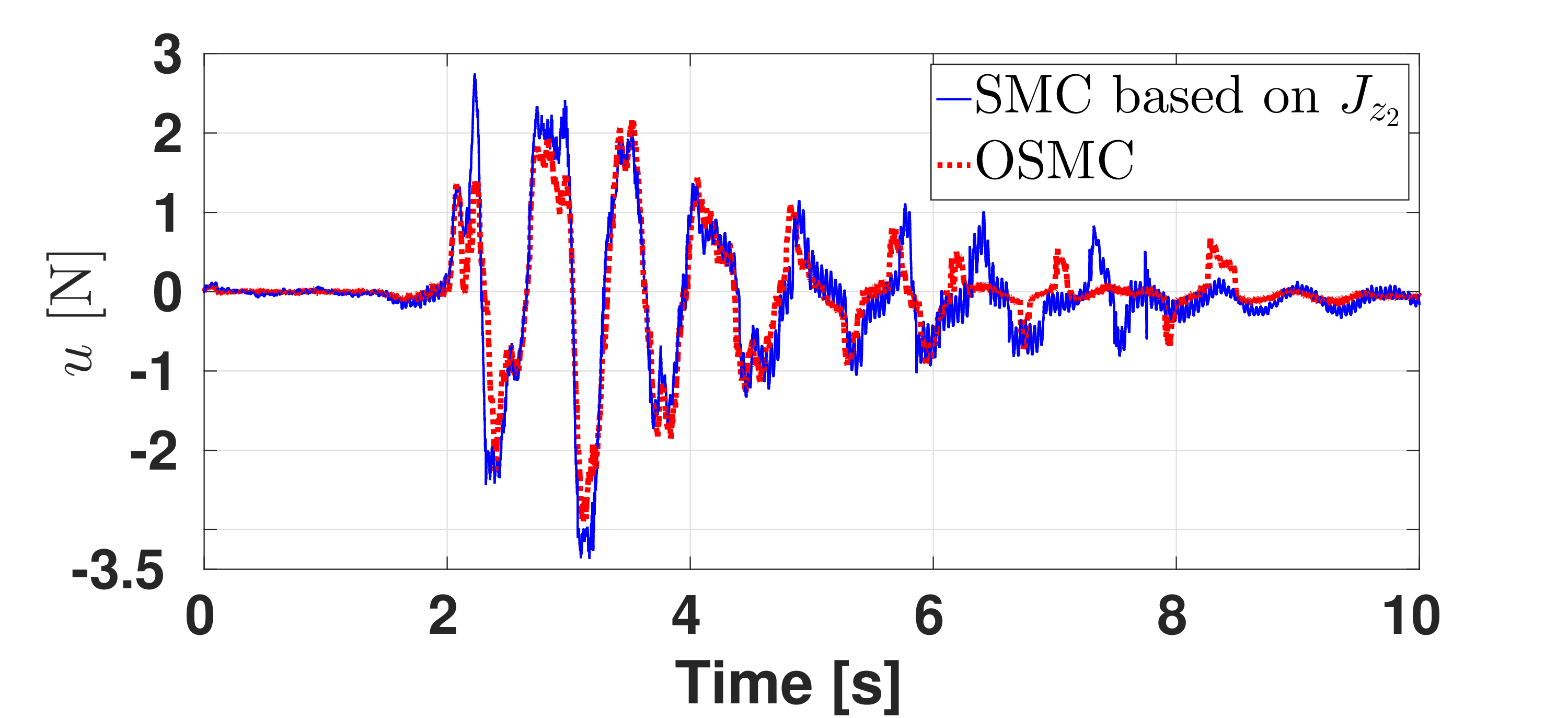}}
\subfigure[Signals $u(t)$.]
	            {\includegraphics[height=3.7cm]{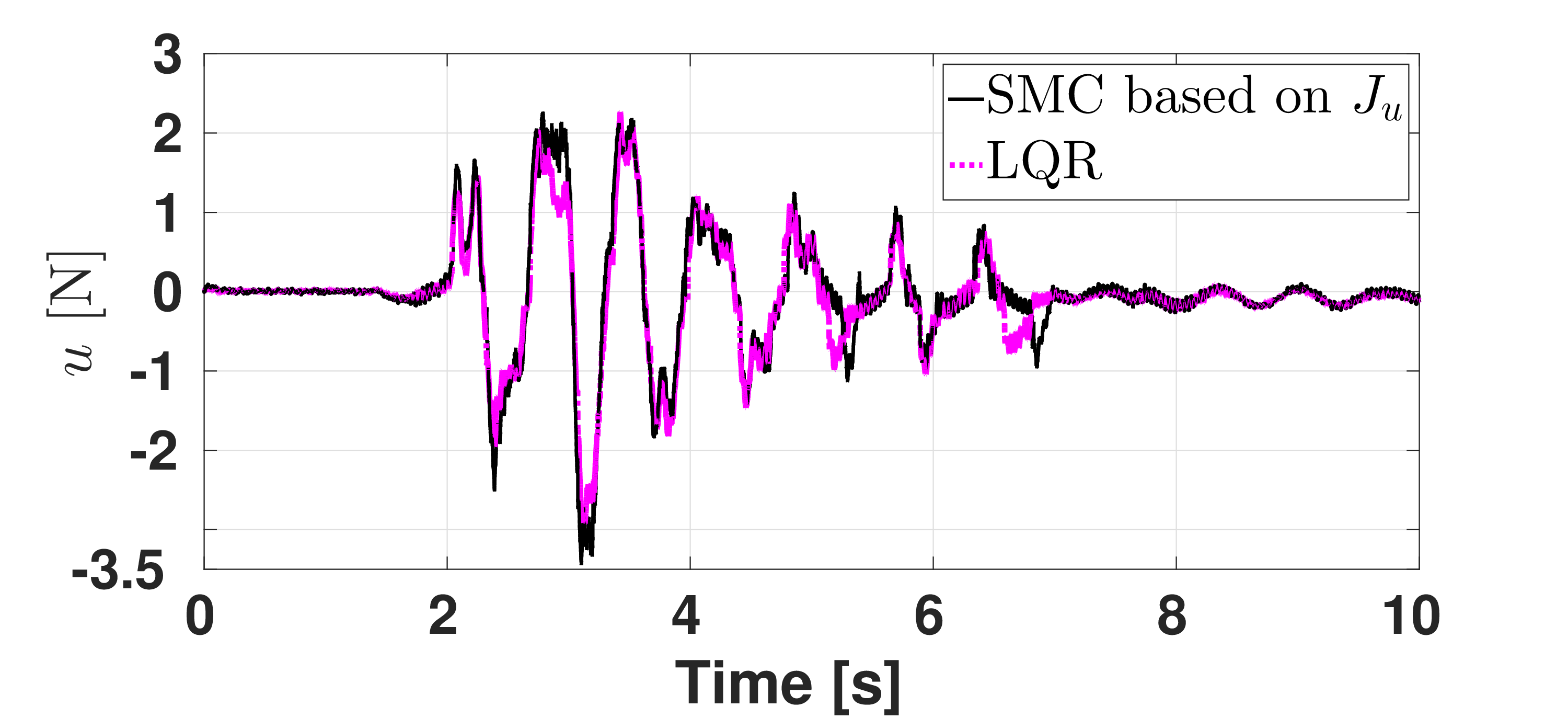}}	
\caption{Control force generated by the controllers.}
\label{fig:13}
\end{figure}          
\end{center}

Figure \ref{fig:7a} depicts the uncontrolled response $z_{2}(t)$ of the experimental structure under the earthquake excitation, which indicates that its maximum magnitude is approximately \SI{3.5}{\cm}. Figures \ref{fig:9}$-$\ref{fig:13} show, respectively, the signals $z_{1}(t)$, $z_{2}(t)$, $z_{3}(t)$, $z_{4}(t)$, and $u(t)$ produced by the LQR, OSMC, and the tuned SMC. From these plots, it is possible to see that the maximum displacement $z_{2}(t)$ corresponding to the SMC, based on either PI $J_{z_{2}}$ or $J_{u}$, is smaller than that produced by the LQR and OSMC. Moreover, the peak value of the signals $u(t)$ and $z_{1}(t)$ corresponding to the SMC is slightly larger than those generated by the LQR and OSMC, but these signals are within the specified limits. 

In addition, Table \ref{tab:rms_val} presents the RMS and peak values of the signals $u(t)$ and $z_{i}$, $i=1,2,3,4$ corresponding to the SMC, LQR and OSMC, during the period of $t=2$ to $t=\SI{6}{\s}$. Note that the relations $\kappa_{i}/z_{i}^{\mathrm{rms}}$, $i=1,2,3,u$, are between 2.15 to 4.54. This table indicates that all the controllers allow attenuating more than three times the peak value of the uncontrolled response of $z_{2}$. Moreover, the results obtained with the LQR and the OSMC are similar to each other. This table also shows that the SMC that minimizes the PI $J_{z_{2}}$ provides the largest attenuation percentages $\mathcal{R}(z_{2}^{\mathrm{rms}})$ and $\mathcal{R}(z_{2}^{\mathrm{peak}})$ of $z_{2}$, as expected. On the other side, these percentages corresponding to the SMC based on the PI $J_{u}$, are larger than those calculated for the LQR and OSMC. In addition, the control efforts $u^{\mathrm{rms}}$ for SMC-$J_{u}$, LQR-$J_{\mathrm{LQR}}$, and OSMC-$J_{\mathrm{OSMC}}$ are similar. These results show that the SMC based on the Ackermann's formula and either of the PIs $J_{z_{2}}$ or $J_{u}$ has great potential for active structural control.

\begin{table*}[ht]
\centering
\caption{RMS and peak values of $z_{i}$, $i=1,2,3,4$ and $u(t)$.}
\label{tab:rms_val}
\begin{adjustbox}{width=1\textwidth}
\footnotesize
\begin{tabular}{c c c c c c c c c c c c c c} \toprule
	\multicolumn{1}{c}{\multirow{2}{*}{Controller}} & \multicolumn{1}{c}{\multirow{2}{*}{PI}}& $z_{1}^{\mathrm{rms}}$ & $z_{1}^{\mathrm{peak}}$ & $z_{2}^{\mathrm{rms}}$& $z_{2}^{\mathrm{peak}}$ & $z_{3}^{\mathrm{rms}}$& $z_{3}^{\mathrm{peak}}$ & $z_{4}^{\mathrm{rms}}$& $z_{4}^{\mathrm{peak}}$ & $u^{\mathrm{rms}}$ & $u^{\mathrm{peak}}$ & $\mathcal{R}(z_{2}^{\mathrm{rms}})$ & $\mathcal{R}(z_{2}^{\mathrm{peak}})$\\
	\cmidrule(lr){3-4} \cmidrule(lr){5-6} \cmidrule(lr){7-8} \cmidrule(lr){9-10} \cmidrule(lr){11-12} \cmidrule(lr){13-14}
 &  &\multicolumn{2}{c}{(cm)} & \multicolumn{2}{c}{(mm)}  & \multicolumn{2}{c}{(cm/s)}   &  \multicolumn{2}{c}{(cm/s)}  & \multicolumn{2}{c}{(N)} & \multicolumn{2}{c}{(\%)} \\ \midrule
  
      Uncontrol   &---  & ---  & ---   & 20.14  & 35.31  & ---  & --- & 17.90   & 34.54  & 0  & 0 &  0 & 0\\
	  SMC  & $J_{z_{2}}$& 0.84 & 2.56 & 2.62  & 7.39 & 9.59& 29.06& 2.52 & 8.49 &1.25 & 3.36 &  86.99 & 79.07\\ 
	  SMC  & $J_{u}$& 0.77 & 2.32 & 3.21  & 9.73 & 9.39 & 29.95 & 2.89 & 10.71 & 1.18 & 3.43 &  84.06 & 72.44 \\ 
      LQR  & $J_{\mathrm{LQR}}$& 0.74 & 2.17 & 4.21  & 11.80 & 8.75& 26.62& 3.50 & 12.68 &1.05 & 2.92  & 79.10 & 66.58\\
      OSMC  & $J_{\mathrm{OSMC}}$& 0.72 & 1.99 & 4.22  & 11.22 & 8.87& 24.18& 3.50 & 13.66 &1.05 & 2.90 & 79.05 & 68.22\\
        \bottomrule
\end{tabular}
\end{adjustbox}
\end{table*}

\section{Conclusions}
\label{sec:conclusions}
This paper proposed a tuning algorithm for a SMC designed for attenuating the structural vibration of a seismically excited building equipped with an ATMD on its top floor. The SMC is based on the Ackermann's formula and its robustness against friction uncertainty was demonstrated. It was proved that the seismic excitation signal is not a coupled disturbance, and as consequence it cannot be eliminated by the SMC; however, its effect can be minimized by designing the SMC with the proposed tuning algorithm.

It was also shown that the responses of the ATMD and building under the SMC can be described by means of dominant second-order filters, whose input is the seismic excitation. Their parameters were automatically tuned by the proposed algorithm in order to minimize the PI $J_{z_{2}}$ or $J_{u}$. The PI $J_{z_{2}}$ is related to the minimization of the top floor displacement, while the PI $J_{u}$ is related to the minimization of the control force applied the ATMD while offering a great attenuation of this displacement. The effectiveness of the designed SMC was demonstrated in a numerical and an experimental structure. Moreover, the experiments showed that the designed SMC has a better vibration attenuation performance than the LQR and OSMC, while ensuring that the structure and the ATMD signals are within the specified limits.

\section*{Acknowledgments}
The authors thank the Programa para el Desarrollo Profesional Docente (PRODEP-SEP) of Mexico for supporting this research. The authors also acknowledge the anonymous reviewers for their helpful comments.

\bibliographystyle{elsarticle-num} 
 \bibliography{paper}

\end{document}